\documentclass[amsmath, amssymb, aps, prl, superscriptaddress, twocolumn, floatfix, 10pt, showpacs, nofootinbib, reprint]{revtex4-1}
\usepackage[utf8]{inputenc}

\usepackage{graphicx,color,dsfont,xcolor}
\usepackage{gensymb}
\usepackage{hyperref}
\usepackage[caption=false]{subfig} 
\usepackage[normalem]{ulem}
\usepackage[titletoc,toc,title]{appendix}

\usepackage{makecell}
\usepackage{multirow}

\usepackage{pifont}
\usepackage{amsthm}
\usepackage{framed}

\newtheorem*{conjecture}{Conjecture}

\newcommand{\refcite}[1]{Ref.\,\onlinecite{#1}}

\newcommand{\eqnref}[1]{Eq.\,(\ref{#1})}

\newcommand{\figref}[1]{Fig.\,\ref{#1}}
\newcommand{\sfigref}[2]{Fig.\,\hyperref[#1]{\ref{#1}(#2)}}

\newcommand{\secref}[1]{Sec.\,\ref{#1}}
\newcommand{\appref}[1]{Appendix\,\ref{#1}}

\definecolor{sea}{RGB}{46,139,87}

\definecolor{darkblue}{RGB}{0,0,127}

\definecolor{kspink}{RGB}{200,0,200}

\newcommand{\bra}[1]{\langle #1 |}
\newcommand{\ket}[1]{|#1\rangle}

\newcommand{\beq}{\begin{equation}}
\newcommand{\eeq}{\end{equation}}
\newcommand{\mP}{\mathcal{P}}

\newcommand{\bpm}{\begin{pmatrix}}
\newcommand{\epm}{\end{pmatrix}}
\newcommand{\bmm}{\begin{matrix}}
\newcommand{\emm}{\end{matrix}}

\newcommand{\nocontentsline}[3]{}
\newcommand{\tocless}[2]{\bgroup\let\addcontentsline=\nocontentsline#1{#2}\egroup}

\usepackage[all]{xy}
\newcommand{\onecell}[4]{%
    \xymatrix@!0{%
 &   &\ar@{-}[dddd]&&\\%
&#1&&#2&           \\%
\ar@{-}[rrrr]&&&&           \\%
&#4&&#3 &      \\%
&&&&
    }
    }
    
\usepackage[all]{xy}
\newcommand{\drawgenerator}[8]{%
\xymatrix@!0{%
& #8 \ar@{-}[ld]\ar@{.}[dd] \ar@{-}[rr] & & #7 \ar@{-}[ld]  \\%
#1 \ar@{-}[rr] \ar@{-}[dd] &  & #2 \ar@{-}[dd] &            \\%
& #6 \ar@{.}[ld] &  & #5 \ar@{-}[uu] \ar@{.}[ll]       \\%
#3 \ar@{-}[rr] &  & #4 \ar@{-}[ru]                       %
}%
}

\hypersetup{
  colorlinks = true,
  urlcolor = blue,
  pdfauthor = {David Aasen, Baniel Bulmash, Abhinav Prem, Kevin Slagle, Dominic J. Williamson},
  pdftitle = {Topological Defect Networks for Fractons of all Types}
}

\newcommand{\cupcapprime}{\mathord{\vcenter{\hbox{\includegraphics[scale=.5]{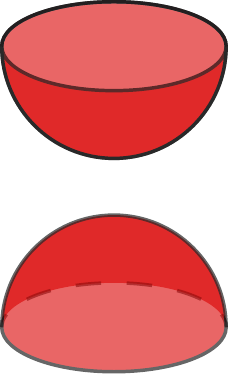}}}}}
\newcommand{\cylinderprime}{\mathord{\vcenter{\hbox{\includegraphics[scale=.5]{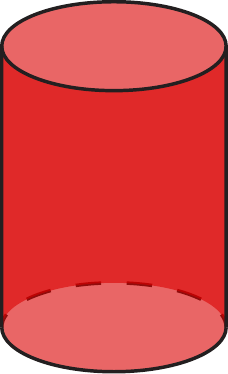}}}}}

\newcommand{\twocellnetbprime}{\mathord{\vcenter{\hbox{\includegraphics[scale=.5]{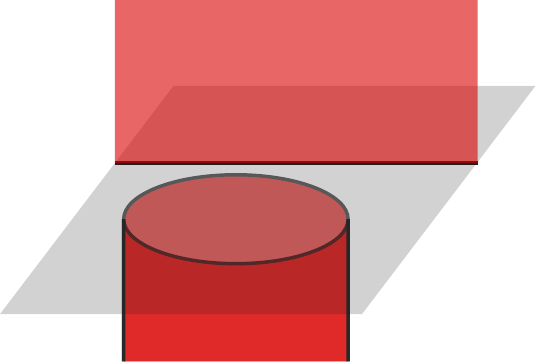}}}}}

\newcommand{\twocellnetaprime}{\mathord{\vcenter{\hbox{\includegraphics[scale=.5]{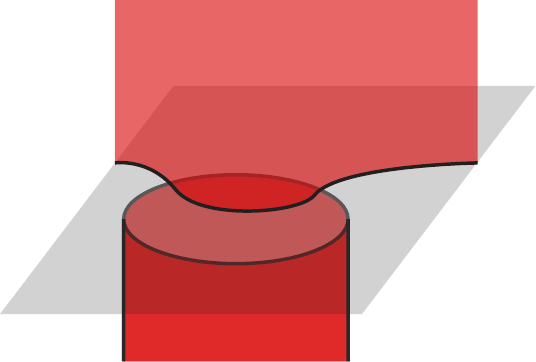}}}}}

\newcommand{\cubeone}{\mathord{\vcenter{\hbox{\includegraphics[scale=1]{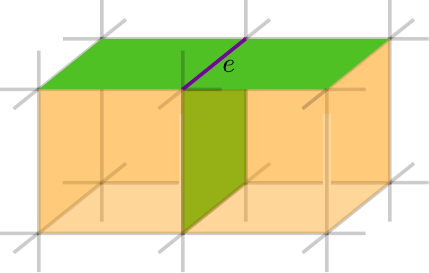}}}}}
\newcommand{\cubetwo}{\mathord{\vcenter{\hbox{\includegraphics[scale=1]{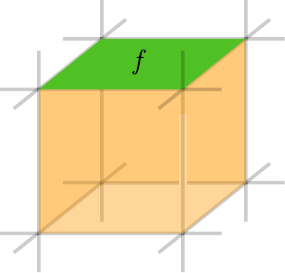}}}}}

\newcommand{\cubetwosmall}{\mathord{\vcenter{\hbox{\includegraphics[scale=.6]{cube2.pdf}}}}}

\newcommand{\cubethree}{\mathord{\vcenter{\hbox{\includegraphics[scale=1]{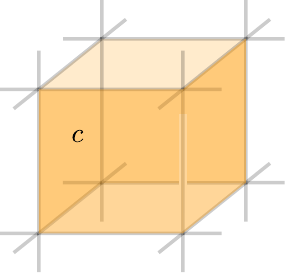}}}}}
\newcommand{\edgethree}{\mathord{\vcenter{\hbox{\includegraphics[scale=1]{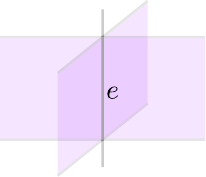}}}}}
\newcommand{\vertexzero}{\mathord{\vcenter{\hbox{\includegraphics[scale=1]{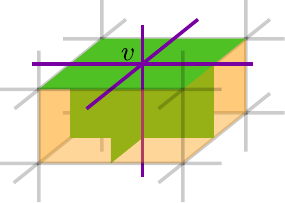}}}}}
\newcommand{\vertexone}{\mathord{\vcenter{\hbox{\includegraphics[scale=1]{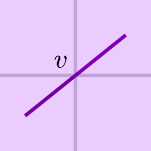}}}}}

\newcommand{\cubethreeblue}{\mathord{\vcenter{\hbox{\includegraphics[scale=1]{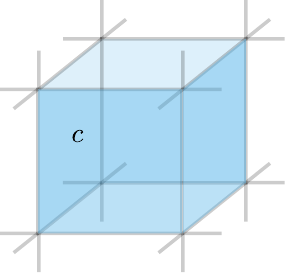}}}}}

\newcommand{\cubetwobluesmall}{\mathord{\vcenter{\hbox{\includegraphics[scale=.6]{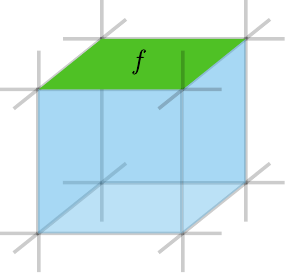}}}}}

\newcommand{\edgethreegrey}{\mathord{\vcenter{\hbox{\includegraphics[scale=1]{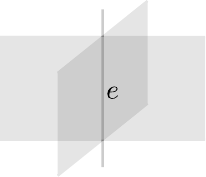}}}}}
\newcommand{\BCodecubeoneone}{\mathord{\vcenter{\hbox{\includegraphics[width=0.20\columnwidth]{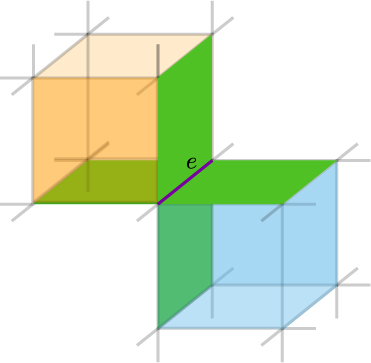}}}}}
\newcommand{\BCodecubeonetwo}{\mathord{\vcenter{\hbox{\includegraphics[width=0.20\columnwidth]{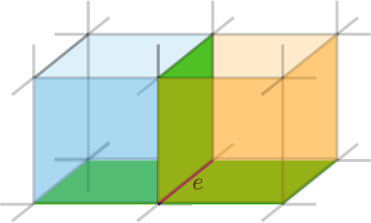}}}}}
\newcommand{\BCodecubeonethree}{\mathord{\vcenter{\hbox{\includegraphics[width=0.20\columnwidth]{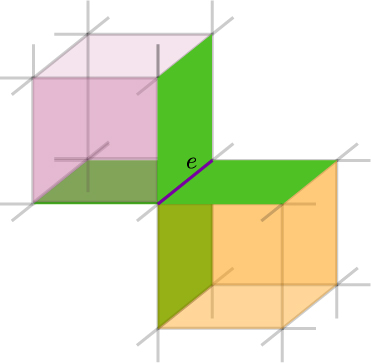}}}}}
\newcommand{\BCodecubeonefour}{\mathord{\vcenter{\hbox{\includegraphics[width=0.15\columnwidth]{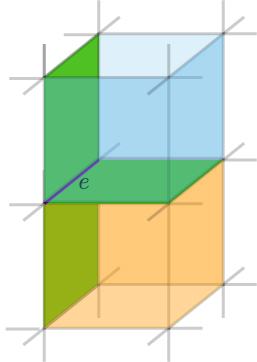}}}}}
\newcommand{\BCodecubeonefive}{\mathord{\vcenter{\hbox{\includegraphics[width=0.15\columnwidth]{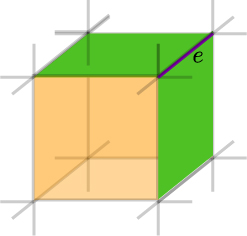}}}}}
\newcommand{\BCodecubeonesix}{\mathord{\vcenter{\hbox{\includegraphics[width=0.15\columnwidth]{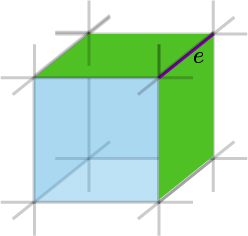}}}}}
\newcommand{\BCodevertexoneone}{\mathord{\vcenter{\hbox{\includegraphics[width=0.12\columnwidth]{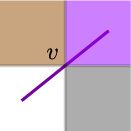}}}}}
\newcommand{\BCodevertexonetwo}{\mathord{\vcenter{\hbox{\includegraphics[width=0.12\columnwidth]{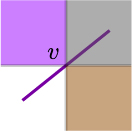}}}}}
\newcommand{\BCodevertexzeroone}{\mathord{\vcenter{\hbox{\includegraphics[width=0.15\columnwidth]{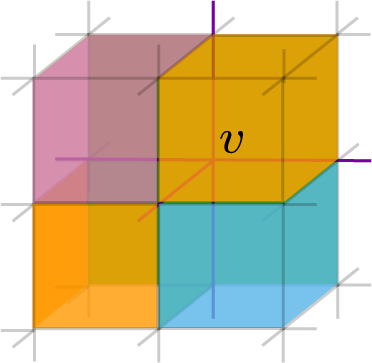}}}}}
\newcommand{\BCodevertexzerotwo}{\mathord{\vcenter{\hbox{\includegraphics[width=0.15\columnwidth]{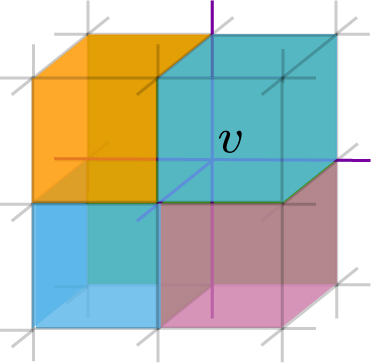}}}}}
\newcommand{\BCodevertexzerothree}{\mathord{\vcenter{\hbox{\includegraphics[width=0.15\columnwidth]{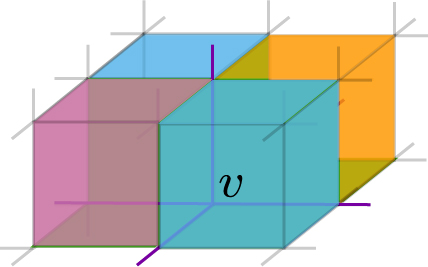}}}}}
\newcommand{\BCodevertexzerofour}{\mathord{\vcenter{\hbox{\includegraphics[width=0.15\columnwidth]{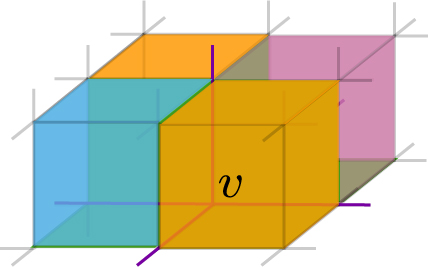}}}}}
\newcommand{\BCodevertexzerofive}{\mathord{\vcenter{\hbox{\includegraphics[width=0.15\columnwidth]{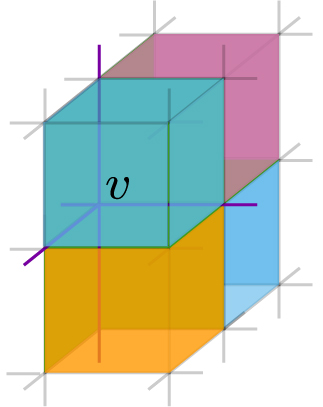}}}}}
\newcommand{\BCodevertexzerosix}{\mathord{\vcenter{\hbox{\includegraphics[width=0.15\columnwidth]{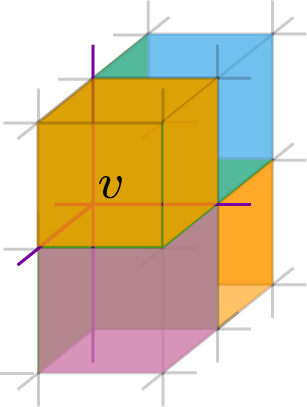}}}}}
\newcommand{\BCodevertexzeroseven}{\mathord{\vcenter{\hbox{\includegraphics[width=0.15\columnwidth]{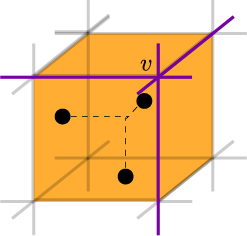}}}}}
\newcommand{\BCodevertexzeroeight}{\mathord{\vcenter{\hbox{\includegraphics[width=0.15\columnwidth]{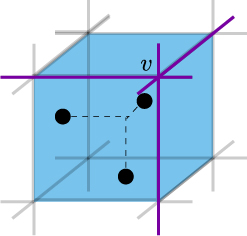}}}}}

\makeatletter
\let\cat@comma@active\@empty
\makeatother

\makeatletter
\def\l@subsubsection#1#2{}
\makeatother

\begin{document}

\title{
Topological Defect Networks for Fractons of all Types
}

\date{\today}

\author{David Aasen}
\affiliation{Microsoft Quantum, Microsoft Station Q, University of California, Santa Barbara, California 93106-6105 USA}
\affiliation{Kavli Institute for Theoretical Physics, University of California, Santa Barbara, California 93106, USA}
\author{Daniel Bulmash}
\affiliation{Condensed Matter Theory Center and Joint Quantum Institute, Department of Physics, University of Maryland, College Park, Maryland 20472 USA}
\author{Abhinav Prem}
\affiliation{Princeton Center for Theoretical Science, Princeton University, NJ 08544, USA}
\author{Kevin Slagle}
\affiliation{Department of Physics and Institute for Quantum Information and Matter, California Institute of Technology, Pasadena, California 91125, USA}
\affiliation{Walter Burke Institute for Theoretical Physics,
California Institute of Technology, Pasadena, California 91125, USA}
\author{Dominic J. Williamson}
\affiliation{Department of Physics, Yale University, New Haven, CT 06511-8499, USA}
\affiliation{Stanford Institute for Theoretical Physics, Stanford University, Stanford, CA 94305, USA}

\begin{abstract} 

Fracton phases exhibit striking behavior which appears to render them beyond the standard topological quantum field theory (TQFT) paradigm for classifying gapped quantum matter.
Here, we explore fracton phases from the perspective of \textit{defect} TQFTs and show that \textit{topological defect networks}---networks of topological defects embedded in stratified 3+1D TQFTs---provide a unified framework for describing various types of gapped fracton phases. In this picture, the sub-dimensional excitations characteristic of fractonic matter are a consequence of mobility restrictions imposed by the defect network. We conjecture that \textit{all} gapped phases, including fracton phases, admit a topological defect network description and support this claim by explicitly providing such a construction for many well-known fracton models, including the X-Cube and Haah's B code. To highlight the generality of our framework, we also provide a defect network construction of a novel fracton phase hosting non-Abelian fractons. As a byproduct of this construction, we obtain a generalized membrane-net description for fractonic ground states as well as an argument that our conjecture implies no type-II topological fracton phases exist in 2+1D gapped systems. Our work also sheds light on new techniques for constructing higher order gapped boundaries of 3+1D TQFTs.
\end{abstract}

\maketitle

{\hypersetup{linkcolor=black} \tableofcontents}


\newpage

\section{Introduction}
\label{sec:intro}

At first a singularly peculiar model displaying behavior vastly different from that expected of well-behaved quantum phases, Haah's code~\cite{haah} by now represents perhaps the best known example of \textit{fractonic matter}: an entire family of 
renegade quantum phases which resist fitting neatly into existing paradigms for classifying quantum matter.
In the relatively short period since receiving the moniker ``fracton," these phases have been subject to intense scrutiny, spurred in part by the discovery of other exactly solvable models exhibiting behavior similar to that of Haah's code\footnote{For the historically inclined, we note that while Chamon's code~\cite{chamon} appeared in the literature earlier than Haah's code, it was only later that the former's fractonic nature was appreciated.}\cite{chamon,castelnovo,kimqupit,yoshida,haah2014bifurcation,fracton1,fracton2}. Chief amongst the features shared by these models is the striking presence of topological excitations with restricted mobility, such as the eponymous fracton, which is strictly immobile in isolation, or subdimensional excitations, which are only mobile along lower dimensional submanifolds.

Although initially of interest for their potential as self-correcting quantum memories~\cite{bravyi,haah2,bravyihaah}, three dimensional (3+1D) gapped fracton models have recently been shown to harbor intriguing connections with topological order~\cite{han,sagar,nonabelian,shirleyfrac,twisted,bulmashboundary,Dua2019}, slow quantum dynamics~\cite{chamon,castelnovo,kimhaah,prem}, subsystem symmetries~\cite{fracton2,williamson,yizhi1,shirleygauging,albertgauge,juvenSSPT,Ibieta,TantivasadakarnSearching}, and quantum information processing~\cite{Brown2019}. Fractonic matter, however, is defined more broadly as including any (not necessarily gapped) quantum phase of matter with restricted mobility excitations (that are not necessarily topological), examples of which by now abound. Prominent amongst these are tensor gauge theories with higher moment conservation laws~\cite{sub,genem,SeibergSymmetry,chiral,emergent,han3,bulmash,bulmash2,williamsonSET,yanholography,wang2019embeddon,wang2019nonabelian}, some of which are dual to familiar theories of elasticity~\cite{leomichael,gromov,pai,potter,RadzihovskyHermele} and the description in terms of which may point towards material realizations of fractonic spin liquids~\cite{pinch,YanPyrochlore,SousPolarons,hughes2020lsm,he2019lsm,FujiLayer,slagle1,balents}. In 1+1D, systems with conserved dipole moment have emerged as platforms for studying the constrained dynamics typically associated with fractonic models~\cite{PaiLocalization,SalaErgodicity,moudgalya2019} and appear to be within experimental reach. We refer the reader to Refs.~[\onlinecite{fractonreview,fractonreview2}] for a comprehensive and current review of fractonic physics.

Fracton models present a novel challenge to the classification of quantum matter, a scheme which has otherwise largely succeeded for topological phases of matter~\cite{wenreview}. This is especially true for topological orders in 2+1D (without any global symmetries), whose classification in terms of modular tensor categories~\cite{Moore1989,kitaev} is widely accepted to be complete. Progress in this direction was aided in part by families of exactly solvable models~\cite{kitaev,Levin2005} which encapsulate the universal features of long-range entangled phases and provide a general framework for studying fractionalized excitations. Likewise, the classification of 3+1D phases admitting a topological quantum field theory (TQFT) description has witnessed ongoing success~\cite{wen2018prx,wen2019prb}, facilitated again by solvable lattice Hamiltonians~\cite{walker2012,wan2015twisted,williamson2017ham}. 

A large class of 3+1D gapped fracton orders can also be described by commuting projector Hamiltonians, whose amenability to exact analytic study has exposed many of their universal features. Much like familiar topologically ordered systems, fracton orders have a gap to all
excitations, support topologically charged excitations which cannot be created locally, and possess long-range entangled ground states~\cite{regnault2,han2,albert,spurious,albertespec,dua2019bifur,ShiEntanglement}. Crucially, however, the number of locally indistinguishable ground states in these models grows exponentially with system size in many cases. It is precisely this sensitivity to the system size, and more generally to the ambient geometry~\cite{SlagleXcubeQFT,slagle3,shirleygeneral,cagenet,symmetric,yanholography,SlagleSMN,gromov2,tian2018}, that ostensibly renders fracton phases ``beyond" the conventional TQFT framework. Although this by no means implies that lessons learned from studying TQFTs do not carry over to fractons, attempts at characterizing fracton phases have led to fundamentally new ideas, including the notion of ``foliated" fracton phases~\cite{chenfoliatedent,foliatedcb,shirleyfrac,shirleygauging,TwistedFoliated} and of statistical processes involving immobile excitations, which do not braid in obvious ways~\cite{twisted,cagenet,bulmashboundary,hermelefusion}. And yet, despite remarkable progress in understanding fracton order, a unified picture akin to the categorical description of TQFTs has thus far proven elusive.  

Besides the fact that gapped fracton phases appear to transcend TQFTs, any underlying mathematical framework is further obscured by their evolving typology; even in the restricted setting of translation invariant commuting projector Hamiltonians, there are a plethora of known examples which fall under the fractonic umbrella but differ in significant ways. Broadly, these models have been classified into type-I and type-II phases in the taxonomy of Ref.~[\onlinecite{fracton2}], with the X-Cube model~\cite{fracton2} and Haah's A~\cite{haah} and B~\cite{haah2014bifurcation} code as representative examples. Unlike type-I models, which host fractons as well as (partially mobile) subdimensional excitations,\footnote{We will refer to subdimensional excitations that can only move along lines (planes) as lineons (planons).} type-II models are distinguished by their lack of \textit{any} string-like operators or any topologically nontrivial mobile particles. This coarse typology has been extended to include fractal type-I models~\cite{yoshida}, which have both fractal and string operators, and the more exotic panoptic type models~\cite{bulmashgauging,premgauging}, which also host fully mobile excitations in addition to subdimensional ones. Finally, while both type-I and panoptic type phases can host non-Abelian subdimensional particles~\cite{nonabelian,twisted,cagenet,bulmashgauging,premgauging,GeneralizedSMN}, it remains unclear whether non-Abelian type-II models exist. 

Despite this breadth of phenomenology, there have been many attempts at taming the fracton zoo, all with varying levels of partial success. Abelian models of all types have in particular been understood from several perspectives.\footnote{As of this writing, we count at least eight different ways of looking at the X-Cube model.} Foremost amongst these is their realization as stabilizer codes, whose classification is complete in 2+1D~\cite{Haah2018b}, remains ongoing in 3+1D~\cite{Dua2019,DuaSorting}, and which has led to key insights regarding the entanglement structure of fracton models~\cite{haah2014bifurcation,han2,albert,chenfoliatedent,spurious,albertespec,dua2019bifur}. Abelian fracton models are also known to be dual to subsystem symmetry protected topological (SSPT) phases~\cite{williamson,fracton2,yizhi1,shirleygauging,strongsspt,albertgauge} (which have been partially classified~\cite{strongsspt,juvenSSPT}) and can additionally be obtained as a result of ``Higgsing" generalized U(1) symmetries~\cite{han3,bulmash,bulmash2,gromov2}. The notion of ``foliated fracton order, and the more general ``bifurcated equivalence," further provide a natural scheme for organizing these models into inequivalent classes and have been successful at sorting Abelian fracton models~\cite{chenfoliatedent,foliatedcb,shirleygauging,shirleyfrac,TwistedFoliated,dua2019bifur}. None of the above ideas, however, have been shown to apply to non-Abelian fracton models; instead, type-I Abelian and non-Abelian models can be understood as a result of ``p-string" condensation~\cite{han,sagar,nonabelian,cagenet}, which drives layers of strongly coupled topological orders into a fracton phase. While this mechanism does not produce any non-Abelian fractons (only non-Abelian lineons), twisting the gauge symmetry of type-I models~\cite{twisted} or gauging the permutation symmetry between copies of type-I or type-II models~\cite{bulmashgauging,premgauging,wang2019nonabelian}, can. Perhaps unsurprisingly, all three of these mechanisms have yet to produce a type-II model.\footnote{For those steeped in the fracton literature, we note that although the panoptic type models host immobile non-Abelian excitations created at the ends of fractal operators, they are not, strictly speaking, of type-II. This is due to the presence of additional fully mobile particles which do not appear in e.g., Haah's code.}

All this to say that it has proven deceptively difficult to unify even the relatively simple-seeming class of translation-invariant, exactly solvable, gapped fracton models. This, then, raises the following natural questions: does there exist a \textit{unified} framework which captures \textit{all} types of gapped fracton phases, and if so, does this framework fit within the existing TQFT landscape?\footnote{Arguments that fractons are ``beyond" TQFT due to their geometric sensitivity beg the question of whether or not TQFTs can be suitably modified to accommodate such behavior.} In this paper, we answer both in the affirmative. Rather than abandoning the TQFT framework, we instead espouse the idea of seeking out modifications to TQFTs which are sensitive to geometry in some fundamental way.
Drawing inspiration from the field of defect TQFTs~\cite{morrison2010,kapustin2010,dkm2011,carqueville2016,carqueville2017,cms2016}, as well as from the recent classification of crystalline SPTs~\cite{song2017defect,huang2017defect,else2019crystalline}, we show that \textit{topological defect networks} are a unified framework for describing all types of gapped fracton phases. 
 
\tocless{\subsection*{Topological Defect Networks} \label{sec:tdn}}

Before introducing the concept of topological defect networks, we briefly review defect TQFTs~\cite{morrison2010,kapustin2010,dkm2011,carqueville2016,carqueville2017,cms2016}, familiar examples of which include topological orders with gapped boundaries~\cite{bravyi1998,BombinDualityDefect,bombin2008,kitaevkong,beigi2011bdry,barkeshli2014symmetry,cong2016gapped,yoshida2017gapped,wang2018gapped} (more mathematically oriented readers are referred to Ref.~[\onlinecite{carqueville2017}].)

A ``topological defect" embedded in a $D$+1 dimensional TQFT corresponds to introducing new interactions (and possibly new degrees of freedom), which are spatially localized on some lower $d<D$ dimensional region, into the system without closing the bulk gap. For a Hamiltonian, this corresponds to modifying its terms near the region specified by the defect while maintaining the energy gap. Consequently, the behavior of bulk topological excitations is modified in the vicinity of the defect. For instance, some bulk excitations may condense \textit{i.e.,} become identified with the vacuum sector, on the defect; or, the defect could nontrivially permute the topological superselection sectors of excitations passing through it. Both kinds of defects---anyon condensing and anyon permuting---have been introduced into the 2+1D toric code Hamiltonian, with the latter of particular interest for its potential applications in topological quantum computation~\cite{bravyi1998,BombinDualityDefect}. 

An $n$+1 dimensional defect TQFT, then, is simply a TQFT with topological defects. More precisely, a defect TQFT is defined in terms of its defect data $\mathbb{D}$, which consists of defect label sets $D_j$ and a set of maps $\mathcal{D}$ between them~\cite{carqueville2016}. Elements of $D_j$ label the $j$-dimensional defects ($j \in [0,n]$) while maps in $\mathcal{D}$ specify how defects of different dimensions are allowed to meet. For instance, these maps encode the allowed $j=1$ dimensional gapped boundaries (or domain walls) between two $n$=2 dimensional regions. A defect TQFT thus naturally defines a hierarchical structure in which lists of elements of $D_{j-1}$ mediate between elements of $D_j$ through the maps in $\mathcal{D}$. Heuristically, one can think of the defect data as the set of distinct topological superselection sectors on each defect and relations between these sets.

\begin{figure}[t]
    \centering
        \includegraphics[width=\columnwidth]{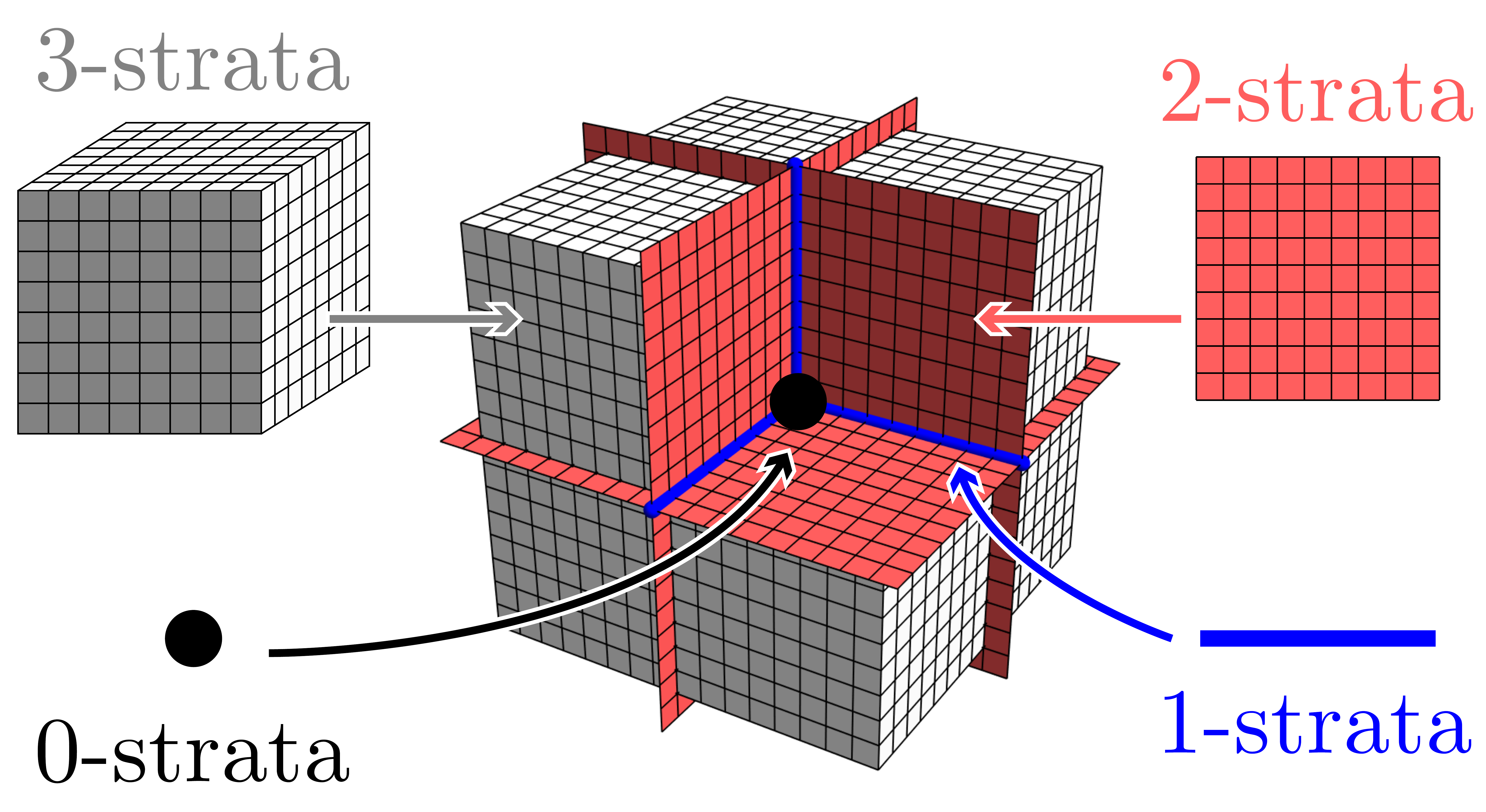}
    \caption{
    A stratification of 3d space into 3-strata (white regions),
      2-strata (red squares), 1-strata (blue lines), and 0-strata (black vertex). A 3+1D TQFT lives on each 3-strata, and they are coupled together along the lower-dimensional 2- and 1-strata defects.
    This defect TQFT could be described by a lattice model with qubits on, e.g., the black cubic lattice.
    }\label{fig:3DTClattice}
\end{figure}

A 3+1D \textit{topological defect network} is a particular instance of a defect TQFT that lives on a ``stratified" 3-manifold $\mathcal{M}$. A stratification consists of collections $S_j$ ($j \in [0,3]$) of $j$-dimensional submanifolds that decompose the manifold, as in Fig.~\ref{fig:3DTClattice}. Elements of $S_j$ are referred to as $j$-strata. We assign a 3+1D TQFT to each 3-strata and associate a topological defect with each $j$-stratum (for $j<3$), thereby coupling together the ambient TQFTs. This coupling, mediated by the defect network, underpins the flexibility of topological defect networks as it directly imposes the mobility constraints characteristic of fracton models. In particular, this coupling dictates the set of topological excitations that condense on a given $j$-strata and hence, along with the braiding data already encoded in the 3+1D TQFT, determines the set of excitations which cannot pass through that strata---any excitations which braid non-trivially with any of the condensed excitations on a defect are prohibited from passing through it.

The Hamiltonian for a topological defect network can be written schematically as 
\beq
H = \sum_{j=0}^3 H_j \, ,
\eeq
where $H_j$ is associated with terms in the Hamiltonian acting on degrees of freedom localized near a $j$-strata. Although we restrict our attention to cubic lattices in this work, topological defect networks are defined for arbitrary stratifications of manifolds; likewise, while we only consider translation invariant models here, we see no obstruction to defining topological defect networks even outside this context. 

While in principle subsumed under defect TQFTs, topological defect networks nevertheless provide a novel framework as they are comprised of an \textit{extensive} network of topological defects enmeshed in a TQFT, and, as we show in this paper, are capable of describing gapped fracton phases. We remark also that our construction differs from earlier layered constructions of fracton models~\cite{han,sagar,nonabelian,cagenet,slagle1}, where layers of coupled 2+1D topological orders were driven into a fracton phase through a condensation transition and the effective Hamiltonian was determined perturbatively. In the defect network picture, no such transition is required since the fracton phase is entirely determined by the choice of 3+1D TQFT and the network of defects it is immersed in. Thus, a significant virtue of this framework is that it allows one to write down \textit{exact} commuting projector Hamiltonians (assuming the strata are non-chiral), thereby providing a constructive approach to discovering new fracton models.

\begin{figure}[t]
    \centering
        \includegraphics[width=.97\linewidth]{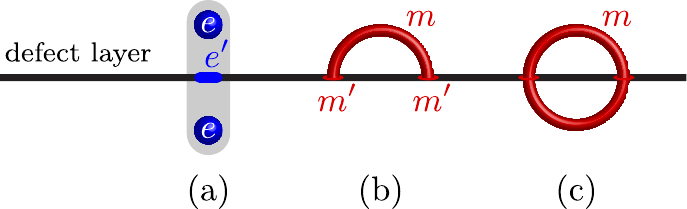}
        \caption{
        A 2-stratum defect layer (black line) in the string-membrane-net model, which is embedded in a 3+1D toric code.
    The 3d toric code has charge (labeled $e$) and flux string (red lines in figure) excitations.
    The defect layer also supports 2d toric code charge (labeled $e'$) and point-like flux (labeled $m'$) excitations.
    The following excitations can be created near a defect layer using local operators:
    {\bf(a)} a 2d toric code charge $e'$ and a 3d toric code charge on each side of the defect layer;
    {\bf(b)} a pair of 2d toric code fluxes $m'$ at the endpoints of a 3d toric code flux string; and
    {\bf(c)} a closed 3d toric code flux string.
    }\label{fig:SMNdefect}
\end{figure}

\begin{figure}[t]
    \centering
        \includegraphics[width=.97\linewidth]{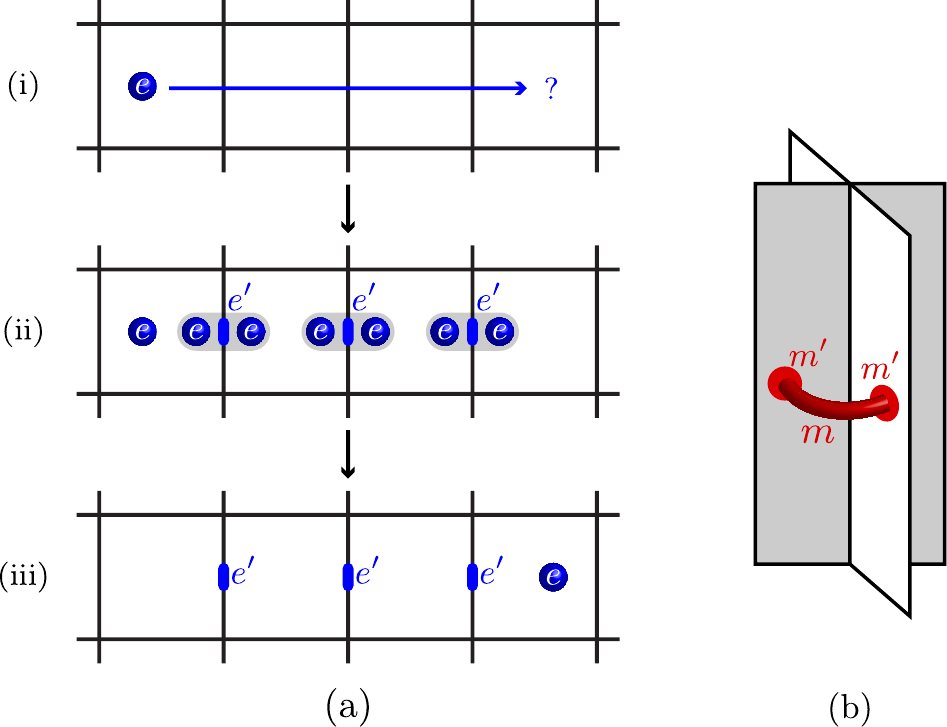}
    \caption{
    Stacks of 2-strata defect layers (defined in \figref{fig:SMNdefect}) result in the fracton and lineon mobility constraints of the X-Cube model.
    {\bf(a)} The defect layers cause the 3d toric code charge (labeled~$e$) to obtain fracton-like mobility.
    To see this, consider (i) trying to move a 3d charge $e$ to move through a bunch of layers.
    (ii) This can only be done by creating the triplets of charges in \sfigref{fig:SMNdefect}{a} on the defect layers.
    (iii) Pairs of 3d charges $e$ can annihilate, but a string of 2d charges $e'$ are left behind.
    This string of excitations linearly confines the 3d charge $e$ to its initial location, just like an X-Cube fracton.
    {\bf(b)} A lineon results from a pair of 2d toric code fluxes $m'$ (which are bound to the endpoints of a 3d toric code flux) on two intersecting defect layers.
    }\label{fig:SMNexcitations}
\end{figure}

The key concepts underlying topological defect networks are captured by a simple example that realizes a phase equivalent to that of the X-Cube model. Consider three stacks (along the $xy$-, $yz$-, and $xz$-planes) of 2-strata layers embedded into a 3+1D toric code. Each 2-strata defect consists of a 2+1D toric code, coupled strongly to the 3+1D toric code living on the 3-strata. The coupling between the 3-- and 2-strata can be characterized in terms of the topological excitations which can locally be created in the vicinity of the 2-strata defects; these excitations are summarized in Fig,~\ref{fig:SMNdefect}. The mobility constraints imposed by this choice of 2-strata defects are described in Fig.~\ref{fig:SMNexcitations}. In fact, the string-membrane-net model~\cite{SlagleSMN} turns out to be secretly describing precisely this topological defect network, as the field theory discussed in that context can be suitably modified to fit the defect picture (see \appref{sec:field theory}). More generally, as we show in this paper, ground state wavefunctions of topological defect networks are fluctuating string-membrane-net configurations.

\vspace{0.6cm}
\tocless{\subsection*{Main Results} \label{sec:mainres}}

In this work, we argue that \textit{all} types of gapped fracton models can be realized by topological defect networks, thereby demonstrating that these models can, in fact, be described in the language of defect TQFT. We proceed mostly by example and construct concrete examples of topological defect networks for well-known gapped fracton models, including the X-Cube and Haah's B code, as well as a fractal type-I model. Besides these, we also present a new non-Abelian fracton model based on 3+1D $D_4$ gauge theory, which hosts fully immobile non-Abelian excitations. 

Although we do not rigorously prove that all gapped fracton phases can be constructed from a topological defect network, we strongly expect this to be true, especially given the plethora of models that fit within this framework.
Furthermore, it has previously been shown that networks of invertible defects can be used to construct crystalline SPT (cSPT) orders \cite{else2019crystalline,song2017defect,huang2017defect},
  which are SPT orders that are protected by spatial symmetries.
This suggests that defect networks should also be capable of describing symmetry enriched topological and fracton orders.
Of course, defect networks can trivially realize any TQFT by simply not embedding any defects within the TQFT.

Since the aforementioned phases appear to exhaust all known zero-temperature gapped phases of matter,
  we are motivated to make the following conjecture:
\begin{framed}
\begin{conjecture}
Topological defect networks realize all zero-temperature gapped phases of matter.
\label{conjecture}
\end{conjecture}
\end{framed}
\noindent That is, for every zero-temperature gapped phase of matter,
  we conjecture that there exists a topological defect network with a Hamiltonian whose ground state is within that phase.
However, we add the caveat that the correct notion of ``phase of matter'' is still somewhat controversial for fracton phases.
Nevertheless, we expect the conjecture to hold for any reasonable definition of a phase, such as phases up to adiabatic deformation or up to local unitary equivalence \cite{ChenWenLU}. Assuming this conjecture, we further argue that there are no stable translation invariant gapped type-II fracton phases of matter in 2+1D.

This paper is organised as follows: In \secref{sec:Xcube}, we build a topological defect network description of the X-Cube model and show that this naturally leads to a membrane-net description of its ground state wavefunction. In \secref{sec:HaahB}, we construct Haah's B code, a type II fracton model with no mobile excitations, within the defect picture. In \secref{sec:nonabelian}, we build on these ideas to construct a new fracton model based on a network of defects embedded in a $D_4$ lattice gauge theory, and we show that the mobility constraints imposed by this network imply the presence of fully immobile non-Abelian fractons. 
So as to keep our discussion accessible to non-experts, various technical details regarding these three models have been relegated to Appendices~\ref{defectXCube},~\ref{app:BCodeLatticeModel}, and~\ref{app:d4lattice} respectively. 
In \secref{sec:classifyingphases}, we present an argument that 2+1D topological defect networks cannot produce stable type-II fracton phases, and conclude with a discussion of open questions and future directions in Sec.~\ref{cncls}.
In Appendix~\ref{sec:field theory} we outline a field theory description of the defect networks for string-membrane-net models; in Appendix~\ref{app:fractallineonmodel} we describe a topological defect network construction of a fractal lineon model~\cite{yoshida,castelnovo}; and in Appendix~\ref{app:trivbulk} we describe a topological defect network construction of a $\mathbb{Z}_3$ type-I model from trivial 3-strata.


\vspace{0.6cm}
\section{X-Cube from Topological Defects in 3+1D Toric Code}
\label{sec:Xcube}

In this section we use the conceptual picture outlined in the introduction to realize the X-Cube model.
This example is similar to the string-membrane-net example discussed in the introduction (and \figref{fig:SMNdefect} and~\ref{fig:SMNexcitations}),
  except now we will not have $e'$ or $m'$ excitations on the boundary layer.
This will result in simpler 2-strata defects, but more complicated 1-strata defects.
The model is realized by three stacks of defect layers, which live in an ambient 3+1D toric code phase (see \figref{fig:3DTClattice}).
In the following three subsections, we describe the construction in two complementary ways.
The first approach defines the construction in terms of the excitations which are condensed on the defects,
  and makes the mobility constraints of the fracton and lineon excitations most explicit.
The second defines the ground state wavefunction as a superposition of allowed membrane-net configurations.
In App.~\ref{defectXCube}, we provide a concrete lattice Hamiltonian that explicitly realizes this defect network.

\subsection{3+1D toric code preliminaries}
\label{sec:3D toric code}

The 3d toric code plays an essential role in the topological defect network construction of the X-Cube model. In this subsection, we briefly review the 3d toric code on the cubic lattice. 

We will take the degrees of freedom to live on the plaquettes, such that the total Hilbert space is a tensor product of one spin-1/2 per plaquette.
The Hamiltonian is a sum of two terms~\cite{chamonfiniteT}:
\begin{align}
\label{3DTCHam}
    H = -J_e\sum_{e} A_e -J_c \sum_c B_c.
\end{align}
Where $\sum_{e}$ runs over all edges and $\sum_c$ runs over all cubes of the cubic lattice.
The first term is given by
\begin{align}
    A_e = \prod_{p \in e} Z_p,
\end{align}
where $p\in e$ denotes all plaquettes $p$ with a common edge $e$.
The second term is given by
\begin{align}
B_c = \sum_{p \in \partial c} X_p,
\end{align}
where $p\in \partial c$ denotes all plaquettes on the boundary of a cube $c$.
All terms in the Hamiltonian \eqref{3DTCHam} commute, and so the model is exactly solvable.
The ground space is given by any state satisfying $A_e \ket{\psi} = B_c \ket{\psi} = \ket{\psi}$ for all edges $e$ and cubes $c$. 

The 3d toric code has particle excitations, denoted $e$, which consist of a single cube where $B_c$ has eigenvalue $-1$. A pair of particle excitations are created at the end points of a string operator which is given by $\prod_{p \in \ell} Z_p$, where $\ell$ is a path on the dual lattice, and $p \in \ell$ are all plaquettes which intersect that path. The 3d toric code also has flux loop excitations, denoted $m$, consisting of a loop of edges where $A_e$ has eigenvalue $-1$. A flux loop is created at the boundary of a membrane operator given by $\prod_{p \in S}X_p $, where $S$ is a surface, and $p \in S$ are all plaquettes contained in that surface.

For later use, we now describe the 3+1D toric code with a boundary. 
There are two natural boundary conditions; the ``rough" boundary which condenses flux loops, and the ``smooth" boundary which condenses the particle excitations.\footnote{In fact, any $\mathbb{Z}_2$ graded fusion category provides a natural boundary condition for the 3+1D toric code.}
In the following, we focus on the flux condensing boundary, typically referred to as the rough boundary since it consists of dangling plaquettes.
That is, we terminate the cubic lattice on a surface (for example, a plane) and remove all plaquettes which reside in that plane.
In the bulk, the Hamiltonian remains un-modified taking the form of \eqref{3DTCHam}.
The sums in \eqref{3DTCHam} are now over all edges which are not in the surface, and all cubes which do not meet the surface.
The boundary Hamiltonian is given by
\begin{align}
    H_{\text{boundary}} = -J_c'\sum_{c \in \text{boundary}} B_c'
\end{align}
where 
\begin{align}
    \label{eqn:roughBoundaryOp}
    B_c' = \prod_{p \in c} X_p,
\end{align}
and `$c \in \text{boundary}$' denotes all cubes which share one (or more) plaquettes with the boundary.
The boundary is said to condense flux loops, since a membrane operator which terminates on the boundary does not violate any terms in the Hamiltonian.
Electric excitations remain as excitations when brought to the boundary.

The ground state wavefunction can be viewed as a condensate of membranes. 
To do so, one associates $Z_p=-1$ with a membrane present on plaquette $p$. The condition $A_e=+1$ enforces that the membranes are closed, and $B_c=+1$ forces all possible closed membrane configurations to appear in the wavefunction with equal amplitude. On a flux condensing boundary, we relax the first of these conditions and allow membranes to terminate on the boundary. The term $B_c'$ then forces the ground state to be an equal weight super position of all such allowed membrane configurations.

In the following section we will use a mild generalization of these gapped boundaries to construct the X-Cube model out of a lattice of 3+1D toric codes coupled together by defects.

\subsection{Condensation on defects}
\label{sec:Xcube defect condensation}

We now describe the topological defects used to construct the X-Cube model by the excitations that condense on them.
For simplicity we write the Hamiltonian on the 3-torus.
We choose a stratification of the 3-torus to be a cubic lattice, where the cubes are the 3-strata, the plaquettes are the 2-strata, the edges are the 1-strata, and the vertices are the 0-strata (see Fig.~\ref{fig:3DTClattice}).
When constructing a lattice model, we use a cell decomposition of the stratified 3-torus given by a smaller lattice, as also shown in Fig.~\ref{fig:3DTClattice}.
The topological defects are used to couple the 3+1D toric codes on the 3-strata in such a way that the resulting theory is equivalent to the X-Cube model.
In the following we describe the defects assigned to the 2- and 1-strata by the particles which condense on them. 

We first look at the 2-strata defects.
Each 2-stratum is neighbored by a pair of 3-strata, and the 2-stratum defect is characterized by the excitations that condense on it. 
With the benefit of hindsight, the excitations which condense on each 2-stratum are generated by the flux loops from the neighboring 3-strata
\begin{align}
    \{ m_+, m_- \}, \label{eq:twocelldefect}
\end{align}
where $m_+$ is from one toric code and $m_-$ is from the other. 
That is, the defect independently condenses the flux loop excitations from both neighboring 3-strata.
This boundary condition forbids the $e$ particles from passing through the 2-strata, thus imposing mobility restrictions necessary for a fracton phase.

Next we zoom in on the 1-strata. 
These live at the junction of four 2- and 3-strata.
Consequently, we need to specify which 3-strata and 2-strata
  excitations condense on the 1-strata.
  In this example, the 2-strata excitations are simply $e$ particles brought from the bulk to the boundary.
Thus we only need to specify how the 3-strata excitations condense on the 1-strata, since this determines how the 2-strata excitations condense on the 1-strata.

\begin{figure}
    \centering
{\includegraphics[width=.97\linewidth]{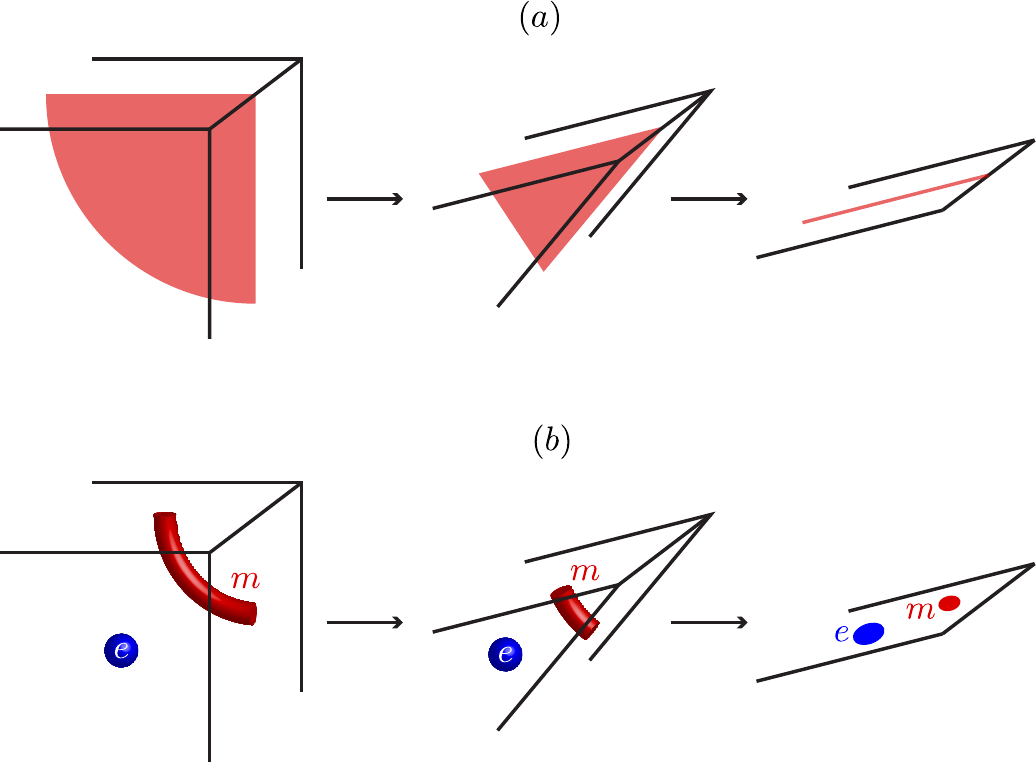}
    \caption{When the 2-stratum condense all fluxes, the 1-strata boundary conditions can be understood using a dimensional reduction trick. In {\bf (a)} we depict how membranes near a 1-strata effectively behave as strings in a dimensionally reduced 2+1D toric code.
    In {\bf (b)} we show how the 3-strata excitations map onto the dimensionally reduced toric code excitations.
    }
    \label{PinchingTrick}
    }
\end{figure}

The 2-strata defects we have chosen are very special and readily allow us to apply a dimensional reduction trick to understand the 1-strata defects.
Near a 1-stratum, each 3-stratum toric code gets pinched into a very thin slab. In the thin slab, the low energy excitations are given by $e$ particles as before, and short $m$-strings connecting both sides of the slab, effectively realizing a 2+1D toric code, as illustrated in \figref{PinchingTrick}.
Hence, we can understand the 1-strata defects from condensation processes of the dimensionally reduced 3d toric codes.
At the 1-stratum defect, we condense quadruples of $e$ particles and all pairs of fluxes coming from the neighboring 3d toric codes. 
These are generated by
\begin{align} 
\{ m_1 m_2, m_2 m_3, m_3m_4, e_1 e_2 e_3 e_4 \} \, ,
\label{onecelldefect}
\end{align}
where 1, 2, 3, and 4 denote the four neighboring 3-strata.
(The ordering does not matter.)
One can check that the algebra object generated by the above particles is a Lagrangian algebra object for the boundary of the four (effectively) 2d toric codes meeting at the 1-stratum,
  and therefore no further particles need to be condensed in order to have a gapped 1-stratum.\footnote{
If $A$ is the algebra object being condensed, then it is Lagrangian if $(\text{dim}\; A)^2 = \text{dim}\; \mathcal{D}(\mathbb{Z}_2^{ 4})$.
There are $2^4 = 16 = \text{dim}\; A$ particles generated by \eqref{onecelldefect}.
We also have $\text{dim}\; \mathcal{D}(\mathbb{Z}_2) = 4$, and so $\text{dim}\; \mathcal{D}(\mathbb{Z}_2^{4}) = 4^4 = (\text{dim} \;A)^2$.
}

The data on higher strata do not uniquely determine a defect on the 0-strata.
For example, the 0-strata could pin topological excitations from the various $3-$, $2-$ and $1-$strata which meet at the 0-strata.
To obtain the X-Cube model, we choose the 0-strata defects to not pin any topological charges from the neighboring 3-, 2- and 1-strata.

We have now finished describing all data needed to specify the defects. 
In the following section, we analyze the character of the topological excitations, and in the subsequent section, we write down an explicit membrane-net condensate.

\subsection{Mobility Constraints}

We now analyze the excitations present in this defect construction and verify that they enjoy mobility constraints typical of fracton phases.
In particular, we show that their mobility constraints are consistent with those of topological excitations in the X-Cube model.

\begin{figure}
    \centering
    \subfloat[]{\includegraphics[width=.4\columnwidth]{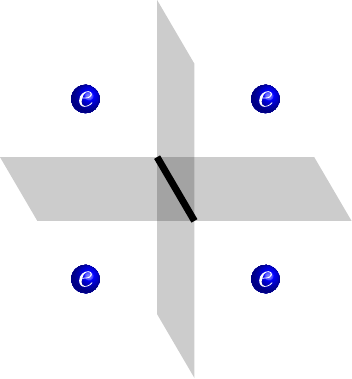}}
    \subfloat[]{\includegraphics[width=.58\columnwidth]{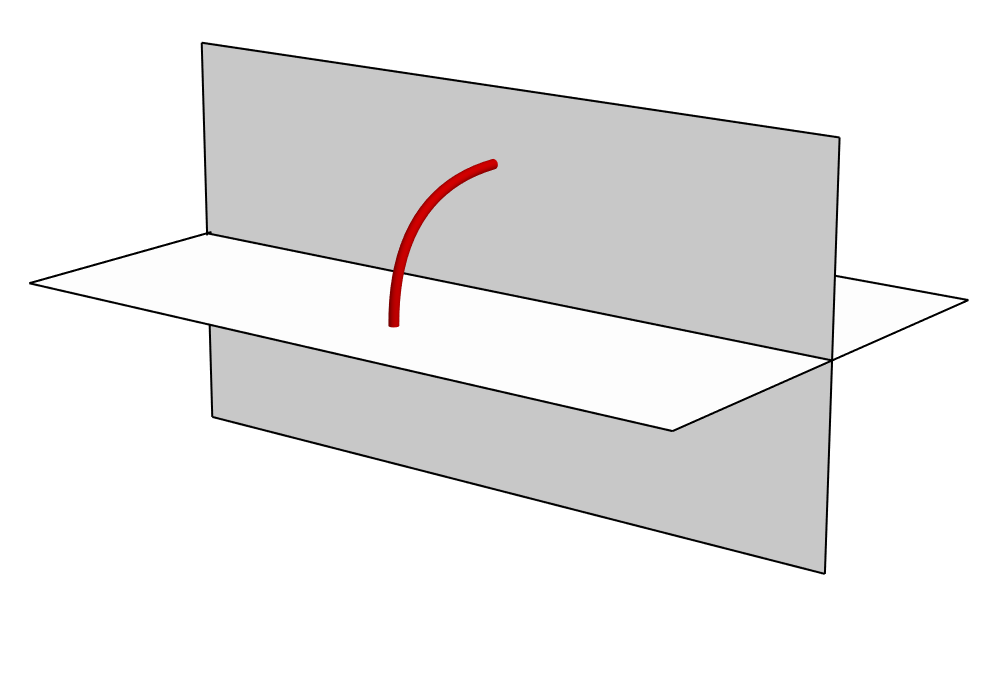}} \\
    \subfloat[]{\includegraphics[width=.4\columnwidth]{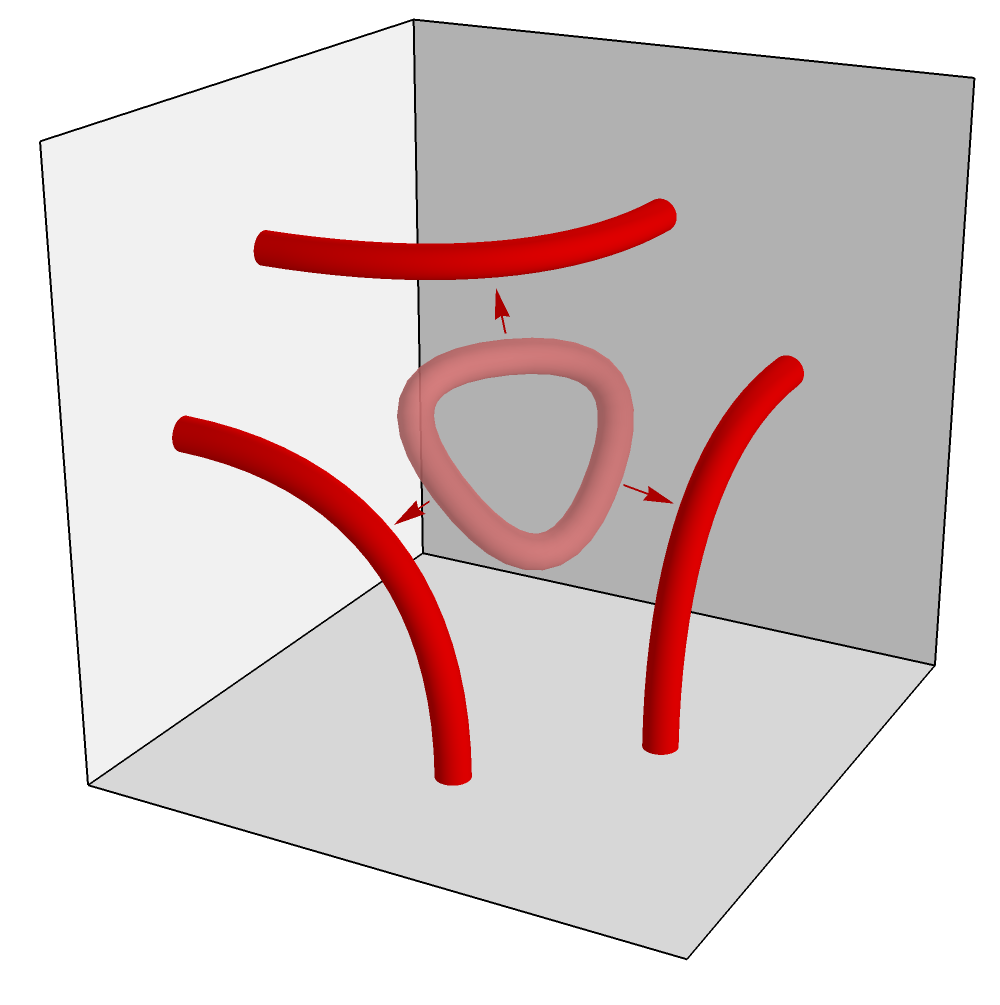}}
    \subfloat[]{\includegraphics[width=.58\columnwidth]{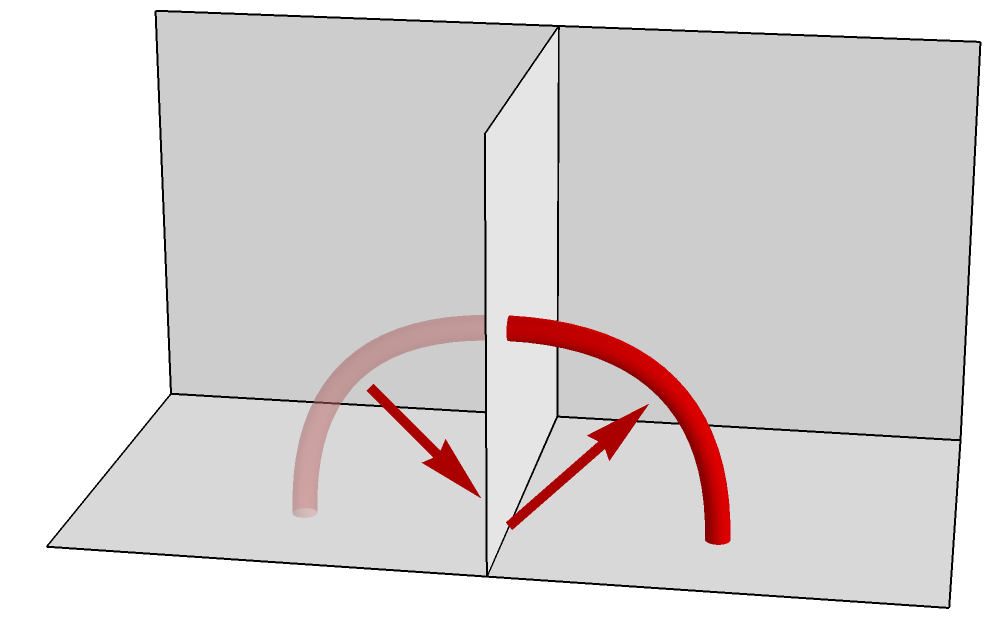}} \\
    \subfloat[]{\includegraphics[width=.49\columnwidth]{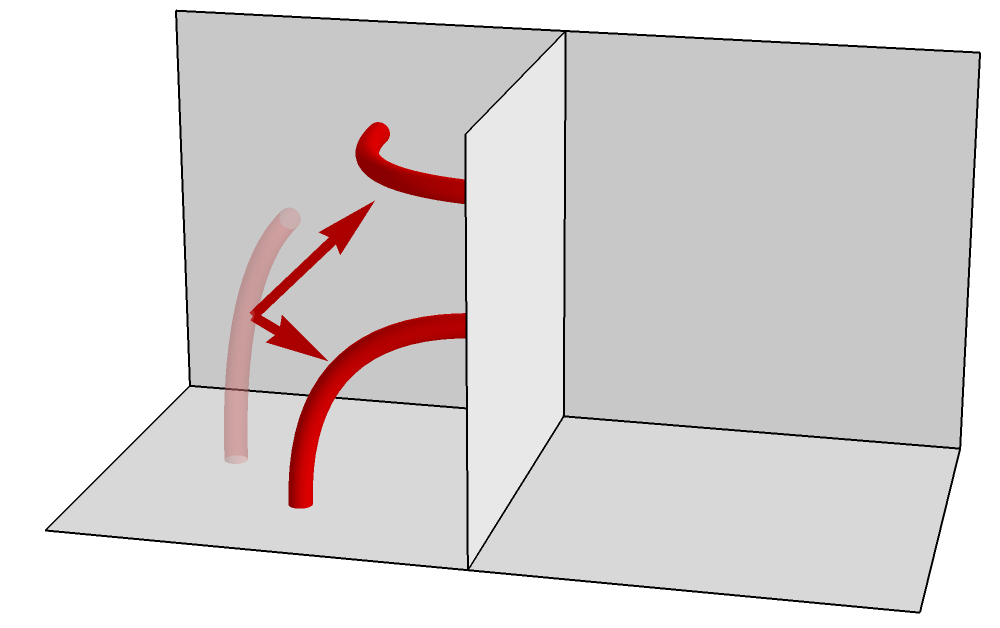}}
    \subfloat[]{\includegraphics[width=.49\columnwidth]{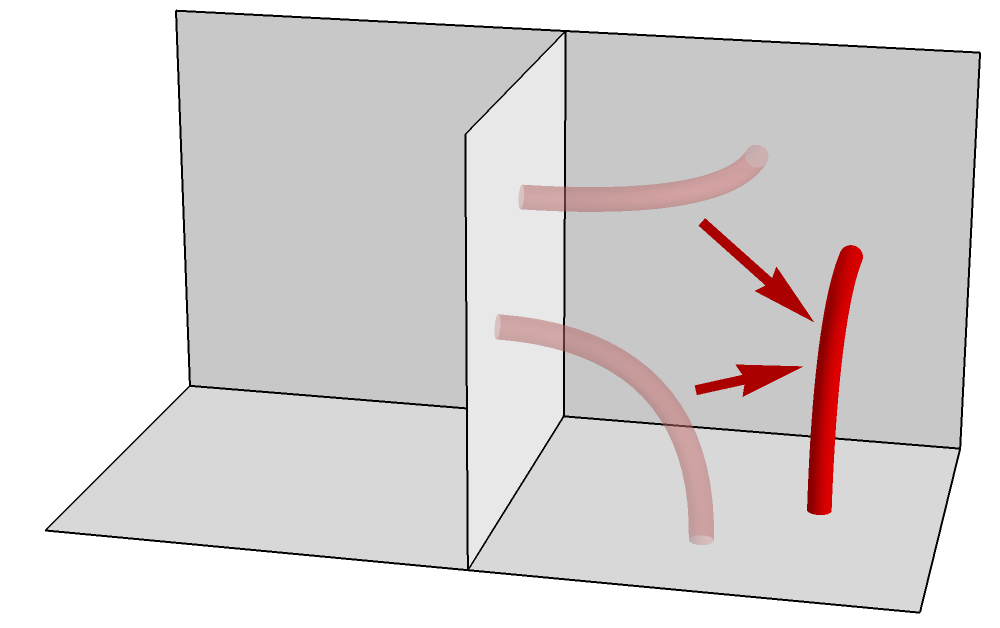}}
    \caption{
    {\bf (a)} The edge condensation [\eqnref{onecelldefect}] allows four charges to be created in four neighboring 3-strata, analogous to X-Cube fractons.
    {\bf (b)} A lineon results from a flux string between two neighboring 2-cells.
    {\bf (c)} Three lineons can be created from the vacuum by creating a small flux loop (light red) near a 0-strata, and then stretching it to the nearby 2-strata.
    This is analogous to the X-Cube model where three lineons can be created at a vertex.
    {\bf (d)} The condensation at 1-strata defects [\eqnref{onecelldefect}] allows the flux loop to move through the 1-strata.
    {\bf (e-f)} A sequence of 4 configurations which show how the lineon can move past a 2-strata by making use of the moves in (c) and (d).
    }\label{fig:Xcube excitations}
\end{figure}

The charge excitations $e$ cannot move through the 2-strata since the flux excitations are condensed there.
However, due to the 1-stratum condensation [\eqnref{onecelldefect}], four charges can be created on four neighboring 3-strata [\sfigref{fig:Xcube excitations}{a}].
As a result, the charge excitations $e$ have exactly the same mobility constraints as the fractons in the X-Cube model.
Furthermore, if two $e$ are in the same $3$-stratum, they can be fused to the identity, in the same way that two fractons on the same cube in the X-Cube model fuse to the identity.

A lineon excitation results from a flux string that ends on two neighboring 2-strata;
  see \sfigref{fig:Xcube excitations}{b}.
This excitation cannot be annihilated by simply shrinking the string since single flux excitations are not condensed on the 1-strata;
  only pairs of flux strings on neighboring 3-strata are condensed [\eqnref{onecelldefect}].
Similar to X-Cube lineons, three lineons can be locally created from the vacuum, as shown in \sfigref{fig:Xcube excitations}{c}.
In \sfigref{fig:Xcube excitations}{d-f}, we show more explicitly how the condensations on the $1$-strata allow this excitation to move like an X-Cube lineon.

Also as in the X-Cube model, planons result from pairs of fractons (charges $e$ in this model) or lineons. 
The lineon pair is equivalent to an m-string which is orthogonal to the mobility plane.

\subsection{Nets and Relations}

In this subsection, we present the X-Cube model as a network of defects in the membrane-net condensate picture of the 3+1D toric code.
In this picture, we present ``allowed" membrane configurations and linear relations amongst those membranes. 
The ground state wavefunction is a weighted superposition of all allowed membrane configurations:
\begin{align}
    |{\Psi}\rangle = \sum_{\text{M}} \phi(\text{M}) |\text{M} \rangle. 
    \label{eq:nets} 
\end{align}
In the above, ``$\text{M}$'' denotes an allowed membrane configuration, and $\phi(\text{M})$ is the weighting which is determined by the linear relations on the membranes.
In the following, $\phi(\text{M})=+1$ for all allowed diagrams.

We specify whether a net diagram is ``allowed" by checking whether it locally satisfies some admissibility constraints. 
We begin by describing the admissible membranes on the interior of a 3-strata, followed by the allowed membranes near the 2-, 1-, and 0-strata. Excitations correspond to violations of the ``allowed" membrane configurations or changes to the phases $\phi(\text{M})$ in \eqref{eq:nets}. 

\subsubsection*{3-strata}
We first specify the allowed membranes which reside on the interior of a 3-stratum. 
These are given by surfaces which are locally closed. 
That is, if we choose a closed path in the interior of a 3-stratum, the path must intersect an even number of surfaces. 
For example,
\begin{align}
\cupcapprime\;=\;\;\cylinderprime \, ,
\end{align}
and so the two configurations appear with equal amplitude in the wavefunction. 

\subsubsection*{2-strata}
The 2-strata defects are given by rough boundary conditions for the the two adjacent 3-strata toric codes. 
That is, we allow membranes from the neighboring 3-strata toric codes to freely terminate on the 2-strata.
Hence, near a 2-stratum, a generic membrane configuration and linear relation looks like:
\begin{align}
\twocellnetaprime = \twocellnetbprime
\end{align}

\subsubsection*{1-strata}
\begin{figure}
    \centering
    \includegraphics[width=.95\columnwidth]{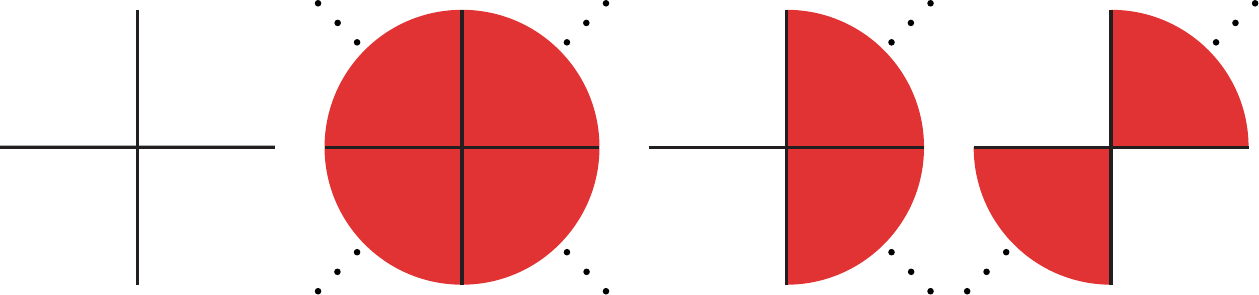}
    \caption{A representative set of allowed membrane configurations near a 1-cell as viewed from above (\textit{i.e.,} down a 1-cell).
    The remaining allowed configurations come from rotating the two diagrams on the right.
}\label{fig:OneCellNets}
\end{figure}

The 1-strata defects can be inferred from the dimensional reduction picture described in Sec.~\ref{sec:Xcube defect condensation}.
The fact that pairs of any two fluxes can condense on the 1-strata tells us that we must have an even number of membranes terminating on each 1-stratum. 
A top view of the allowed membrane configurations is shown in Fig.~\ref{fig:OneCellNets}.

\subsubsection*{0-strata}
A membrane configuration near a 0-stratum is ``allowed" if it satisfies the constraints imposed by the 1- and 2-strata defects when pulled away from the 0-stratum. 
That is, membranes near the 0-strata do not pick up any additional constraints not already imposed by the 1- and 2-strata defects. 

In App.~\ref{defectXCube}, we explicitly write out a Hamiltonian whose ground state can be viewed as an equal weight superposition of the allowed membrane configurations described in this section. 


\section{Haah's B Code from Topological Defects}
\label{sec:HaahB}

In this section, we show that Haah's B code~\cite{haah2014bifurcation}, which is a so-called ``type-II" fracton model in which all topologically non-trivial excitations are immobile and created by operators with fractal support, can also be realized using a defect network in an ambient 3+1D $\mathbb{Z}_2 \times \mathbb{Z}_2$ toric code. We first briefly review the B code before describing the defect network.

\subsection{Review of Haah's B code}

\begin{figure}
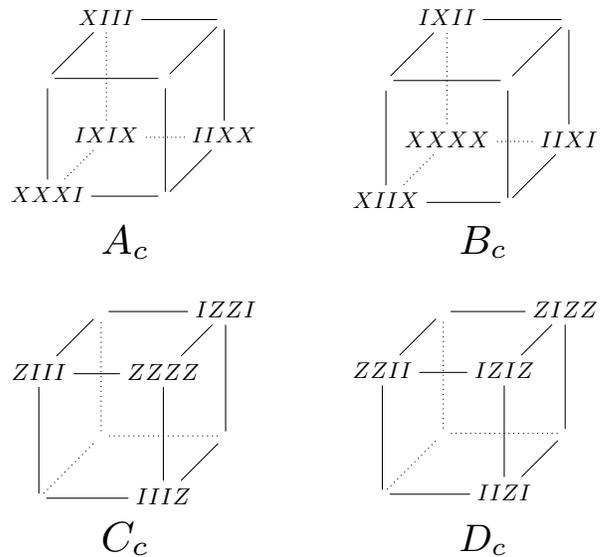

    \centering
        \resizebox{0.4\columnwidth}{!}{\drawgenerator{}{}{XXXI}{}{IIXX}{IXIX}{}{XIII}} \hspace{0.1\columnwidth}
        \resizebox{0.4\columnwidth}{!}{\drawgenerator{}{}{XIIX}{}{IIXI}{XXXX}{}{IXII}}\\ 
        \vspace{0.1cm}
        \resizebox{0.65cm}{!}{$A_c$} \hspace{0.45\columnwidth} \resizebox{0.65cm}{!}{$B_c$}\\
        \vspace{0.4cm}
        \resizebox{0.4\columnwidth}{!}{\drawgenerator{ZIII}{ZZZZ}{}{IIIZ}{}{}{IZZI}{}}
        \hspace{0.1\columnwidth}
        \resizebox{0.4\columnwidth}{!}{\drawgenerator{ZZII}{IZIZ}{}{IIZI}{}{}{ZIZZ}{}} \\
     \vspace{0.1cm}
        \resizebox{0.65cm}{!}{$C_c$} \hspace{0.45\columnwidth} \resizebox{0.65cm}{!}{$D_c$}
    \caption{Terms in the Hamiltonian for Haah's B code, which has four qubits per site of the cubic lattice. $X$ and $Z$ are Pauli operators, $I$ is the identity, and the shorthand $XXXX$ (for example) means a product of Pauli X operators on all four qubits on the site in question.}
    \label{fig:BCodeH}
\end{figure}

The Hilbert space of Haah's B code consists of four qubits per site of a cubic lattice. We use a shorthand where, for example, $XZXZ$ means applying a Pauli $X$ operator to the first and third qubit on a given site and a Pauli $Z$ operator to the second and fourth qubits on that site. The Hamiltonian is
\begin{equation}
    H = -\sum_c \left(A_c+B_c+C_c+D_c\right)
\end{equation}
where $c$ is an elementary cube in the lattice and $A_c$, $B_c$, $C_c$, and $D_c$ are products of Pauli operators shown in Fig.~\ref{fig:BCodeH}.

It is straightforward to check that all terms in the Hamiltonian mutually commute. Hence, in the ground state(s), every term has eigenvalue $+1$. Elementary excitations consist of cubes on which one or more of these four terms have eigenvalue $-1$.

\begin{figure}
    \centering
    \subfloat[\label{fig:BCode_LocalOps}]{\includegraphics[width=0.7\columnwidth]{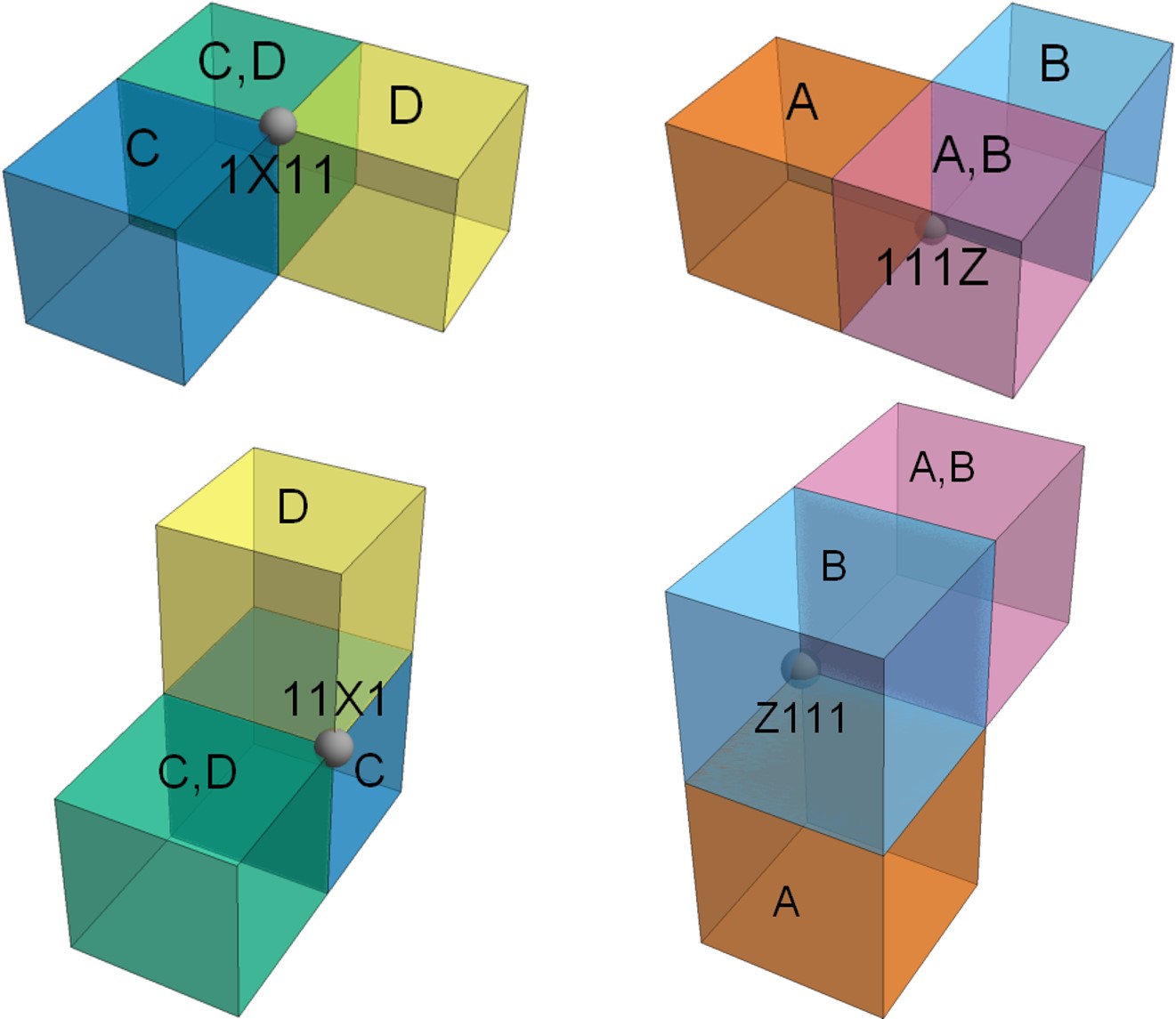}}\\
    \subfloat[\label{fig:BCode_ABFractal}]{\includegraphics[width=0.6\columnwidth]{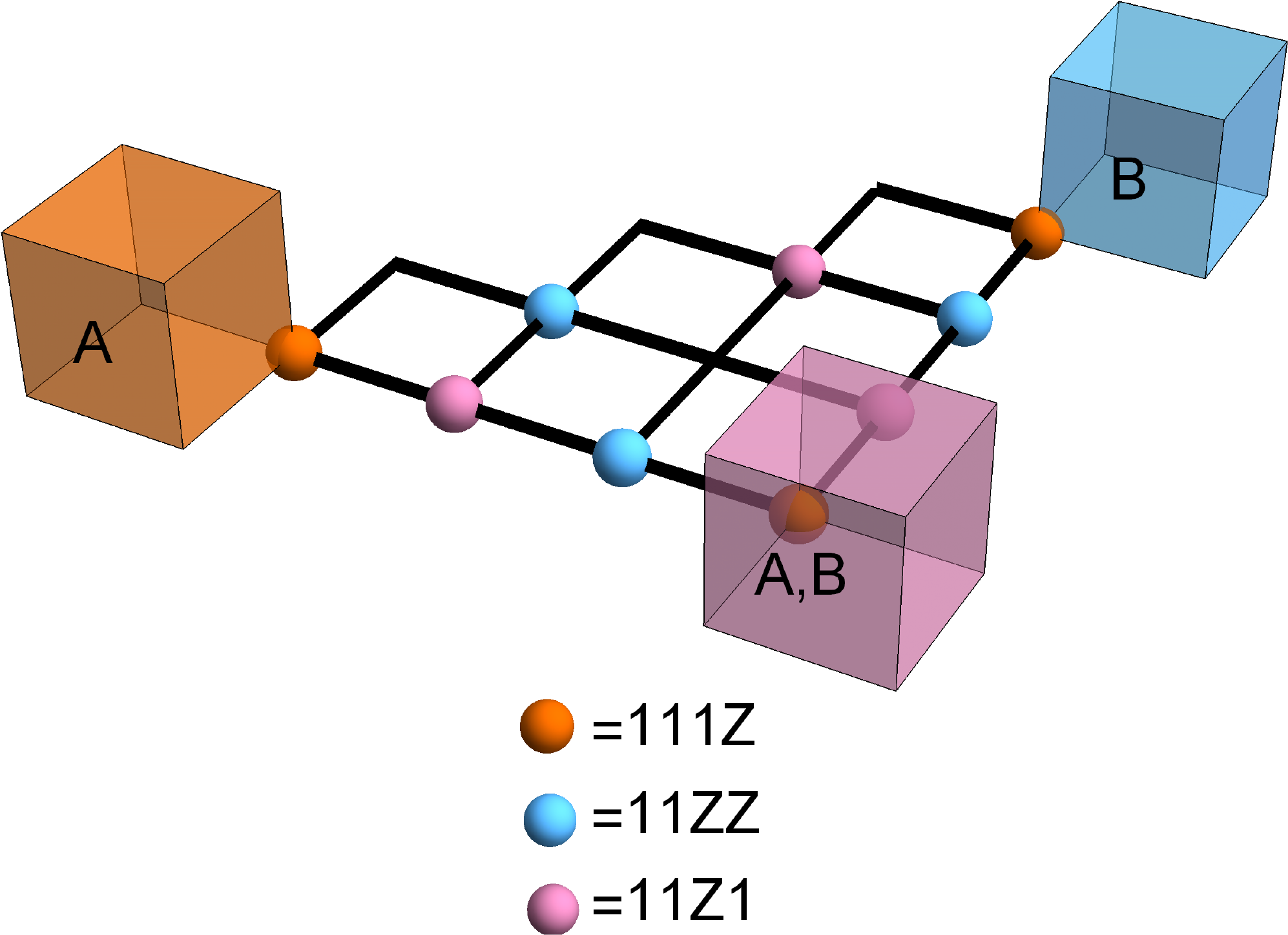}}
    \caption{Excitations in Haah's B code. (a) Excitations created by some local operators (spheres) on a ground state of the B code. Colored cubes indicate elementary cubes where the labeled Hamiltonian term has eigenvalue $-1$ (AB means both $A_c$ and $B_c$ have eigenvalue $-1$). (b) An operator with fractal-like support (the product of local operators indicated by the colored spheres) creates widely separated excitations (cubes) in Haah's B code.}
\end{figure}

Local operators acting on ground states can easily be checked to create excitation configurations such as those in Fig.~\ref{fig:BCode_LocalOps}. These patterns are such that certain operators with fractal-like support, such as the one in Fig.~\ref{fig:BCode_ABFractal}, can create isolated excitations at their corners. We will show in the next subsection that a defect construction can admit local operators which create excitations in the same patterns as the ones shown here.

\subsection{Condensation on defects}

The defect construction uses the same stratification of space shown in Fig.~\ref{fig:3DTClattice}, that is, a cubic lattice of 3-strata. Each 3-stratum contains a $\mathbb{Z}_2 \times \mathbb{Z}_2$ gauge theory, that is, two copies of the 3+1D toric code. Each 3-stratum therefore contains two types of electric charges, which we label $e^A$ and $e^B$; two types of magnetic string excitations $m^A$ and $m^B$; and the bound states of these objects. Intuitively, a 3-stratum in the defect picture containing an odd number of $e^A$ (resp. $e^B$) will correspond to a unit cube with $A_c=-1$ (resp. $B_c=-1$) in Haah's B code, while the $m^A$ (resp. $m^B$) excitations will, in a more subtle way, correspond to unit cubes with $C_c=-1$ (resp. $D_c=-1$). This is why we choose to start from \textit{two} copies of the 3+1D toric code in each unit cell instead of one.

Each 2-stratum borders two 3-strata containing a $\mathbb{Z}_2 \times \mathbb{Z}_2$ 3d toric code. We declare that all $m$ excitations in either 3-stratum can condense on the 2-strata, \textit{i.e.,} a generating set of local excitations on the defects is
\begin{equation}
    \lbrace m^A_+,m^A_-,m^B_+, m^B_- \rbrace
\end{equation}
where the $+$ excitations are from one $\mathbb{Z}_2 \times \mathbb{Z}_2$ toric code and $-$ is from the other. This ensures that no point-like objects can pass freely through the 2-strata.

\begin{figure}
    \centering
    \includegraphics[width=0.6\columnwidth]{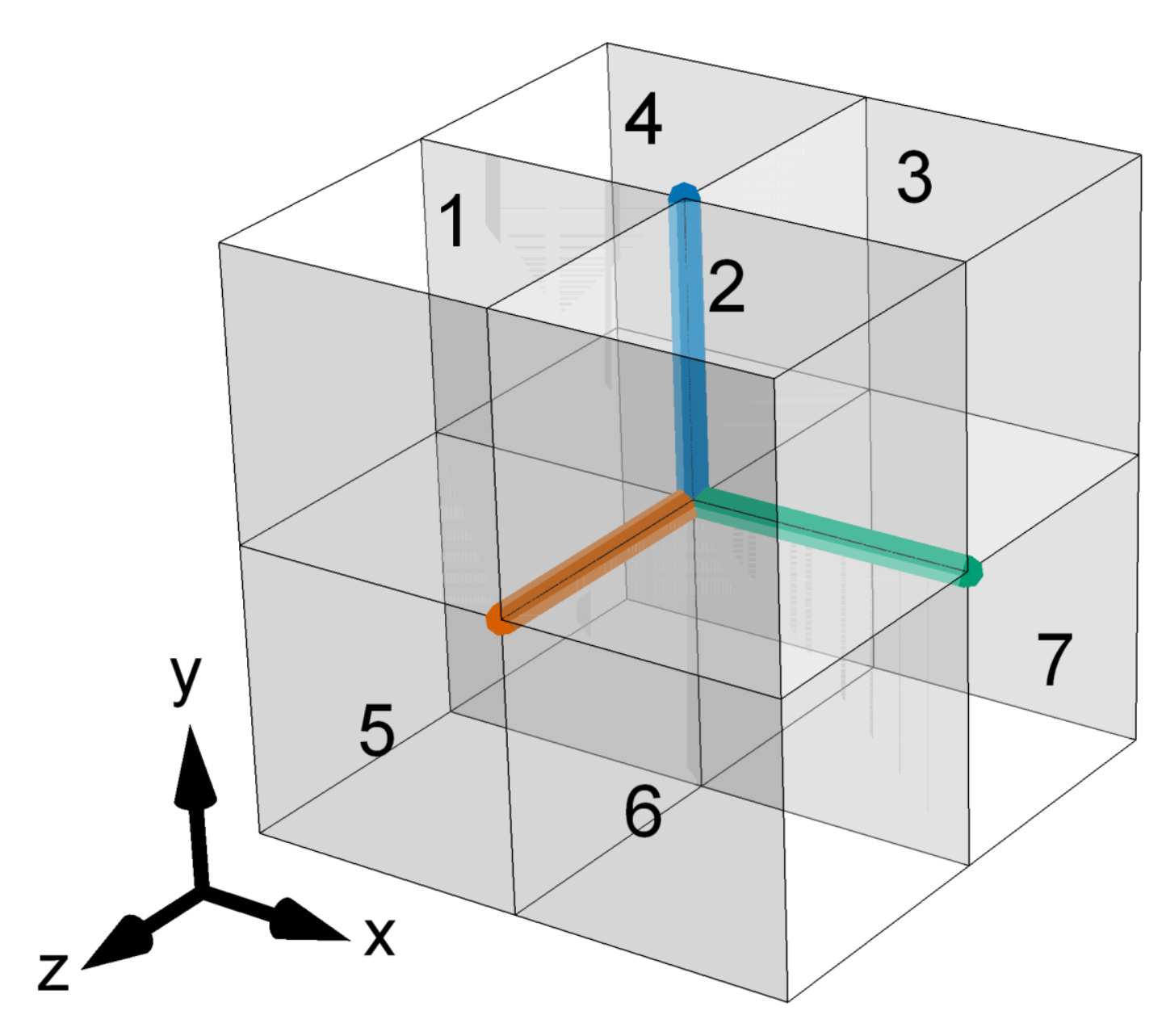}
    \caption{Labeling scheme for 3-strata used in defining condensed objects on the colored 1-strata for the defect construction of Haah's B code.}
    \label{fig:BCode_CubeLabels}
\end{figure}

On the 1-strata, we allow some bulk excitations to condense. Generating sets of the condensed excitations consist of: 

\begin{widetext}
\begin{align}
  \lbrace e^B_2 e^A_3 &e^B_3 e^A_6, e^A_2 e^B_2 e^A_3 e^B_6, m^A_2 m^B_6, m^B_2 m^A_3, m^B_3 m^A_6,
  m^B_2 m^A_6 m^B_6,  m^A_7, m^B_7 \rbrace  \hspace{0.5cm}\text{ ($x$-oriented 1-stratum)} \label{eqn:BCodeXCondensations}\\
  \lbrace e^A_1 e^B_1 &e^B_2 e^A_3, e^A_1 e^A_2 e^B_2 e^B_3, m^A_1 m^B_2, m^B_1 m^A_3, m^A_2 m^B_3, 
  m_1^A m_1^B m_3^B, m^A_4, m^B_4\rbrace  \hspace{0.5cm}  \text{ ($y$-oriented 1-stratum)}\label{eqn:BCodeYCondensations}\\
  \lbrace e^A_1 e^B_1 &e^A_2 e^B_6, e^A_1 e^B_2 e^A_6 e^B_6, m^A_1 m^B_6, m^B_1 m^A_2, m^B_2 m^A_6, 
  m^A_1 m^B_1 m^A_6, m^A_5, m^B_5\rbrace  \hspace{0.5cm}  \text{ ($z$-oriented 1-stratum)} \label{eqn:BCodeZCondensations}
\end{align}
\end{widetext}
The subscripts label 3-strata according to  Fig.~\ref{fig:BCode_CubeLabels}, in which the links on which we are specifying condensations are colored green, blue, and orange, respectively. Using the dimensional reduction trick (see Fig.~\ref{PinchingTrick}), it is easy to check that these condensations yield a gapped boundary. Alternatively, this will be clear when discussing the nets and relations picture of the B code defect construction.

Finally, the 0-strata defects are chosen to not pin any topological charges from the neighboring higher strata.

\begin{figure}
    \centering
    \subfloat[\label{fig:BCode_electricCorrespondence}]{\includegraphics[width=0.6\columnwidth]{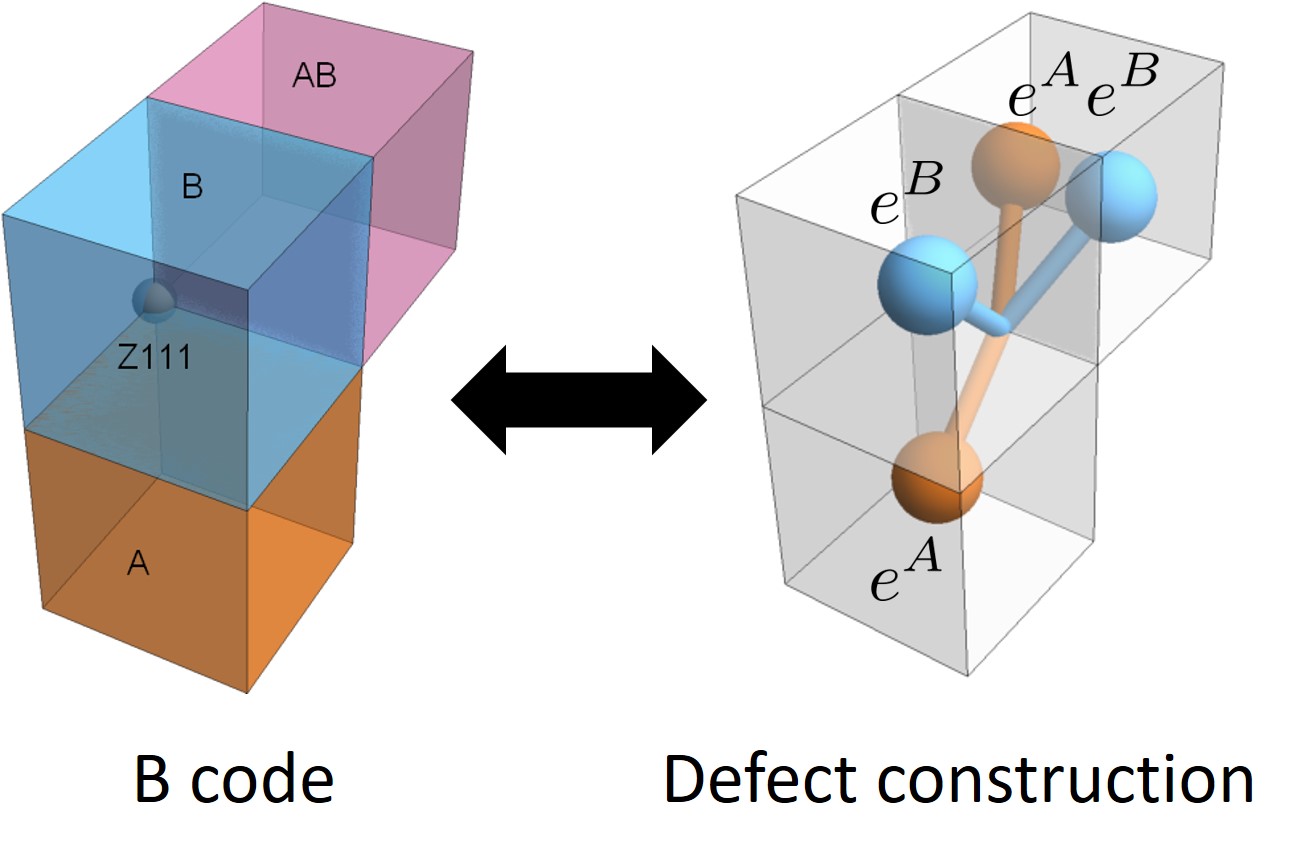}}\\
    \subfloat[\label{fig:BCode_ElectricFractal}]{\includegraphics[width=0.6\columnwidth]{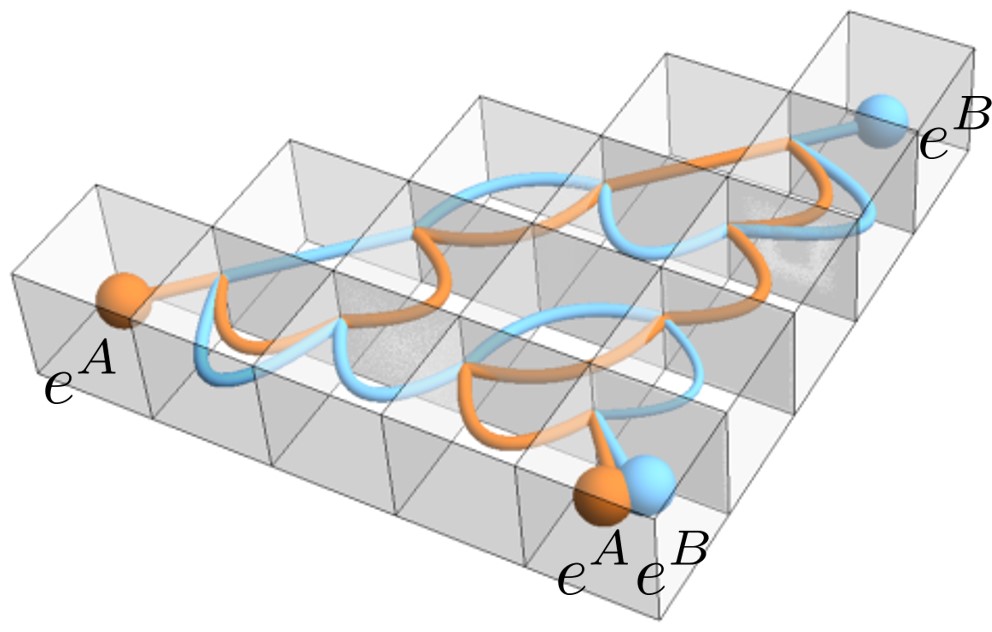}}
    \caption{Electric excitations in the defect network construction and their correspondence to $A$ and $B$ excitations in the B code. Orange (resp. blue) lines are electric string operators in the A (resp. B) 3+1D toric code layer, and correspondingly colored spheres are point-like electric excitations in the appropriate layer. Perspective is the same as Fig.~\ref{fig:BCode_CubeLabels}. (a) Correspondence between the action of local Pauli $Z$ operators in the B code (see Fig.~\ref{fig:BCode_LocalOps}) and the condensation of electric charges on 1-strata in the defect network construction. (b) Network of $\mathbb{Z}_2 \times \mathbb{Z}_2$ 3+1D toric code string operators which create isolated electric excitations (spheres) in the defect network construction of the B code. The choice of condensations in Eq.~\eqref{eqn:BCodeYCondensations} ensures that no additional excitations are created at the 1-strata. Compare the excitation configuration to that in the B code in Fig.~\ref{fig:BCode_ABFractal}.}
\end{figure}

The data on the 1-strata is fairly complicated, so we presently explain it by examining the excitations in this model. We have chosen the condensed electric charges to match the action of local operators in Haah's B code in the following sense. Under the aforementioned correspondence where a 3-stratum containing an odd number of $e^A$ (resp. $e^B$) is to be thought of as a cube in the B code with $A_c=-1$ (resp. $B_c=-1$), the statement that $e^B_2e^A_3e^B_3e^A_6$ is condensed (i.e. can be created locally) at an $x$-oriented 1-stratum corresponds to the fact that, in Haah's B code, the operator $ZIII$ creates the pattern of excitations in Fig. ~\ref{fig:BCode_LocalOps}. This correspondence is shown pictorially in Fig.~\ref{fig:BCode_electricCorrespondence}. Likewise, the condensation of $e^A_2e^B_2e^A_3e^B_6$ and $e^A_2e^B_3e^A_6e^B_6$, respectively, correspond to the actions of $IZII$ and $ZZII$ in Haah's B code. It is straightforward to check that every condensed electric charge in the defect construction corresponds to a local product of Pauli $Z$ operators in Haah's B code. These condensations also permit the creation of widely separated  excitations using networks of string operators whose support is fractal-like. An example which corresponds to the B code excitation pattern created in Fig.~\ref{fig:BCode_ABFractal}   is shown in Fig.~\ref{fig:BCode_ElectricFractal}.

\begin{figure}
    \centering
    \subfloat[\label{fig:BCode_membrane}]{\includegraphics[width=0.35\columnwidth]{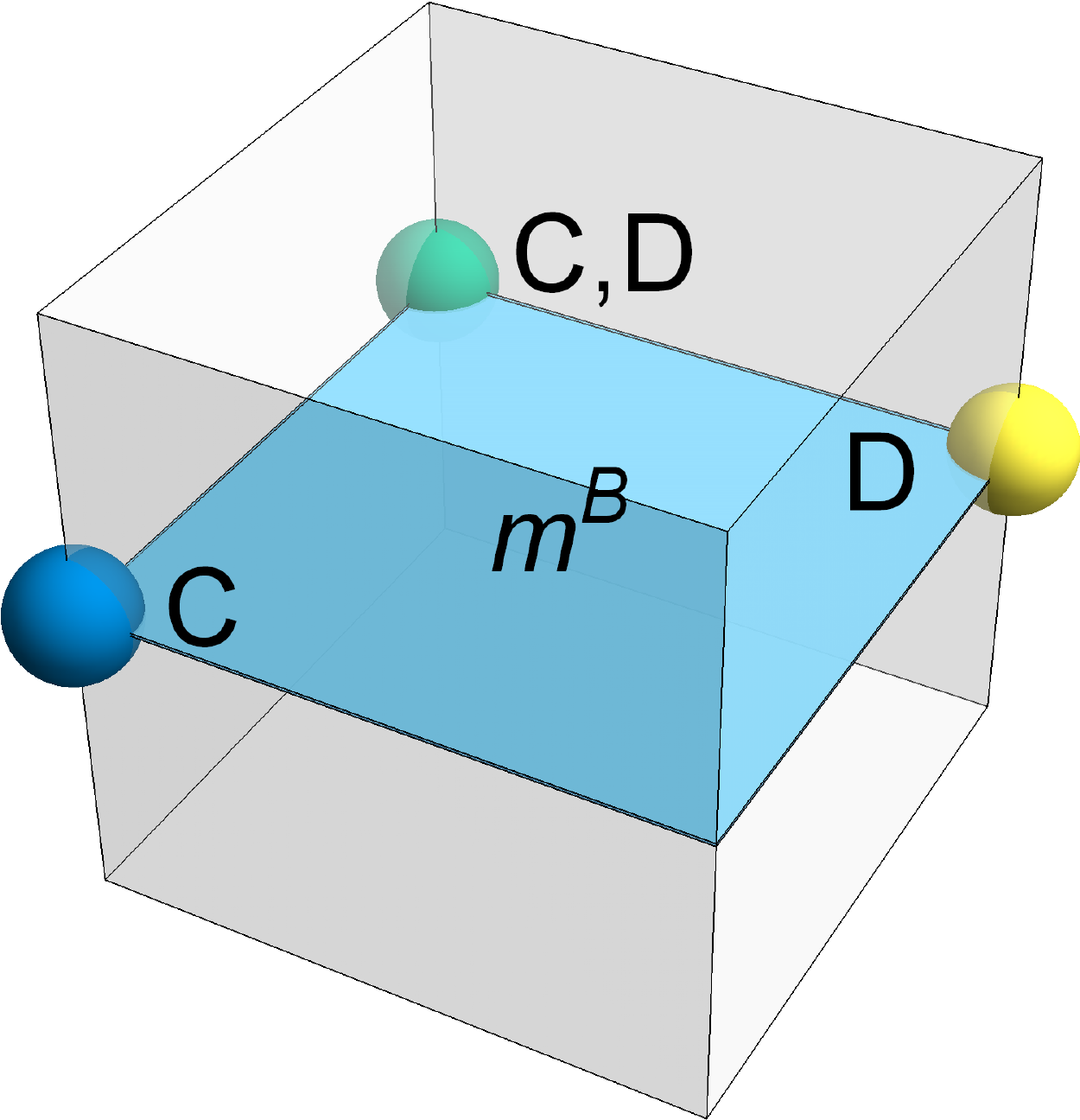}}
    \subfloat[\label{fig:BCode_magneticFractal}]{\includegraphics[width=0.6\columnwidth]{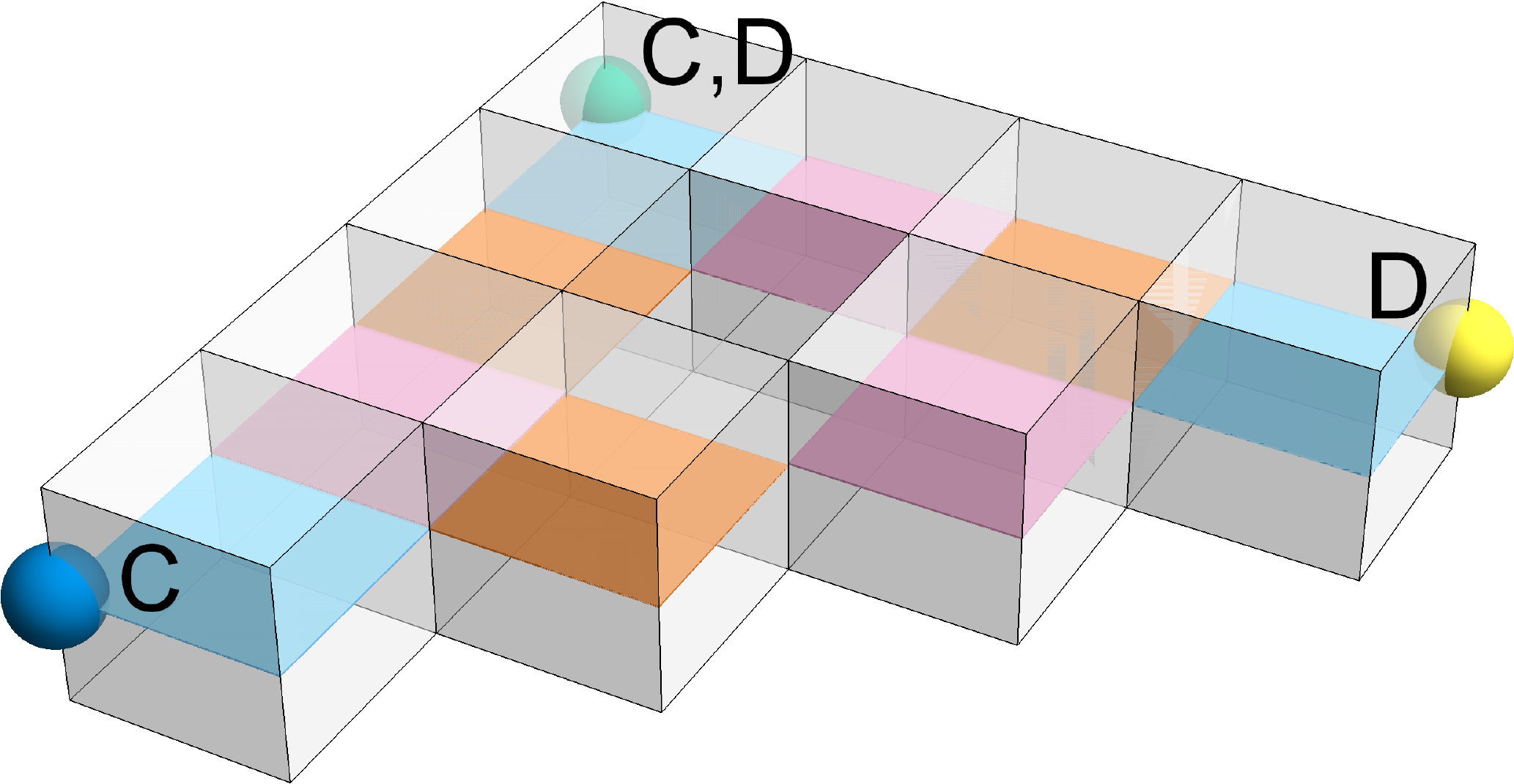}}\\
    \subfloat[\label{fig:BCode_CandDOperators}]{\includegraphics[width=0.6\columnwidth]{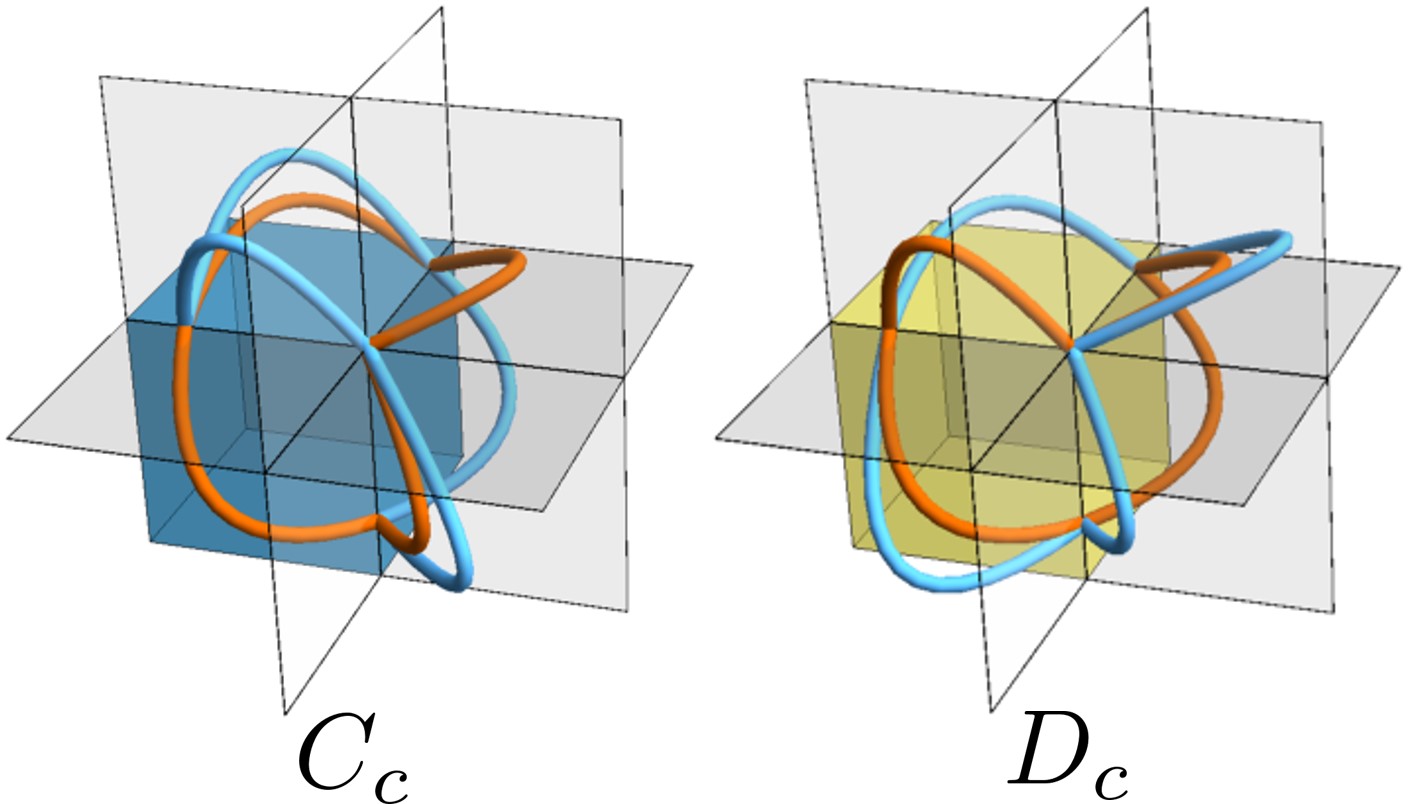}}
    \caption{Magnetic excitations in the defect network construction for the B code. (a) A B-layer magnetic membrane operator (blue sheet) which creates excitations only at three of its corners (colored spheres). (b) A magnetic membrane operator with fractal-like support which creates isolated, widely separated excitations. Orange (resp. blue) is a membrane operator in the A (resp. B) toric code layer, and purple is a product of membrane operators in both layers. 
    (c) The B code $C_c$ and $D_c$ operators on the blue and yellow cubes respectively represented as 3+1D toric code string operators (orange and blue are A and B layer string operators respectively) using the correspondence of local operators with condensed objects shown in Fig.~\ref{fig:BCode_electricCorrespondence}. The excitations in (a) are labeled by considering which membrane operators commute and anticommute with these string operators.}
\end{figure}

The magnetic excitations are created by applying open membrane operators whose boundaries lie entirely on the boundaries of 3-strata. Since $m$ excitations are condensed on the 2-strata, excitations only occur when the membrane operators intersect 1-strata in certain patterns. The most convenient membrane operators to investigate are those which lie parallel to a 2-stratum, an example of which is shown in Fig.~\ref{fig:BCode_membrane}. Due to the condensation of $m^A_4$ and $m^B_4$ in Eq.~\eqref{eqn:BCodeYCondensations} (see Fig.~\ref{fig:BCode_CubeLabels} for labeling), the front-most $y-$oriented link in Fig.~\ref{fig:BCode_membrane} does \textit{not} have an excitation on it, while the other three $y$-oriented links do because single $m$ excitations in this 3-stratum are not condensed on those links. (We will shortly explain why we have labeled the excitations with B code labels $C$ and $D$.) However, since some $m$ bound states are condensed on the 1-strata, applying several such membrane operators on neighboring 3-strata can remove these excitations provided the layers of the membrane operators are chosen carefully. For example, the fractal-like membrane operator in Fig.~\ref{fig:BCode_magneticFractal} creates three widely separated excitations with nothing in between. These patterns of excitations are exactly the ones which can be created in the B code using patterns of Pauli $X$ operators analogous to those shown for Pauli $Z$ operators in Fig.~\ref{fig:BCode_ABFractal}.

We now explain how we identify the excitations  created by magnetic membranes on the 1-strata. Using the correspondence between local Pauli $Z$ operators and the creation of condensed electric particles on the 1-strata, illustrated for example in Fig.~\ref{fig:BCode_electricCorrespondence}, we can represent the action of $C_c$ and $D_c$ as creating certain condensed sets of electric particles at the 1-strata and annihilating pairs inside the 3-strata, that is, as a complicated set of closed toric code electric string operators. 

These string operators are shown in Fig.~\ref{fig:BCode_CandDOperators}. These string operators anticommute with some of the membrane operators in question. For example, if the B-layer membrane operator in Fig.~\ref{fig:BCode_membrane} lives in cube 5 (using the same labeling as Fig.~\ref{fig:BCode_CubeLabels}) in Fig.~\ref{fig:BCode_CandDOperators}, it will anticommute with the $D_c$ operator on the highlighted cube (cube 8) but will commute with the $C_c$ operator on that cube. Hence the $m^B$ operator in question creates a $D_c$ excitation on cube 8. To obtain the labeling in Fig.~\ref{fig:BCode_membrane}, we must associate this B code excitation in a 3-stratum to a 1-stratum. Such a correspondence arises from the fact that in Fig.~\ref{fig:BCode_CandDOperators}, string operators only act on three of the twelve 1-strata that border the highlighted 3-stratum, one of each orientation. It is easy to check that each 1-stratum is involved only in a single $C_c$ operator and a single $D_c$ operator. In this sense, magnetic excitations on a 1-stratum can be uniquely assigned to a 3-stratum. In this way, the excitation labeled $D$ in Fig.~\ref{fig:BCode_membrane} in the defect construction language can be associated to a $D_c$ excitation on cube 8 in the B code language.  One can further check that there exist membrane operators which move magnetic excitations between the three 1-strata associated to the same cube (but not to any other 1-strata) without creating additional excitations, which is why the correspondence between 3-strata and 1-strata is not one-to-one.

\subsection{Nets and relations}

\begin{figure*}
    \centering
    \includegraphics[width=1.4\columnwidth]{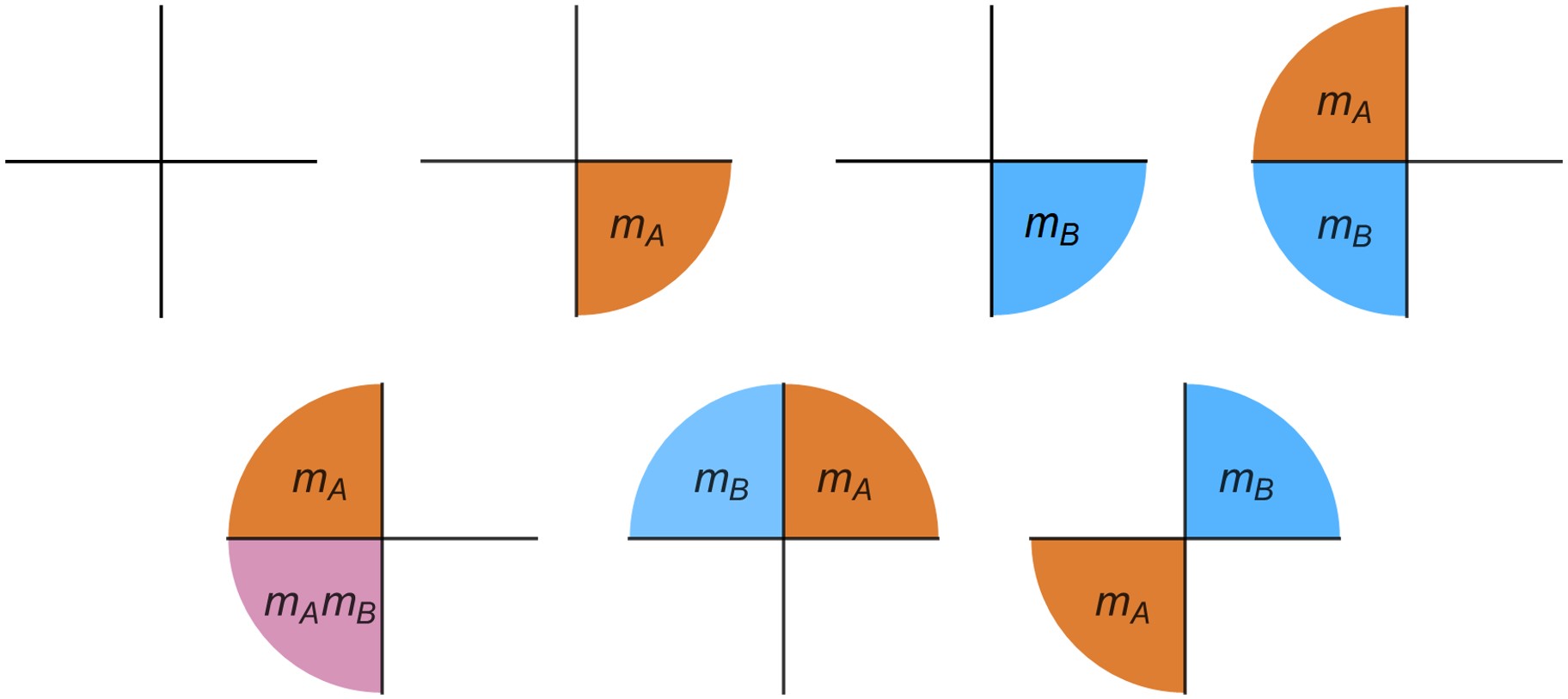}
    \caption{Allowed membrane configurations on the $x$-oriented 1-strata for the defect construction of Haah's B code. The perspective is looking down the 1-strata from the positive $x$-axis towards the origin. Orange (resp. blue) membranes are nets from the $A$ (resp. $B$) toric code layers. 
    }\label{fig:BCode_1StrataMembranes}
\end{figure*}

In this subsection, we explain the condensation picture of the previous subsection in terms of membrane-net diagrams and relations on them.

The ground state wavefunction in our construction can be written as a superposition
\begin{equation}
     |{\Psi}\rangle = \sum_{\text{M}} \phi(\text{M}) |\text{M} \rangle.
\end{equation}
where $M$ is a membrane configuration and $\phi(M)$ is the weighting determined by linear relations on the membranes. For our case, $\phi(M)=+1$ for allowed net-diagrams and $\phi(M)=0$ otherwise. Since the ambient topological order is the bilayer 3+1D toric code, there are two membrane colors, which we refer to as orange and blue. As before, ``allowed" net-diagrams are those which satisfy constraints which we specify presently.

The 3-stratum constraint is that each color of membrane is locally closed independently, that is, closed paths on the interior of a 3-stratum must intersect an even number of surfaces of each color. In the presence of a 2-stratum defect, membranes of either color may end freely on the 2-stratum without constraint. That is, we independently impose the same 3- and 2-stratum constraints used in the X-Cube construction on each color of 3d toric code.

The 1-strata have constraints which couple the two membrane colors. Each 1-stratum interfaces with four 3-stratum bilayer toric codes. The set of membranes which may freely terminate on a 1-stratum is given by the set of $m$ particles which are condensed in Eqs.~\eqref{eqn:BCodeXCondensations}-\eqref{eqn:BCodeZCondensations}, where $m^A$ (resp. $m^B$) corresponds to an orange (resp. blue) net. A generating set of allowed configurations for the $x$-oriented 1-strata are shown in Fig.~\ref{fig:BCode_1StrataMembranes}; the allowed configurations on the other 1-strata are obtained analogously.

Any membrane configuration surrounding a 0-stratum which satisfies the 1-, 2-, and 3-stratum constraints is admissible.

In Appendix~\ref{app:BCodeLatticeModel}, we construct an exactly solvable Hamiltonian for our defect network construction of Haah's B code. Our arguments from this section, along with the exactly solvable model, demonstrate the fact that our construction can indeed realize type-II fracton models; we will demonstrate in the next section that we can also realize models with non-Abelian fractons.


\section{Non-Abelian fracton model from topological defects in 3+1D \texorpdfstring{$D_4$}{D4} gauge theory}
\label{sec:nonabelian}

The defect picture developed for the X-Cube model in Section~\ref{sec:Xcube} provides a roadmap for building new fracton models, including those hosting non-Abelian excitations with restricted mobility. In this section, we discuss a defect network construction for a non-Abelian fracton model based on $D_4$ gauge theory, distinct from non-Abelian fracton models that have previously appeared in the literature. 
Similar to the X-Cube defect network, the model is constructed from three stacks of coupled defect layers within an ambient 3+1D $D_4$ gauge theory bulk. The defect layers serve to restrict the mobility of the non-Abelian bulk particle, thereby promoting it to a fracton. We follow the previous sections by first describing the construction in terms of excitations condensing on defects, and secondly in terms of superpositions of allowed configurations appearing in the ground state wavefunction. In App.~\ref{app:d4lattice} we describe a local lattice Hamiltonian that realizes the non-Abelian topological defect network construction of this section. 

\subsection{Review of \texorpdfstring{$D_4$}{D4} gauge theory} 
\label{sec:repd4}

The group $D_4$ corresponds to the symmetries of a square and is specified by 
\begin{align}
    D_4 = \left\langle r,s \, \large| \, r^2=s^4=rsrs=1
    \right\rangle 
    \, ,
\end{align}
with center given by $\langle s^2 \rangle$.
There are five conjugacy classes $\{1\}$,  $\{s^2\}$,  $\{s,s^3\}$,  $\{r,r s^2\}$,  $\{rs,rs^3\}$, and hence five irreducible representations (irreps): four of dimension one and one of dimension two. 
The character table is given by 
\begin{align}
\begin{tabular}{ l r r r r r }
 & $1$  & $s^2$ & $s$ & $r$ & $rs$  \\
 \hline
$\chi_{00}$ & 1  & 1 & 1 & 1 & 1 \\
$\chi_{01}$ & 1 & 1 & $- 1$ & 1 & $-1$  \\
$\chi_{10}$ & 1 & 1 & 1 & $-1$  & $-1$  \\
 $\chi_{11}$ & 1 & 1  & $-1$  & $- 1$ & 1 \\
$\chi_\sigma$ & 2 & $-2$ & 0 & 0 & 0
\end{tabular}
\label{chartbl}
\end{align}
Notice that the first four irreps obey the obvious $\mathbb{Z}_2 \times \mathbb{Z}_2$ fusion rules, while 
\begin{align}
   {ij} \otimes \sigma = \sigma \, ,
    &&
    \sigma \otimes \sigma = \sum_{i,j} ij \label{eq:sigma fusion}
    \, ,
\end{align}
obey a $\mathbb{Z}_2$ grading, where $i,j\in \mathbb{Z}_2$. Physically, the $\mathbb{Z}_2$ grading on the particles is induced by their braiding with the Abelian $s^2$ loop excitation. 
The non-trivial $F$-symbols of the fusion category Rep$(D_4)$ are given by 
\begin{align}
    [F_{k\ell}^{\sigma\, ij \, \sigma}]^{\sigma}_{\sigma} = [F_\sigma^{ij\, \sigma \, k \ell}]^{\sigma}_{\sigma} = 
    2 \, [F_\sigma^{\sigma \sigma \sigma}]^{i j}_{k \ell} = (-1)^{ik + j \ell} \, .
\end{align}
Rep$(D_4)$ furthermore admits a trivial braiding.

In three spatial dimensions, the pointlike gauge charges of $D_4$ gauge theory are described by Rep$(D_4)$ and the looplike flux excitations are locally labelled by the conjugacy classes of $D_4$~\cite{dijkgraaf}. 
The pointlike charges can be measured by braiding flux loops around a charge, while the looplike fluxes can be measured by braiding charges around a loop.  The braiding phase between an irrep labelled charge and a conjugacy class labelled flux is given by the phase of the corresponding entry in the the character table above.\footnote{For example, the wave function picks up a minus sign if the $11$ charge is braided with the $\{s,s^3\}$ flux, or if the $\sigma$ charge is braided with the $s^2$ flux.}

\subsection{Condensation on defects} 

In this section, we specify the defects used to construct our non-Abelian fracton model in terms of the excitations that condense on them. We again consider a cubic lattice stratification of the 3-torus with degrees of freedom living on a much finer grained cubic lattice. 
Cubes of $D_4$ gauge theory on the 3-strata with gapped boundaries are coupled together in a similar way to the X-Cube defect TQFT to achieve a model where the non-Abelian $D_4$ gauge charge is promoted to a fracton. 

\begin{figure}[t]
    \centering
        \includegraphics[width=.97\linewidth]{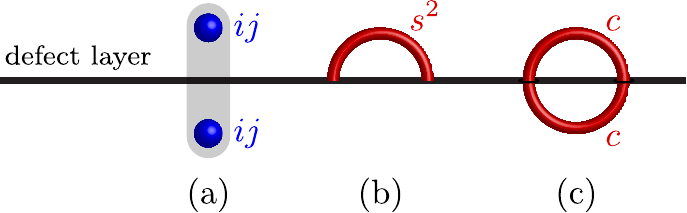}
        \caption{
        A 2-stratum defect layer (black line) in the non-Abelian $D_4$ model described in the main text.
The defect allows $s^2$ membranes to terminate on the 2-strata. 
    The following excitations can be created near a defect layer using local operators:
    {\bf(a)} pairs of Abelian charges labeled $ij$ on opposite sides of the defect;
    {\bf(b)} a flux arc labeled by the conjugacy class $s^2$, and {\bf (c)} a flux loop on opposing sides of the defect labeled by the same conjugacy class $c$.
    }\label{fig:D42strata}
\end{figure}

On the 2-strata, we introduce a topological defect on which the flux loop excitations labelled by $s^2$ condense, but no other topological excitations from single 3-strata condense. This has the effect of restricting the non-Abelian $\sigma$ particle from moving across the 2-strata (since $\sigma$ braids non-trivially with $s^2$ flux loops), while other uncondensed topological excitations are free to pass through. 
The full set of excitations condensing on the 2-strata are labelled
\begin{align}
\label{eq:cond2strat}
    \{ s^2_+, \; s^2_-, \; c_+ c_-, \; ij_+ ij_- \}
    \, ,
\end{align}
where $c_\pm$ runs over all conjugacy classes, $ij_\pm$ runs over the Abelian charges, and $\pm$ indicates which adjacent cube the topological excitations originate from, see \figref{fig:D42strata}. 

\begin{figure}[t]
    \centering
    \subfloat[]{\includegraphics[width=.4\linewidth]{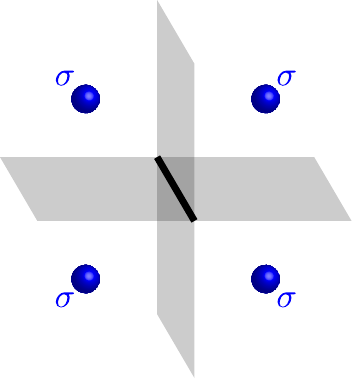}}
    \subfloat[]{\includegraphics[width=.4\linewidth]{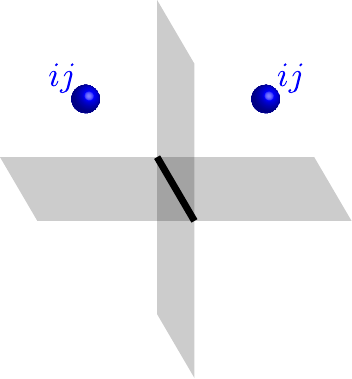}} \\
    \subfloat[]{\includegraphics[width=.4\linewidth]{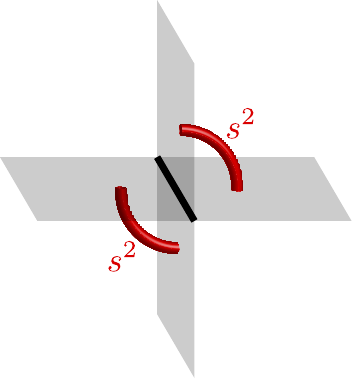}}
    \subfloat[]{\includegraphics[width=.4\linewidth]{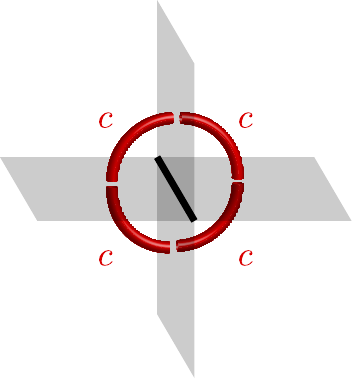}}
\caption{A representative set of 1-stratum condensations described by Eq.~\eqref{eq:cond1strat}. 
{\bf (a)} Four non-Abelian $\sigma$ charges can simultaneously condense to the vacuum on a 1-stratum.
{\bf (b)} The Abelian charges labeled $ij$ are allowed to freely pass through the 2- and 1-strata. 
{ \bf (c) } Any pair of $s^2$ flux strings can condense at the 1-strata. 
{\bf (d) } Any quadruple of flux strings (labeled by a conjugacy class $c$) meeting at a 1-stratum can condense into vacuum. }
\label{fig:D41s}
\end{figure}

On the 1-strata, we pick a topological defect where the following excitations condense 
\begin{align}
\label{eq:cond1strat}
  \{ s^2_\alpha s^2_\beta, \; c_1 c_2 c_3 c_4, \; {ij}_\alpha {ij}_\beta, \; \sigma_1 \sigma_2 \sigma_3 \sigma_4 \} 
    \, .
\end{align}
Here, $c$ runs over all conjugacy classes; $ij$ runs over the Abelian irreps; and $\alpha\neq\beta$ run over the quadrants 1,2,3,4, see \figref{fig:D41s}. 
We remark that when the non-Abelian excitation $\sigma_1 \sigma_2 \sigma_3 \sigma_4$ is brought onto the 1-stratum, it can fuse into any of the Abelian topological charges labeled by $ij$. 
For $ij\neq 00$, the leftover particle is a 1-stratum excitation (however we note that such excitations can be moved off the 1-stratum as the $ij$ particles do not have any mobility constraints). 
Thus, only the vacuum channel of $\sigma_1 \sigma_2 \sigma_3 \sigma_4$ is condensed at the 1-stratum.

We remark that we are forced to include $c_1 c_2 c_3 c_4$ and ${ij}_\alpha {ij}_\beta$ in the condensate for consistency with the adjacent 2-strata. 
While the above condensing particles may not fully specify the defect, they suffice to understand the properties of the fracton model thus constructed.
We provide a full description of the defect in terms of nets and relations in Sec.~\ref{sec:netsandrelationsd4}, and an explicit lattice model in App.~\ref{app:d4lattice}. 

The looplike excitations, $c$, have a well-defined braiding with a 0-stratum that can detect the presence of a single (or an odd numbers of like) point charges there. Furthermore, there is a braiding process of an arclike $s^2$ excitation tracing out a hemisphere about the 0-stratum for each oriented axis $\pm \hat{x}, \pm \hat{y},\pm \hat{z},$ that can detect an odd number of $\sigma$ charges above or below the ${yz},{xz},{xy}$ plane, respectively. 
Together, these braiding processes can detect $\sigma_\alpha \sigma_\beta$ pairs for $\alpha\neq \beta$, but braid trivially with the $\sigma_1 \sigma_2 \sigma_3 \sigma_4 $ quadruples that condense on the 1-strata. 
On the 0-strata, we choose a topological defect that is trivial under the aformentioned braiding processes. This ensures that the defect does not induce any further condensation nor pin any pointlike topological excitations, including $\sigma_\alpha \sigma_\beta$. This is easy to see for the Abelian particles, and for the non-Abelian particles utilizes the fact that a single $\sigma_\alpha$ particle picks up a sign under the $s^2$ braiding processes that pass through its octant.

\subsection{Mobility constraints} 

In this section, we analyze the mobility of the topological excitations in the $D_4$ defect network model. 
We demonstrate that the non-Abelian $\sigma$ particle, and arcs of the Abelian $s^2$ flux excitation obey similar constraints to the particle and flux excitations in the X-Cube defect network, respectively. On the other hand, the Abelian particles and non-Abelian flux loops remain fully mobile. 

The non-Abelian $\sigma$ charge excitations cannot freely pass through the 2-strata since they braid non-trivially with the Abelian $s^2$ flux loops that condense there. Similar to the X-Cube defect network, they may be created in quadruples on the four 3-strata adjacent to a 1-strata, see \sfigref{fig:D41s}{a}. Furthermore, no additional mobility is endowed to the $\sigma$ particles by the defects on the 0-strata. Hence, the $\sigma$ excitations become non-Abelian fractons with analogous mobility to those in the X-Cube model. 

An arc of Abelian $s^2$ flux between adjacent 2-strata can move along a line via the hopping process shown in \sfigref{fig:Xcube excitations}{e-f}. 
Shrinking any arc from a single quadrant onto the adjacent 1-strata leads to an equivalent excitation on the 1-strata.  
Three such excitations can be created in a 3-strata along the $\hat{x},\hat{y},$ and $\hat{z}$ axes, see \sfigref{fig:Xcube excitations}{c}. These properties lead to the same mobility as the X-Cube lineon excitations. 

The Abelian $ij$ charges are fully mobile, as they can pass through the 2-strata due to the condensation of $ij_+ij_-$ there. Similarly, the non-Abelian $c$ flux loops excitations are fully mobile as they can pass through the 2-strata, due to the condensation of $c_+ c_-$ there, and the 1-strata, due to the condensation of $c_1 c_2 c_3 c_4$ there.

\subsection{Nets and relations} 
\label{sec:netsandrelationsd4}
In this section we present a defect network construction of the non-Abelian $D_4$ fracton model, introduced in the previous subsections, in terms of a membrane-net condensate. 
Following Eq.~\eqref{eq:nets} we describe allowed configurations of $D_4$ membranes on the various strata and the local relations between allowed configurations with the same coefficient in the wavefunction. 

\subsubsection*{3-strata} 

Within the 3-strata the allowed configurations are oriented membranes labelled by elements of $D_4$ that satisfy the group multiplication rule at their junctions, see Eq.~\eqref{eq:lgtedge}. 
The local relations are generated on elementary volumes by fusing spheres, labelled by single group elements, into the bounding membranes on the lattice,  see Eq.~\eqref{eq:lgtcube}. 

\subsubsection*{2-strata}

On the 2-strata only membranes labelled by $s^2$ are allowed to end; all other membranes must pass through while maintaining the same label (up to multiplication with $s^2$). 
The local relations include those from the 3-strata, while also allowing hemispheres labelled by $s^2$ that end on either side of the 2-strata to be created and fused into the existing membrane-net, see Fig.~\ref{fig:schematic1}. Furthermore, the $s^2$ membranes ending on the 2-strata from opposite sides may freely commute past one another. This gapped domain wall is specified by the subgroup 
\beq
K = \langle g_+ g_-, s_+^2 \, |\,  \forall g\in  D_4 \rangle \cong D_4 \times \mathbb{Z}_2 \, , 
\eeq
of membranes that may end on the 2-strata. 

\subsubsection*{1-strata} 

Choosing a 1-stratum defect consists of choosing a generating set of membranes that are allowed to terminate on the 1-strata. This set must be consistent with the 2-strata defects, which means that we may only choose among membranes labeled by $s^2$ in any number of quadrants and any other label spread over all four quadrants. The defect we choose allows pairs of $s^2$ membranes in adjacent quadrants and membranes of any other label matching over all four quadrants, corresponding to the subgroup 
\beq
M = \langle g_1 g_2 g_3 g_4, s_1^2s_2^2, s_2^2s_3^2, \, |\,  \forall g\in D_4 \rangle \cong D_4 \times \mathbb{Z}_2^2 \, .
\eeq
The local relations again include those of the 3-strata, while also allowing a hemisphere labelled by $s^2$ over two adjacent quadrants to be fused into the existing membrane-net configuration. 

\subsubsection*{0-strata}

Around a 0-stratum the allowed configurations are closed membrane-nets with any group labels, and hemispheres labelled by $s^2$ over four 3-strata adjacent to a single 1-stratum. The local relations once more include those of the 3-strata, while also allowing a hemisphere labelled by $s^2$ spread over four adjacent 3-strata to be fused into the membrane-net. 

\subsection{Relation to other models} 
\label{sec:remarks}

Here, we briefly discuss how our $D_4$ fracton model relates to other non-Abelian fracton models. First, We remark that the coexistence of non-Abelian fracton excitations with fully mobile Abelian excitations in our model is similar to those in the swap-gauged bilayer X-Cube model of Refs.~\onlinecite{bulmashgauging,premgauging}, although the models are not the same.\footnote{One way to see this is the fact that our model contains three distinct, fully mobile non-trivial point particles, while the gauged bilayer X-cube model contains only one.} Next, 
we note that a distinct non-Abelian fracton model based on $D_4$ gauge theory can also be realized through $p$-string condensation~\cite{han,sagar,nonabelian,cagenet}. In this picture, isotropic layers of 2+1D $D_4$ are driven into a fracton phase by condensing $p$-strings composed of point-like $s^2$ fluxes, with the end points of open strings becoming fractons in the flux-condensed phase. However, the fractons would necessarily be Abelian in this case, being labelled by $s^2$. Indeed, due to their loop-like nature, it does not appear to be possible to obtain non-Abelian fractons by condensing $p$-strings unless the 2+1D layers are embedded in an ambient 3+1D TQFT~\cite{GeneralizedSMN}. This is in contrast with the $D_4$ topological defect network, which hosts intrinsically non-Abelian fractons, and does not arise as the consequence of a condensation driven phase transition. 

In this section we have explained how a non-Abelian fracton model can be constructed from a topological defect network. This example confirms that topological defect networks are sufficiently general to capture both non-Abelian and panoptic fracton topological order. 
An explicit lattice Hamiltonian realization of this non-Abelian model is described in Appendix~\ref{app:d4lattice}. 


\section{Classifying phases with topological defect networks}
\label{sec:classifyingphases}

Topological defect networks provide a framework for the classification of all gapped phases of matter, particularly fracton topological orders, in terms of purely topological data associated to defects. This recasts the classification problem into the language of TQFT, which has proven extremely useful for the classification of topological phases in 2+1D. 
This is advantageous as it brings the mathematical tools of TQFT to bear upon fracton models. 
For example, the ground space degeneracy of a fracton model expressed as a topological defect network can be calculated by counting the inequivalent fusion channels to vacuum upon tensoring together all defects in the network with some given finite boundary conditions~\cite{Lan2015,wan2015twisted}. In this section, we collect some remarks on the construction, classification, and equivalence of fracton topological phases in the defect network formalism. 

\subsection{No type-II fractons in 2+1D}

\begin{figure}[t]
    \centering
    \subfloat[\label{fig:No2DFractonsa}]{\includegraphics[height=0.42\columnwidth]{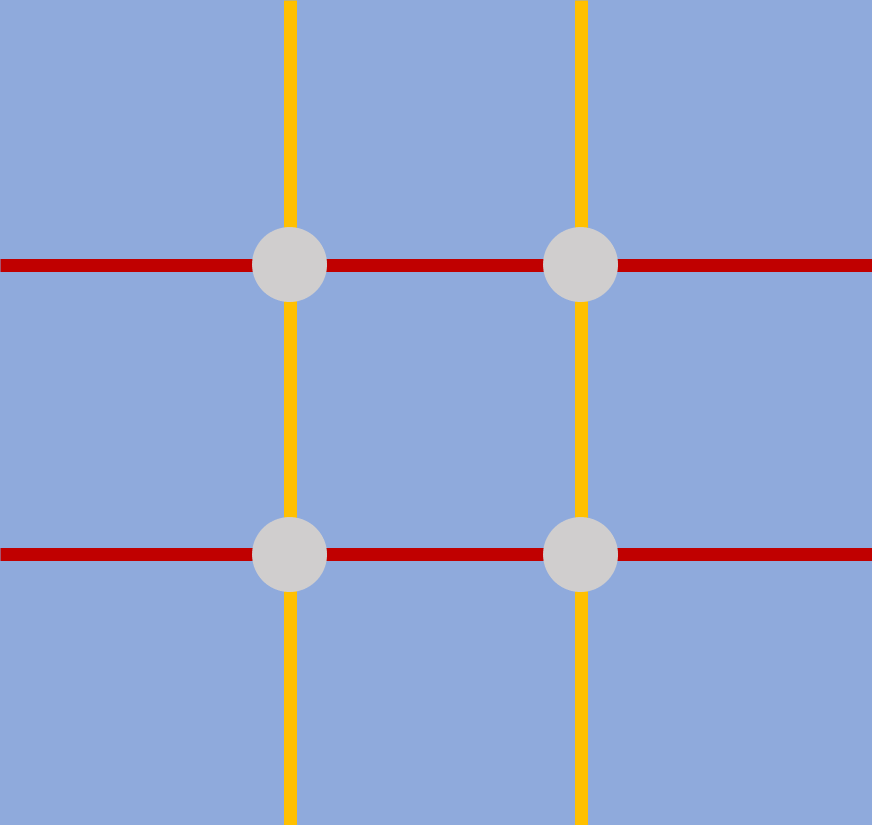}}
    \hspace{.5cm}
    \subfloat[\label{fig:No2DFractonsb}]{\includegraphics[height=0.42\columnwidth]{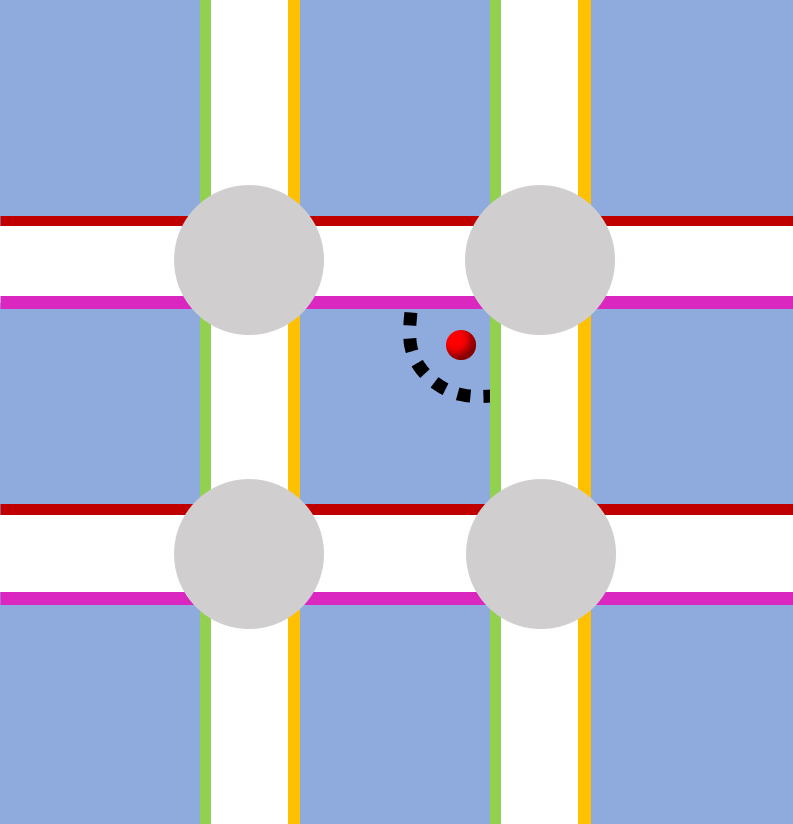}}
    \\
    \subfloat[\label{fig:No2DFractonsc}]{\includegraphics[height=0.25\columnwidth]{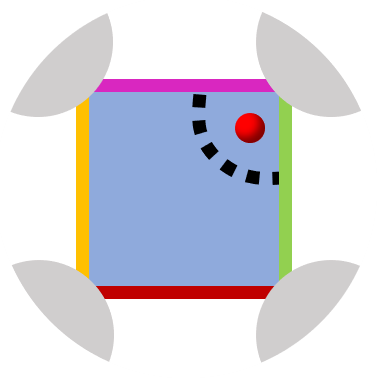}}
    \hspace{.5cm}
    \subfloat[\label{fig:No2DFractonsd}]{\includegraphics[height=0.25\columnwidth]{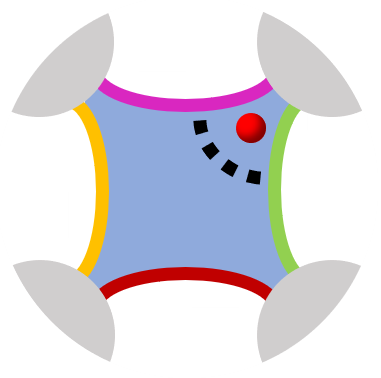}}
    \hspace{.5cm}
    \subfloat[\label{fig:No2DFractonse}]{\includegraphics[height=0.25\columnwidth]{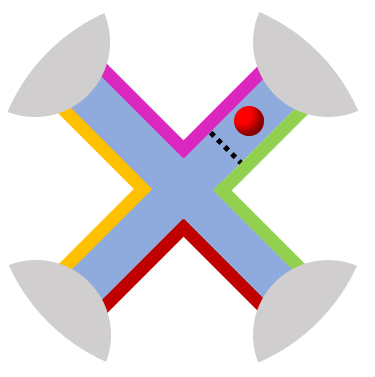}}
    \caption{(a) A topological defect network in 2+1D. (b) Gapped boundary to vacuum domain walls that prevent particle mobility and a topological charge (red) being measured by a conjugate string operator near a corner. (c-e) Applying the inflation trick to a 2-stratum from (b).  }
\end{figure}

Before considering 2+1D, we point out it has previously been shown that there are no (intrinsic) topological phases of matter in 1+1D~\cite{Chen2011,Schuch2011}. 
Therefore, there are no fracton topological phases of matter in 1+1D.

In 2+1D, the question of existence of fracton topological phases is less trivial.
It has already been shown that translation invariant Pauli stabilizer models cannot be fractonic~\cite{Haah2018b}.
However, it has been argued that higher rank $U(1)$ gauge theories in 2+1D can give rise to chiral fracton phases~\cite{chiral,emergent,gromov}. 

In this subsection, we argue that the translation invariant topological defect network construction gives rise to no stable type-II fracton topological phases of matter in 2+1D. 
To be precise, we are referring to gapped phases that are stable to all local perturbations and have deconfined topological excitations with mobility constraints. This is to be contrasted with SSPTs or subsystem symmetry breaking phases, which require symmetry to be non-trivial. 
Taken together with our main \hyperref[conjecture]{conjecture}, this implies that there are no stable gapped translation invariant fracton orders in 2+1D. 
It is worth noting that this no-go result also excludes the possibility of any chiral translation invariant gapped fracton type-II topological phases, as we do not assume the existence of a commuting projector Hamiltonian in our argument.

The starting point for our argument is a translation invariant square lattice defect network (see Fig.~\ref{fig:No2DFractonsa}) where the unit cell is a single square. Any other translation invariant lattice can be mapped to this case after some finite coarse-graining (which does not affect the topological phase of matter produced). 
For non-trivial topological order to emerge, the 2-strata must be chosen to contain some non-trivial topological order.\footnote{Conversely, if the 2-strata were in the trivial phase, all excitations on the 1- and 0-strata would be in the trivial superselection sector, and the resulting phase would be trivial.} In our argument we make reference to a microscopic parent Hamiltonian for the defect network, but this need not be a commuting projector Hamiltonian. 

We next consider the gapped domain walls on the 1-strata. In general, some topological charges could pass through the domain walls. Notice that these charges could also be permuted when they pass through a domain wall. But the overall permutation action must have a finite order since the symmetry group of any anyon theory is finite~\cite{barkeshli2014symmetry}. Hence, after a finite amount of coarse graining, the permutation action becomes trivial. 
The particles within a 2-stratum that do not condense on its boundaries can be divided according to their 2d, 1d, or 0d mobility via the 1-strata.

We now argue that there are no defect networks where all the uncondensed particles in the 2-strata are stable fractons, i.e. have 0D mobility. 
That is, we are ruling out models where every superselection sector supported on a single 2-strata is a fracton. 
In particular, this excludes type-II fracton phases.\footnote{%
In addition to excluding type II phases, the argument also excludes the possibility of a phase where all 2-strata superselection sectors are fractons, but where a composite of fractons on nearby 2-strata has 1d or 2d mobility.} To begin the argument, let us assume that there exists a defect network where all the particles in the 2-strata are fractons.
We proceed by first noting that such defect networks must have domain walls that are equivalent to pairs of gapped boundaries to vacuum, as depicted in Fig.~\ref{fig:No2DFractonsb}. If either the horizontal or vertical domain walls are not equivalent to pairs of gapped boundaries to vacuum, then some particles may pass through them, thus picking up 1d or 2d mobility, which contradicts our initial assumption. 
Thus if we have a 2+1D fracton phase as assumed, the 1-strata must be very special gapped boundaries; namely, there must be a thin slice of vacuum between the 2-strata.
In this case, we can inflate the 1-strata into the 2-strata as shown in \figref{fig:No2DFractonse}. The 2-strata phase has now been squeezed into a 1+1D phase, which can not support any stable deconfined topological excitations \cite{Chen2011,Schuch2011}, and therefore also no fractons.\footnote{%
  Under this dimensional reduction, anyons from the 2-strata are mapped to domain walls (of a symmetry breaking phase) in the squeezed 1+1D phase. These domain walls are not stable topological excitations since they are confined when explicit symmetry-breaking perturbations are added to the Hamiltonian.}
This completes the argument.

We remark that different choices of 0-strata can only produce various unstable subsystem symmetry breaking models from the defect networks discussed in the preceding paragraph, see Fig.~\ref{fig:No2DFractonsb}.

A direct generalization of the above argument to 3+1D implies that topological defect networks constructed by coupling together 3-strata (with gapped boundaries to vacuum on the 2-- and 1-strata) via their corners at 0-strata alone must also be trivial. 
The non-trivial topological defect network constructions we have presented avoid this restriction by involving non-trivial topological defects along the 1-strata.

\subsection{Phase preserving defect network equivalences} 
\label{defect_deformation}

\begin{figure}[t]
    \centering
\includegraphics[width=0.97\linewidth]{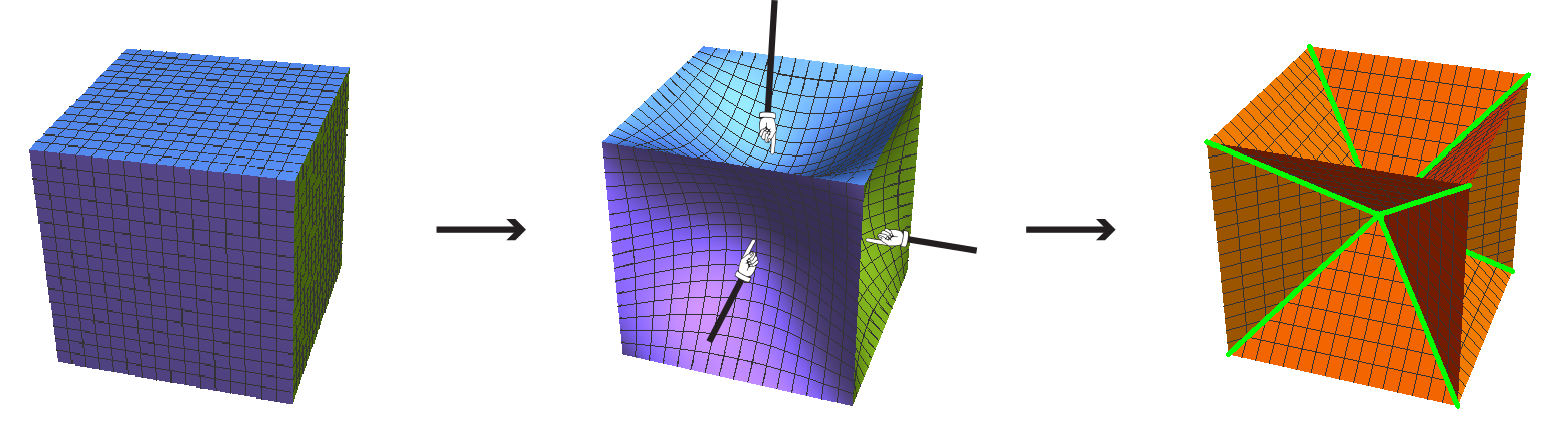}
\caption{Inflation trick: Certain 2-strata defects are effectively gapped boundaries to vacuum. 
In this case, we can deform the 2-strata bounding a given 3-stratum into the center of the 3-stratum.
The 3+1D topological order in each 3-stratum undergoes a dimensional reduction to sheets of (effectively) 2+1D topological order, and one finds a purely 2+1D defect network for the same phase.  }
    \label{fig:inflation}
\end{figure}

We have proposed that defect networks provide a useful framework for classifying fracton phases. In order to do so, it is important to determine if fracton phases are in one-to-one correspondence with defect networks, or at least with some subset of the data used to define a defect network. 
In this subsection, we show that this is not the case; for both the X-Cube model and Haah's B code, there is a deformation of the defect network which changes both the stratification of space and the topological phase on the 3-strata, but preserves the defect network's phase of matter. 

The basic idea is to notice that some types of 2-strata gapped boundaries can be expanded into the 3-strata.
In particular, notice that for the defect construction of the X-Cube model described in Sec.~\ref{sec:Xcube}, the 2-strata are fully flux condensing. 
One can think of this boundary condition as having the 3+1D toric codes on either side of the 2-strata as separated by a thin slab of the trivial phase. 
We can then inflate this thin slab as shown in \figref{fig:inflation}, until the 2-strata bounding a 3-stratum all meet. 
At this point the 3d toric codes become (effectively) 2d toric codes; recall \figref{PinchingTrick}. 
One arrives at a defect network of 2d topological orders on a different stratification of space, but with the same emergent excitations.
One can further describe the defects using purely 2+1D gapped boundaries, which can be inferred from the dimensional reduction.
The old 1-strata are still present, and have the same condensations as in \eqref{onecelldefect}.
We remark that the same procedure can be used for the Haah B code described in Sec.~\ref{sec:HaahB}, as the 2-strata boundary conditions are fully flux-condensing boundaries of $\mathbb{Z}_2 \times \mathbb{Z}_2$ gauge theory.


\section{Conclusion}
\label{cncls}

In this work, we have demonstrated that a comprehensive variety of gapped fracton phases can be described by topological defect networks. Apart from capturing well-known phases representing different types of fracton models, we have also shown how defect networks naturally provide a constructive framework for finding new fracton phases. Based on our ability to fit the broad typology of gapped fractonic matter into our framework, we expect that all fracton phases admit a defect network description. As such, we \hyperref[conjecture]{conjecture} that topological defect networks realize all zero-temperature gapped phases of matter. As a byproduct of this conjecture, we have also argued that no type-II fracton phases exist in 2+1D gapped systems, thereby demonstrating the potential of topological defect networks as tools for the classification of phases of matter. 

Proving our conjecture is an important future direction. A first, and important, step in this direction is the construction of all known fracton models via topological defect networks. Preliminary results in this direction have been promising, as topological defect networks can capture all string-membrane-net models~\cite{SlagleSMN}, including their non-Abelian generalizations~\cite{GeneralizedSMN} which contain the cage-net models~\cite{cagenet} as a special case. This construction proceeds by first filling the 3-strata with an untwisted gauge theory based on a group containing an Abelian subgroup; next, one determines the 2-strata by coupling the 3+1D gauge theory to three stacks of a 2+1D topological order that contains an Abelian boson; this in turns determines the appropriate 1- and 0-strata. This approach also allows one to capture the non-Abelian fracton models of Ref.[~\onlinecite{GaugedLayers}] by coupling a stack of a 2d topological order that contains an Abelian boson to two stacks of 2d gauge theory via an appropriate choice of 1- and 0-strata. This is somewhat similar to the example discussed in App.~\ref{app:trivbulk}, which also has trivial topological orders on the 3-strata. 

On the other hand, a general recipe for converting simple translation-invariant Pauli stabilizer fracton models into defect networks is not yet known to us. This includes the prominent example of Haah's cubic code~\cite{haah}. 
As another example, we speculate that the recently introduced type-I gauged strong SSPT models~\cite{juvenSSPT} can be captured by making the appropriate choice of 3-cocycle on the gapped 2-strata boundaries in the X-Cube defect network, as this can reproduce the lineon braiding properties of such models. 
Further, we anticipate that the twisted fracton models introduced in Ref.[\onlinecite{twisted}] can be captured with a topological defect network construction similar to the X-Cube construction, but with an appropriate choice of non-trivial 3-cocycle on the gapped 2-strata boundaries. Since none of the ideas underlying topological defect networks rely on commuting projector Hamiltonians, it would also be useful to find defect network descriptions of chiral fracton phases~\cite{nonabelian,FujiLayer}.

An important issue, inherent to the search for defect network constructions of various models, is calculating the emergent phase of matter from defect data. 
In particular, even confirming that a given defect network results in a stable topological order can be non-trivial. This is because deciding what topological phase a model is in requires considering all possible topologically trivial operators. Such operators may span over many adjacent 3-strata, despite the strata being large compared to the lattice scale. 

A further related difficulty lies in determining equivalence relations on defect networks that lead to the same emergent topological phase of matter. In the previous section, we saw that defect networks on different stratifications could lead to the same emergent fracton order. Even seemingly different defect networks on the same stratification may lead to the same emergent phase of matter. 
Technically this is related to Morita equivalence of the chosen defects. 
As a simple example, we point out that any choice of defects that are related by fusing an invertible domain wall into the boundary of each 3-strata are equivalent. Developing an understanding of this generalized Morita equivalence relation on topological defect networks is an important open problem. 

As a natural extension of this work, it would also be interesting to systematically study gapped boundaries of fracton models within the defect network framework. While boundaries of the X-Cube and of Haah's code have previously been addressed~\cite{bulmashboundary,albertespec}, a general understanding is currently lacking. Similarly, incorporating global symmetries into our picture offers a route towards describing the allowed patterns of symmetry fractionalization on excitations with restricted mobility, and therefore also the class of symmetry enriched fracton orders.

The defect TQFT construction draws an interesting connection between fracton orders and crystalline symmetry protected topological (SPT) order.
In Refs.~[\onlinecite{else2019crystalline,song2017defect,huang2017defect}], it was shown that crystalline SPT phases can also be described by a network of defects.
However, the crucial difference is that while the defects that compose crystalline SPT phases are invertible defects, which are defects that can be cancelled out by bringing another defect next to it,\footnote{%
  For example, in toric code one can consider the so-called duality defect \cite{BombinDualityDefect}, which interchanges the $e$ and $m$ anyons. But two duality defects cancel each other out because swapping $e$ and $m$ twice results in no change. Therefore, the duality defect is invertible. The simplest example of a noninvertible defect is embedding a decoupled 2d toric code layer into any 3+1D TQFT. The 2d toric code can not be cancelled out; it is therefore a noninvertible defect.} the defects considered in this work are \emph{non-invertible}.
In Ref.~[\onlinecite{else2019crystalline}], the defect network construction was shown to be useful in the classification of crystalline SPT phases; it remains to be seen if gapped fracton phases could be classified with the help of defect networks.

\textbf{Note added:}  During the preparation of this manuscript, we became aware of a recent work~\cite{Wen2020} by Xiao-Gang Wen. While our work has some overlap with this article, the results contained in both were obtained independently.


\tocless{\section*{Acknowledgments}}

It is a pleasure to thank Maissam Barkeshli, Dominic Else, Jeongwan Haah, Michael Hermele, Sheng-Jie Huang, Zhu-Xi Luo, Wilbur Shirley, and Zhenghan Wang for stimulating discussions and correspondence. 
This work was initiated and performed in part at the Aspen Center for Physics, which is supported by National Science Foundation grant PHY-1607611.
D.A. is supported by a postdoctoral fellowship from the the Gordon and Betty Moore Foundation, under the EPiQS initiative, Grant GBMF4304.
D.B. is supported by JQI-PFC-UMD.
A.P. acknowledges support through a PCTS fellowship at Princeton University. 
K.S. is supported by the Walter Burke Institute for Theoretical Physics at Caltech.
D.W. acknowledges support from the Simons Foundation.


\bibliography{defectNetworks}

\begin{thebibliography}{121}%
\makeatletter
\providecommand \@ifxundefined [1]{%
 \@ifx{#1\undefined}
}%
\providecommand \@ifnum [1]{%
 \ifnum #1\expandafter \@firstoftwo
 \else \expandafter \@secondoftwo
 \fi
}%
\providecommand \@ifx [1]{%
 \ifx #1\expandafter \@firstoftwo
 \else \expandafter \@secondoftwo
 \fi
}%
\providecommand \natexlab [1]{#1}%
\providecommand \enquote  [1]{``#1''}%
\providecommand \bibnamefont  [1]{#1}%
\providecommand \bibfnamefont [1]{#1}%
\providecommand \citenamefont [1]{#1}%
\providecommand \href@noop [0]{\@secondoftwo}%
\providecommand \href [0]{\begingroup \@sanitize@url \@href}%
\providecommand \@href[1]{\@@startlink{#1}\@@href}%
\providecommand \@@href[1]{\endgroup#1\@@endlink}%
\providecommand \@sanitize@url [0]{\catcode `\\12\catcode `\$12\catcode
  `\&12\catcode `\#12\catcode `\^12\catcode `\_12\catcode `\%12\relax}%
\providecommand \@@startlink[1]{}%
\providecommand \@@endlink[0]{}%
\providecommand \url  [0]{\begingroup\@sanitize@url \@url }%
\providecommand \@url [1]{\endgroup\@href {#1}{\urlprefix }}%
\providecommand \urlprefix  [0]{URL }%
\providecommand \Eprint [0]{\href }%
\providecommand \doibase [0]{http://dx.doi.org/}%
\providecommand \selectlanguage [0]{\@gobble}%
\providecommand \bibinfo  [0]{\@secondoftwo}%
\providecommand \bibfield  [0]{\@secondoftwo}%
\providecommand \translation [1]{[#1]}%
\providecommand \BibitemOpen [0]{}%
\providecommand \bibitemStop [0]{}%
\providecommand \bibitemNoStop [0]{.\EOS\space}%
\providecommand \EOS [0]{\spacefactor3000\relax}%
\providecommand \BibitemShut  [1]{\csname bibitem#1\endcsname}%
\let\auto@bib@innerbib\@empty
\bibitem [{\citenamefont {Haah}(2011)}]{haah}%
  \BibitemOpen
  \bibfield  {author} {\bibinfo {author} {\bibfnamefont {J.}~\bibnamefont
  {Haah}},\ }\href {\doibase 10.1103/PhysRevA.83.042330} {\bibfield  {journal}
  {\bibinfo  {journal} {Phys. Rev. A}\ }\textbf {\bibinfo {volume} {83}},\
  \bibinfo {pages} {042330} (\bibinfo {year} {2011})}\BibitemShut {NoStop}%
\bibitem [{\citenamefont {Chamon}(2005)}]{chamon}%
  \BibitemOpen
  \bibfield  {author} {\bibinfo {author} {\bibfnamefont {C.}~\bibnamefont
  {Chamon}},\ }\href {\doibase 10.1103/PhysRevLett.94.040402} {\bibfield
  {journal} {\bibinfo  {journal} {Phys. Rev. Lett.}\ }\textbf {\bibinfo
  {volume} {94}},\ \bibinfo {pages} {040402} (\bibinfo {year}
  {2005})}\BibitemShut {NoStop}%
\bibitem [{\citenamefont {{Castelnovo}}\ and\ \citenamefont
  {{Chamon}}(2012)}]{castelnovo}%
  \BibitemOpen
  \bibfield  {author} {\bibinfo {author} {\bibfnamefont {C.}~\bibnamefont
  {{Castelnovo}}}\ and\ \bibinfo {author} {\bibfnamefont {C.}~\bibnamefont
  {{Chamon}}},\ }\href {\doibase 10.1080/14786435.2011.609152} {\bibfield
  {journal} {\bibinfo  {journal} {Philosophical Magazine}\ }\textbf {\bibinfo
  {volume} {92}},\ \bibinfo {pages} {304} (\bibinfo {year} {2012})},\ \Eprint
  {http://arxiv.org/abs/1108.2051} {arXiv:1108.2051} \BibitemShut {NoStop}%
\bibitem [{\citenamefont {{Kim}}(2012)}]{kimqupit}%
  \BibitemOpen
  \bibfield  {author} {\bibinfo {author} {\bibfnamefont {I.~H.}\ \bibnamefont
  {{Kim}}},\ }\href@noop {} {\  (\bibinfo {year} {2012})},\ \Eprint
  {http://arxiv.org/abs/1202.0052} {arXiv:1202.0052} \BibitemShut {NoStop}%
\bibitem [{\citenamefont {Yoshida}(2013)}]{yoshida}%
  \BibitemOpen
  \bibfield  {author} {\bibinfo {author} {\bibfnamefont {B.}~\bibnamefont
  {Yoshida}},\ }\href {\doibase 10.1103/PhysRevB.88.125122} {\bibfield
  {journal} {\bibinfo  {journal} {Phys. Rev. B}\ }\textbf {\bibinfo {volume}
  {88}},\ \bibinfo {pages} {125122} (\bibinfo {year} {2013})}\BibitemShut
  {NoStop}%
\bibitem [{\citenamefont {Haah}(2014)}]{haah2014bifurcation}%
  \BibitemOpen
  \bibfield  {author} {\bibinfo {author} {\bibfnamefont {J.}~\bibnamefont
  {Haah}},\ }\href {\doibase 10.1103/PhysRevB.89.075119} {\bibfield  {journal}
  {\bibinfo  {journal} {Phys. Rev. B}\ }\textbf {\bibinfo {volume} {89}},\
  \bibinfo {pages} {075119} (\bibinfo {year} {2014})}\BibitemShut {NoStop}%
\bibitem [{\citenamefont {Vijay}\ \emph {et~al.}(2015)\citenamefont {Vijay},
  \citenamefont {Haah},\ and\ \citenamefont {Fu}}]{fracton1}%
  \BibitemOpen
  \bibfield  {author} {\bibinfo {author} {\bibfnamefont {S.}~\bibnamefont
  {Vijay}}, \bibinfo {author} {\bibfnamefont {J.}~\bibnamefont {Haah}}, \ and\
  \bibinfo {author} {\bibfnamefont {L.}~\bibnamefont {Fu}},\ }\href {\doibase
  10.1103/PhysRevB.92.235136} {\bibfield  {journal} {\bibinfo  {journal} {Phys.
  Rev. B}\ }\textbf {\bibinfo {volume} {92}},\ \bibinfo {pages} {235136}
  (\bibinfo {year} {2015})}\BibitemShut {NoStop}%
\bibitem [{\citenamefont {Vijay}\ \emph {et~al.}(2016)\citenamefont {Vijay},
  \citenamefont {Haah},\ and\ \citenamefont {Fu}}]{fracton2}%
  \BibitemOpen
  \bibfield  {author} {\bibinfo {author} {\bibfnamefont {S.}~\bibnamefont
  {Vijay}}, \bibinfo {author} {\bibfnamefont {J.}~\bibnamefont {Haah}}, \ and\
  \bibinfo {author} {\bibfnamefont {L.}~\bibnamefont {Fu}},\ }\href {\doibase
  10.1103/PhysRevB.94.235157} {\bibfield  {journal} {\bibinfo  {journal} {Phys.
  Rev. B}\ }\textbf {\bibinfo {volume} {94}},\ \bibinfo {pages} {235157}
  (\bibinfo {year} {2016})}\BibitemShut {NoStop}%
\bibitem [{\citenamefont {{Bravyi}}\ \emph {et~al.}(2011)\citenamefont
  {{Bravyi}}, \citenamefont {{Leemhuis}},\ and\ \citenamefont
  {{Terhal}}}]{bravyi}%
  \BibitemOpen
  \bibfield  {author} {\bibinfo {author} {\bibfnamefont {S.}~\bibnamefont
  {{Bravyi}}}, \bibinfo {author} {\bibfnamefont {B.}~\bibnamefont
  {{Leemhuis}}}, \ and\ \bibinfo {author} {\bibfnamefont {B.~M.}\ \bibnamefont
  {{Terhal}}},\ }\href {\doibase 10.1016/j.aop.2010.11.002} {\bibfield
  {journal} {\bibinfo  {journal} {Annals of Physics}\ }\textbf {\bibinfo
  {volume} {326}},\ \bibinfo {pages} {839} (\bibinfo {year} {2011})},\ \Eprint
  {http://arxiv.org/abs/1006.4871} {arXiv:1006.4871} \BibitemShut {NoStop}%
\bibitem [{\citenamefont {Bravyi}\ and\ \citenamefont {Haah}(2013)}]{haah2}%
  \BibitemOpen
  \bibfield  {author} {\bibinfo {author} {\bibfnamefont {S.}~\bibnamefont
  {Bravyi}}\ and\ \bibinfo {author} {\bibfnamefont {J.}~\bibnamefont {Haah}},\
  }\href {\doibase 10.1103/PhysRevLett.111.200501} {\bibfield  {journal}
  {\bibinfo  {journal} {Phys. Rev. Lett.}\ }\textbf {\bibinfo {volume} {111}},\
  \bibinfo {pages} {200501} (\bibinfo {year} {2013})}\BibitemShut {NoStop}%
\bibitem [{\citenamefont {Bravyi}\ and\ \citenamefont
  {Haah}(2011)}]{bravyihaah}%
  \BibitemOpen
  \bibfield  {author} {\bibinfo {author} {\bibfnamefont {S.}~\bibnamefont
  {Bravyi}}\ and\ \bibinfo {author} {\bibfnamefont {J.}~\bibnamefont {Haah}},\
  }\href {\doibase 10.1103/PhysRevLett.107.150504} {\bibfield  {journal}
  {\bibinfo  {journal} {Phys. Rev. Lett.}\ }\textbf {\bibinfo {volume} {107}},\
  \bibinfo {pages} {150504} (\bibinfo {year} {2011})}\BibitemShut {NoStop}%
\bibitem [{\citenamefont {Ma}\ \emph {et~al.}(2017)\citenamefont {Ma},
  \citenamefont {Lake}, \citenamefont {Chen},\ and\ \citenamefont
  {Hermele}}]{han}%
  \BibitemOpen
  \bibfield  {author} {\bibinfo {author} {\bibfnamefont {H.}~\bibnamefont
  {Ma}}, \bibinfo {author} {\bibfnamefont {E.}~\bibnamefont {Lake}}, \bibinfo
  {author} {\bibfnamefont {X.}~\bibnamefont {Chen}}, \ and\ \bibinfo {author}
  {\bibfnamefont {M.}~\bibnamefont {Hermele}},\ }\href {\doibase
  10.1103/PhysRevB.95.245126} {\bibfield  {journal} {\bibinfo  {journal} {Phys.
  Rev. B}\ }\textbf {\bibinfo {volume} {95}},\ \bibinfo {pages} {245126}
  (\bibinfo {year} {2017})}\BibitemShut {NoStop}%
\bibitem [{\citenamefont {{Vijay}}(2017)}]{sagar}%
  \BibitemOpen
  \bibfield  {author} {\bibinfo {author} {\bibfnamefont {S.}~\bibnamefont
  {{Vijay}}},\ }\href@noop {} {\  (\bibinfo {year} {2017})},\ \Eprint
  {http://arxiv.org/abs/1701.00762} {arXiv:1701.00762} \BibitemShut {NoStop}%
\bibitem [{\citenamefont {{Vijay}}\ and\ \citenamefont
  {{Fu}}(2017)}]{nonabelian}%
  \BibitemOpen
  \bibfield  {author} {\bibinfo {author} {\bibfnamefont {S.}~\bibnamefont
  {{Vijay}}}\ and\ \bibinfo {author} {\bibfnamefont {L.}~\bibnamefont {{Fu}}},\
  }\href@noop {} {\  (\bibinfo {year} {2017})},\ \Eprint
  {http://arxiv.org/abs/1706.07070} {arXiv:1706.07070} \BibitemShut {NoStop}%
\bibitem [{\citenamefont {{Shirley}}\ \emph
  {et~al.}(2019{\natexlab{a}})\citenamefont {{Shirley}}, \citenamefont
  {{Slagle}},\ and\ \citenamefont {{Chen}}}]{shirleyfrac}%
  \BibitemOpen
  \bibfield  {author} {\bibinfo {author} {\bibfnamefont {W.}~\bibnamefont
  {{Shirley}}}, \bibinfo {author} {\bibfnamefont {K.}~\bibnamefont {{Slagle}}},
  \ and\ \bibinfo {author} {\bibfnamefont {X.}~\bibnamefont {{Chen}}},\ }\href
  {\doibase 10.1016/j.aop.2019.167922} {\bibfield  {journal} {\bibinfo
  {journal} {Annals of Physics}\ }\textbf {\bibinfo {volume} {410}},\ \bibinfo
  {eid} {167922} (\bibinfo {year} {2019}{\natexlab{a}})},\ \Eprint
  {http://arxiv.org/abs/1806.08625} {arXiv:1806.08625} \BibitemShut {NoStop}%
\bibitem [{\citenamefont {Song}\ \emph {et~al.}(2019)\citenamefont {Song},
  \citenamefont {Prem}, \citenamefont {Huang},\ and\ \citenamefont
  {Martin-Delgado}}]{twisted}%
  \BibitemOpen
  \bibfield  {author} {\bibinfo {author} {\bibfnamefont {H.}~\bibnamefont
  {Song}}, \bibinfo {author} {\bibfnamefont {A.}~\bibnamefont {Prem}}, \bibinfo
  {author} {\bibfnamefont {S.-J.}\ \bibnamefont {Huang}}, \ and\ \bibinfo
  {author} {\bibfnamefont {M.~A.}\ \bibnamefont {Martin-Delgado}},\ }\href
  {\doibase 10.1103/PhysRevB.99.155118} {\bibfield  {journal} {\bibinfo
  {journal} {Phys. Rev. B}\ }\textbf {\bibinfo {volume} {99}},\ \bibinfo
  {pages} {155118} (\bibinfo {year} {2019})}\BibitemShut {NoStop}%
\bibitem [{\citenamefont {{Bulmash}}\ and\ \citenamefont
  {{Iadecola}}(2019)}]{bulmashboundary}%
  \BibitemOpen
  \bibfield  {author} {\bibinfo {author} {\bibfnamefont {D.}~\bibnamefont
  {{Bulmash}}}\ and\ \bibinfo {author} {\bibfnamefont {T.}~\bibnamefont
  {{Iadecola}}},\ }\href {\doibase 10.1103/PhysRevB.99.125132} {\bibfield
  {journal} {\bibinfo  {journal} {Phys. Rev. B}\ }\textbf {\bibinfo {volume}
  {99}},\ \bibinfo {pages} {125132} (\bibinfo {year} {2019})}\BibitemShut
  {NoStop}%
\bibitem [{\citenamefont {Dua}\ \emph {et~al.}(2019)\citenamefont {Dua},
  \citenamefont {Williamson}, \citenamefont {Haah},\ and\ \citenamefont
  {Cheng}}]{Dua2019}%
  \BibitemOpen
  \bibfield  {author} {\bibinfo {author} {\bibfnamefont {A.}~\bibnamefont
  {Dua}}, \bibinfo {author} {\bibfnamefont {D.~J.}\ \bibnamefont {Williamson}},
  \bibinfo {author} {\bibfnamefont {J.}~\bibnamefont {Haah}}, \ and\ \bibinfo
  {author} {\bibfnamefont {M.}~\bibnamefont {Cheng}},\ }\href
  {https://link.aps.org/doi/10.1103/PhysRevB.99.245135} {\bibfield  {journal}
  {\bibinfo  {journal} {Phys. Rev. B}\ }\textbf {\bibinfo {volume} {99}},\
  \bibinfo {pages} {245135} (\bibinfo {year} {2019})}\BibitemShut {NoStop}%
\bibitem [{\citenamefont {Kim}\ and\ \citenamefont {Haah}(2016)}]{kimhaah}%
  \BibitemOpen
  \bibfield  {author} {\bibinfo {author} {\bibfnamefont {I.~H.}\ \bibnamefont
  {Kim}}\ and\ \bibinfo {author} {\bibfnamefont {J.}~\bibnamefont {Haah}},\
  }\href {\doibase 10.1103/PhysRevLett.116.027202} {\bibfield  {journal}
  {\bibinfo  {journal} {Phys. Rev. Lett.}\ }\textbf {\bibinfo {volume} {116}},\
  \bibinfo {pages} {027202} (\bibinfo {year} {2016})}\BibitemShut {NoStop}%
\bibitem [{\citenamefont {Prem}\ \emph {et~al.}(2017)\citenamefont {Prem},
  \citenamefont {Haah},\ and\ \citenamefont {Nandkishore}}]{prem}%
  \BibitemOpen
  \bibfield  {author} {\bibinfo {author} {\bibfnamefont {A.}~\bibnamefont
  {Prem}}, \bibinfo {author} {\bibfnamefont {J.}~\bibnamefont {Haah}}, \ and\
  \bibinfo {author} {\bibfnamefont {R.}~\bibnamefont {Nandkishore}},\ }\href
  {\doibase 10.1103/PhysRevB.95.155133} {\bibfield  {journal} {\bibinfo
  {journal} {Phys. Rev. B}\ }\textbf {\bibinfo {volume} {95}},\ \bibinfo
  {pages} {155133} (\bibinfo {year} {2017})}\BibitemShut {NoStop}%
\bibitem [{\citenamefont {Williamson}(2016)}]{williamson}%
  \BibitemOpen
  \bibfield  {author} {\bibinfo {author} {\bibfnamefont {D.~J.}\ \bibnamefont
  {Williamson}},\ }\href {\doibase 10.1103/PhysRevB.94.155128} {\bibfield
  {journal} {\bibinfo  {journal} {Phys. Rev. B}\ }\textbf {\bibinfo {volume}
  {94}},\ \bibinfo {pages} {155128} (\bibinfo {year} {2016})}\BibitemShut
  {NoStop}%
\bibitem [{\citenamefont {You}\ \emph {et~al.}(2018)\citenamefont {You},
  \citenamefont {Devakul}, \citenamefont {Burnell},\ and\ \citenamefont
  {Sondhi}}]{yizhi1}%
  \BibitemOpen
  \bibfield  {author} {\bibinfo {author} {\bibfnamefont {Y.}~\bibnamefont
  {You}}, \bibinfo {author} {\bibfnamefont {T.}~\bibnamefont {Devakul}},
  \bibinfo {author} {\bibfnamefont {F.~J.}\ \bibnamefont {Burnell}}, \ and\
  \bibinfo {author} {\bibfnamefont {S.~L.}\ \bibnamefont {Sondhi}},\ }\href
  {\doibase 10.1103/PhysRevB.98.035112} {\bibfield  {journal} {\bibinfo
  {journal} {Phys. Rev. B}\ }\textbf {\bibinfo {volume} {98}},\ \bibinfo
  {pages} {035112} (\bibinfo {year} {2018})}\BibitemShut {NoStop}%
\bibitem [{\citenamefont {{Shirley}}\ \emph
  {et~al.}(2019{\natexlab{b}})\citenamefont {{Shirley}}, \citenamefont
  {{Slagle}},\ and\ \citenamefont {{Chen}}}]{shirleygauging}%
  \BibitemOpen
  \bibfield  {author} {\bibinfo {author} {\bibfnamefont {W.}~\bibnamefont
  {{Shirley}}}, \bibinfo {author} {\bibfnamefont {K.}~\bibnamefont {{Slagle}}},
  \ and\ \bibinfo {author} {\bibfnamefont {X.}~\bibnamefont {{Chen}}},\ }\href
  {\doibase 10.21468/SciPostPhys.6.4.041} {\bibfield  {journal} {\bibinfo
  {journal} {SciPost Physics}\ }\textbf {\bibinfo {volume} {6}},\ \bibinfo
  {eid} {041} (\bibinfo {year} {2019}{\natexlab{b}})}\BibitemShut {NoStop}%
\bibitem [{\citenamefont {{Schmitz}}(2019)}]{albertgauge}%
  \BibitemOpen
  \bibfield  {author} {\bibinfo {author} {\bibfnamefont {A.~T.}\ \bibnamefont
  {{Schmitz}}},\ }\href {\doibase 10.1016/j.aop.2019.167927} {\bibfield
  {journal} {\bibinfo  {journal} {Annals of Physics}\ }\textbf {\bibinfo
  {volume} {410}},\ \bibinfo {eid} {167927} (\bibinfo {year} {2019})},\ \Eprint
  {http://arxiv.org/abs/1809.10151} {arXiv:1809.10151} \BibitemShut {NoStop}%
\bibitem [{\citenamefont {{Devakul}}\ \emph {et~al.}(2019)\citenamefont
  {{Devakul}}, \citenamefont {{Shirley}},\ and\ \citenamefont
  {{Wang}}}]{juvenSSPT}%
  \BibitemOpen
  \bibfield  {author} {\bibinfo {author} {\bibfnamefont {T.}~\bibnamefont
  {{Devakul}}}, \bibinfo {author} {\bibfnamefont {W.}~\bibnamefont
  {{Shirley}}}, \ and\ \bibinfo {author} {\bibfnamefont {J.}~\bibnamefont
  {{Wang}}},\ }\href@noop {} {\  (\bibinfo {year} {2019})},\ \Eprint
  {http://arxiv.org/abs/1910.01630} {arXiv:1910.01630} \BibitemShut {NoStop}%
\bibitem [{\citenamefont {{Ibieta-Jimenez}}\ \emph {et~al.}(2019)\citenamefont
  {{Ibieta-Jimenez}}, \citenamefont {{Queiroz Xavier}}, \citenamefont
  {{Petrucci}},\ and\ \citenamefont {{Teotonio-Sobrinho}}}]{Ibieta}%
  \BibitemOpen
  \bibfield  {author} {\bibinfo {author} {\bibfnamefont {J.~P.}\ \bibnamefont
  {{Ibieta-Jimenez}}}, \bibinfo {author} {\bibfnamefont {L.~N.}\ \bibnamefont
  {{Queiroz Xavier}}}, \bibinfo {author} {\bibfnamefont {M.}~\bibnamefont
  {{Petrucci}}}, \ and\ \bibinfo {author} {\bibfnamefont {P.}~\bibnamefont
  {{Teotonio-Sobrinho}}},\ }\href@noop {} {\  (\bibinfo {year} {2019})},\
  \Eprint {http://arxiv.org/abs/1908.07601} {arXiv:1908.07601} \BibitemShut
  {NoStop}%
\bibitem [{\citenamefont {{Tantivasadakarn}}\ and\ \citenamefont
  {{Vijay}}(2019)}]{TantivasadakarnSearching}%
  \BibitemOpen
  \bibfield  {author} {\bibinfo {author} {\bibfnamefont {N.}~\bibnamefont
  {{Tantivasadakarn}}}\ and\ \bibinfo {author} {\bibfnamefont {S.}~\bibnamefont
  {{Vijay}}},\ }\href@noop {} {\  (\bibinfo {year} {2019})},\ \Eprint
  {http://arxiv.org/abs/1912.02826} {arXiv:1912.02826} \BibitemShut {NoStop}%
\bibitem [{\citenamefont {{Brown}}\ and\ \citenamefont
  {{Williamson}}(2019)}]{Brown2019}%
  \BibitemOpen
  \bibfield  {author} {\bibinfo {author} {\bibfnamefont {B.~J.}\ \bibnamefont
  {{Brown}}}\ and\ \bibinfo {author} {\bibfnamefont {D.~J.}\ \bibnamefont
  {{Williamson}}},\ }\href@noop {} {\  (\bibinfo {year} {2019})},\ \Eprint
  {http://arxiv.org/abs/1901.08061} {arXiv:1901.08061} \BibitemShut {NoStop}%
\bibitem [{\citenamefont {Pretko}(2017{\natexlab{a}})}]{sub}%
  \BibitemOpen
  \bibfield  {author} {\bibinfo {author} {\bibfnamefont {M.}~\bibnamefont
  {Pretko}},\ }\href {\doibase 10.1103/PhysRevB.95.115139} {\bibfield
  {journal} {\bibinfo  {journal} {Phys. Rev. B}\ }\textbf {\bibinfo {volume}
  {95}},\ \bibinfo {pages} {115139} (\bibinfo {year}
  {2017}{\natexlab{a}})}\BibitemShut {NoStop}%
\bibitem [{\citenamefont {Pretko}(2017{\natexlab{b}})}]{genem}%
  \BibitemOpen
  \bibfield  {author} {\bibinfo {author} {\bibfnamefont {M.}~\bibnamefont
  {Pretko}},\ }\href {\doibase 10.1103/PhysRevB.96.035119} {\bibfield
  {journal} {\bibinfo  {journal} {Phys. Rev. B}\ }\textbf {\bibinfo {volume}
  {96}},\ \bibinfo {pages} {035119} (\bibinfo {year}
  {2017}{\natexlab{b}})}\BibitemShut {NoStop}%
\bibitem [{\citenamefont {{Seiberg}}(2019)}]{SeibergSymmetry}%
  \BibitemOpen
  \bibfield  {author} {\bibinfo {author} {\bibfnamefont {N.}~\bibnamefont
  {{Seiberg}}},\ }\href@noop {} {\  (\bibinfo {year} {2019})},\ \Eprint
  {http://arxiv.org/abs/1909.10544} {arXiv:1909.10544} \BibitemShut {NoStop}%
\bibitem [{\citenamefont {Pretko}(2017{\natexlab{c}})}]{chiral}%
  \BibitemOpen
  \bibfield  {author} {\bibinfo {author} {\bibfnamefont {M.}~\bibnamefont
  {Pretko}},\ }\href {\doibase 10.1103/PhysRevB.96.125151} {\bibfield
  {journal} {\bibinfo  {journal} {Phys. Rev. B}\ }\textbf {\bibinfo {volume}
  {96}},\ \bibinfo {pages} {125151} (\bibinfo {year}
  {2017}{\natexlab{c}})}\BibitemShut {NoStop}%
\bibitem [{\citenamefont {Prem}\ \emph
  {et~al.}(2018{\natexlab{a}})\citenamefont {Prem}, \citenamefont {Pretko},\
  and\ \citenamefont {Nandkishore}}]{emergent}%
  \BibitemOpen
  \bibfield  {author} {\bibinfo {author} {\bibfnamefont {A.}~\bibnamefont
  {Prem}}, \bibinfo {author} {\bibfnamefont {M.}~\bibnamefont {Pretko}}, \ and\
  \bibinfo {author} {\bibfnamefont {R.~M.}\ \bibnamefont {Nandkishore}},\
  }\href {\doibase 10.1103/PhysRevB.97.085116} {\bibfield  {journal} {\bibinfo
  {journal} {Phys. Rev. B}\ }\textbf {\bibinfo {volume} {97}},\ \bibinfo
  {pages} {085116} (\bibinfo {year} {2018}{\natexlab{a}})}\BibitemShut
  {NoStop}%
\bibitem [{\citenamefont {Ma}\ \emph {et~al.}(2018{\natexlab{a}})\citenamefont
  {Ma}, \citenamefont {Hermele},\ and\ \citenamefont {Chen}}]{han3}%
  \BibitemOpen
  \bibfield  {author} {\bibinfo {author} {\bibfnamefont {H.}~\bibnamefont
  {Ma}}, \bibinfo {author} {\bibfnamefont {M.}~\bibnamefont {Hermele}}, \ and\
  \bibinfo {author} {\bibfnamefont {X.}~\bibnamefont {Chen}},\ }\href {\doibase
  10.1103/PhysRevB.98.035111} {\bibfield  {journal} {\bibinfo  {journal} {Phys.
  Rev. B}\ }\textbf {\bibinfo {volume} {98}},\ \bibinfo {pages} {035111}
  (\bibinfo {year} {2018}{\natexlab{a}})}\BibitemShut {NoStop}%
\bibitem [{\citenamefont {Bulmash}\ and\ \citenamefont
  {Barkeshli}(2018)}]{bulmash}%
  \BibitemOpen
  \bibfield  {author} {\bibinfo {author} {\bibfnamefont {D.}~\bibnamefont
  {Bulmash}}\ and\ \bibinfo {author} {\bibfnamefont {M.}~\bibnamefont
  {Barkeshli}},\ }\href {\doibase 10.1103/PhysRevB.97.235112} {\bibfield
  {journal} {\bibinfo  {journal} {Phys. Rev. B}\ }\textbf {\bibinfo {volume}
  {97}},\ \bibinfo {pages} {235112} (\bibinfo {year} {2018})}\BibitemShut
  {NoStop}%
\bibitem [{\citenamefont {{Bulmash}}\ and\ \citenamefont
  {{Barkeshli}}(2018)}]{bulmash2}%
  \BibitemOpen
  \bibfield  {author} {\bibinfo {author} {\bibfnamefont {D.}~\bibnamefont
  {{Bulmash}}}\ and\ \bibinfo {author} {\bibfnamefont {M.}~\bibnamefont
  {{Barkeshli}}},\ }\href@noop {} {\  (\bibinfo {year} {2018})},\ \Eprint
  {http://arxiv.org/abs/1806.01855} {arXiv:1806.01855} \BibitemShut {NoStop}%
\bibitem [{\citenamefont {Williamson}\ \emph
  {et~al.}(2019{\natexlab{a}})\citenamefont {Williamson}, \citenamefont {Bi},\
  and\ \citenamefont {Cheng}}]{williamsonSET}%
  \BibitemOpen
  \bibfield  {author} {\bibinfo {author} {\bibfnamefont {D.~J.}\ \bibnamefont
  {Williamson}}, \bibinfo {author} {\bibfnamefont {Z.}~\bibnamefont {Bi}}, \
  and\ \bibinfo {author} {\bibfnamefont {M.}~\bibnamefont {Cheng}},\ }\href
  {\doibase 10.1103/PhysRevB.100.125150} {\bibfield  {journal} {\bibinfo
  {journal} {Phys. Rev. B}\ }\textbf {\bibinfo {volume} {100}},\ \bibinfo
  {pages} {125150} (\bibinfo {year} {2019}{\natexlab{a}})}\BibitemShut
  {NoStop}%
\bibitem [{\citenamefont {Yan}(2019)}]{yanholography}%
  \BibitemOpen
  \bibfield  {author} {\bibinfo {author} {\bibfnamefont {H.}~\bibnamefont
  {Yan}},\ }\href {\doibase 10.1103/PhysRevB.99.155126} {\bibfield  {journal}
  {\bibinfo  {journal} {Phys. Rev. B}\ }\textbf {\bibinfo {volume} {99}},\
  \bibinfo {pages} {155126} (\bibinfo {year} {2019})}\BibitemShut {NoStop}%
\bibitem [{\citenamefont {{Wang}}\ and\ \citenamefont
  {{Xu}}(2019)}]{wang2019embeddon}%
  \BibitemOpen
  \bibfield  {author} {\bibinfo {author} {\bibfnamefont {J.}~\bibnamefont
  {{Wang}}}\ and\ \bibinfo {author} {\bibfnamefont {K.}~\bibnamefont {{Xu}}},\
  }\href@noop {} {\  (\bibinfo {year} {2019})},\ \Eprint
  {http://arxiv.org/abs/1909.13879} {arXiv:1909.13879} \BibitemShut {NoStop}%
\bibitem [{\citenamefont {{Wang}}\ \emph {et~al.}(2019)\citenamefont {{Wang}},
  \citenamefont {{Xu}},\ and\ \citenamefont {{Yau}}}]{wang2019nonabelian}%
  \BibitemOpen
  \bibfield  {author} {\bibinfo {author} {\bibfnamefont {J.}~\bibnamefont
  {{Wang}}}, \bibinfo {author} {\bibfnamefont {K.}~\bibnamefont {{Xu}}}, \ and\
  \bibinfo {author} {\bibfnamefont {S.-T.}\ \bibnamefont {{Yau}}},\ }\href@noop
  {} {\  (\bibinfo {year} {2019})},\ \Eprint {http://arxiv.org/abs/1911.01804}
  {arXiv:1911.01804} \BibitemShut {NoStop}%
\bibitem [{\citenamefont {Pretko}\ and\ \citenamefont
  {Radzihovsky}(2018)}]{leomichael}%
  \BibitemOpen
  \bibfield  {author} {\bibinfo {author} {\bibfnamefont {M.}~\bibnamefont
  {Pretko}}\ and\ \bibinfo {author} {\bibfnamefont {L.}~\bibnamefont
  {Radzihovsky}},\ }\href {\doibase 10.1103/PhysRevLett.120.195301} {\bibfield
  {journal} {\bibinfo  {journal} {Phys. Rev. Lett.}\ }\textbf {\bibinfo
  {volume} {120}},\ \bibinfo {pages} {195301} (\bibinfo {year}
  {2018})}\BibitemShut {NoStop}%
\bibitem [{\citenamefont {Gromov}(2019{\natexlab{a}})}]{gromov}%
  \BibitemOpen
  \bibfield  {author} {\bibinfo {author} {\bibfnamefont {A.}~\bibnamefont
  {Gromov}},\ }\href {\doibase 10.1103/PhysRevLett.122.076403} {\bibfield
  {journal} {\bibinfo  {journal} {Phys. Rev. Lett.}\ }\textbf {\bibinfo
  {volume} {122}},\ \bibinfo {pages} {076403} (\bibinfo {year}
  {2019}{\natexlab{a}})}\BibitemShut {NoStop}%
\bibitem [{\citenamefont {Pai}\ and\ \citenamefont {Pretko}(2018)}]{pai}%
  \BibitemOpen
  \bibfield  {author} {\bibinfo {author} {\bibfnamefont {S.}~\bibnamefont
  {Pai}}\ and\ \bibinfo {author} {\bibfnamefont {M.}~\bibnamefont {Pretko}},\
  }\href {\doibase 10.1103/PhysRevB.97.235102} {\bibfield  {journal} {\bibinfo
  {journal} {Phys. Rev. B}\ }\textbf {\bibinfo {volume} {97}},\ \bibinfo
  {pages} {235102} (\bibinfo {year} {2018})}\BibitemShut {NoStop}%
\bibitem [{\citenamefont {Kumar}\ and\ \citenamefont {Potter}(2019)}]{potter}%
  \BibitemOpen
  \bibfield  {author} {\bibinfo {author} {\bibfnamefont {A.}~\bibnamefont
  {Kumar}}\ and\ \bibinfo {author} {\bibfnamefont {A.~C.}\ \bibnamefont
  {Potter}},\ }\href {\doibase 10.1103/PhysRevB.100.045119} {\bibfield
  {journal} {\bibinfo  {journal} {Phys. Rev. B}\ }\textbf {\bibinfo {volume}
  {100}},\ \bibinfo {pages} {045119} (\bibinfo {year} {2019})}\BibitemShut
  {NoStop}%
\bibitem [{\citenamefont {Radzihovsky}\ and\ \citenamefont
  {Hermele}(2020)}]{RadzihovskyHermele}%
  \BibitemOpen
  \bibfield  {author} {\bibinfo {author} {\bibfnamefont {L.}~\bibnamefont
  {Radzihovsky}}\ and\ \bibinfo {author} {\bibfnamefont {M.}~\bibnamefont
  {Hermele}},\ }\href {\doibase 10.1103/PhysRevLett.124.050402} {\bibfield
  {journal} {\bibinfo  {journal} {Phys. Rev. Lett.}\ }\textbf {\bibinfo
  {volume} {124}},\ \bibinfo {pages} {050402} (\bibinfo {year}
  {2020})}\BibitemShut {NoStop}%
\bibitem [{\citenamefont {Prem}\ \emph
  {et~al.}(2018{\natexlab{b}})\citenamefont {Prem}, \citenamefont {Vijay},
  \citenamefont {Chou}, \citenamefont {Pretko},\ and\ \citenamefont
  {Nandkishore}}]{pinch}%
  \BibitemOpen
  \bibfield  {author} {\bibinfo {author} {\bibfnamefont {A.}~\bibnamefont
  {Prem}}, \bibinfo {author} {\bibfnamefont {S.}~\bibnamefont {Vijay}},
  \bibinfo {author} {\bibfnamefont {Y.-Z.}\ \bibnamefont {Chou}}, \bibinfo
  {author} {\bibfnamefont {M.}~\bibnamefont {Pretko}}, \ and\ \bibinfo {author}
  {\bibfnamefont {R.~M.}\ \bibnamefont {Nandkishore}},\ }\href {\doibase
  10.1103/PhysRevB.98.165140} {\bibfield  {journal} {\bibinfo  {journal} {Phys.
  Rev. B}\ }\textbf {\bibinfo {volume} {98}},\ \bibinfo {pages} {165140}
  (\bibinfo {year} {2018}{\natexlab{b}})}\BibitemShut {NoStop}%
\bibitem [{\citenamefont {{Yan}}\ \emph {et~al.}(2019)\citenamefont {{Yan}},
  \citenamefont {{Benton}}, \citenamefont {{Jaubert}},\ and\ \citenamefont
  {{Shannon}}}]{YanPyrochlore}%
  \BibitemOpen
  \bibfield  {author} {\bibinfo {author} {\bibfnamefont {H.}~\bibnamefont
  {{Yan}}}, \bibinfo {author} {\bibfnamefont {O.}~\bibnamefont {{Benton}}},
  \bibinfo {author} {\bibfnamefont {L.~D.~C.}\ \bibnamefont {{Jaubert}}}, \
  and\ \bibinfo {author} {\bibfnamefont {N.}~\bibnamefont {{Shannon}}},\
  }\href@noop {} {\  (\bibinfo {year} {2019})},\ \Eprint
  {http://arxiv.org/abs/1902.10934} {arXiv:1902.10934} \BibitemShut {NoStop}%
\bibitem [{\citenamefont {{Sous}}\ and\ \citenamefont
  {{Pretko}}(2019)}]{SousPolarons}%
  \BibitemOpen
  \bibfield  {author} {\bibinfo {author} {\bibfnamefont {J.}~\bibnamefont
  {{Sous}}}\ and\ \bibinfo {author} {\bibfnamefont {M.}~\bibnamefont
  {{Pretko}}},\ }\href@noop {} {\  (\bibinfo {year} {2019})},\ \Eprint
  {http://arxiv.org/abs/1904.08424} {arXiv:1904.08424} \BibitemShut {NoStop}%
\bibitem [{\citenamefont {{Dubinkin}}\ \emph {et~al.}(2020)\citenamefont
  {{Dubinkin}}, \citenamefont {{May-Mann}},\ and\ \citenamefont
  {{Hughes}}}]{hughes2020lsm}%
  \BibitemOpen
  \bibfield  {author} {\bibinfo {author} {\bibfnamefont {O.}~\bibnamefont
  {{Dubinkin}}}, \bibinfo {author} {\bibfnamefont {J.}~\bibnamefont
  {{May-Mann}}}, \ and\ \bibinfo {author} {\bibfnamefont {T.~L.}\ \bibnamefont
  {{Hughes}}},\ }\href@noop {} {\  (\bibinfo {year} {2020})},\ \Eprint
  {http://arxiv.org/abs/2001.04477} {arXiv:2001.04477} \BibitemShut {NoStop}%
\bibitem [{\citenamefont {{He}}\ \emph {et~al.}(2019)\citenamefont {{He}},
  \citenamefont {{You}},\ and\ \citenamefont {{Prem}}}]{he2019lsm}%
  \BibitemOpen
  \bibfield  {author} {\bibinfo {author} {\bibfnamefont {H.}~\bibnamefont
  {{He}}}, \bibinfo {author} {\bibfnamefont {Y.}~\bibnamefont {{You}}}, \ and\
  \bibinfo {author} {\bibfnamefont {A.}~\bibnamefont {{Prem}}},\ }\href@noop {}
  {\  (\bibinfo {year} {2019})},\ \Eprint {http://arxiv.org/abs/1912.10520}
  {arXiv:1912.10520} \BibitemShut {NoStop}%
\bibitem [{\citenamefont {Fuji}(2019)}]{FujiLayer}%
  \BibitemOpen
  \bibfield  {author} {\bibinfo {author} {\bibfnamefont {Y.}~\bibnamefont
  {Fuji}},\ }\href {\doibase 10.1103/PhysRevB.100.235115} {\bibfield  {journal}
  {\bibinfo  {journal} {Phys. Rev. B}\ }\textbf {\bibinfo {volume} {100}},\
  \bibinfo {pages} {235115} (\bibinfo {year} {2019})}\BibitemShut {NoStop}%
\bibitem [{\citenamefont {Slagle}\ and\ \citenamefont
  {Kim}(2017{\natexlab{a}})}]{slagle1}%
  \BibitemOpen
  \bibfield  {author} {\bibinfo {author} {\bibfnamefont {K.}~\bibnamefont
  {Slagle}}\ and\ \bibinfo {author} {\bibfnamefont {Y.~B.}\ \bibnamefont
  {Kim}},\ }\href {\doibase 10.1103/PhysRevB.96.165106} {\bibfield  {journal}
  {\bibinfo  {journal} {Phys. Rev. B}\ }\textbf {\bibinfo {volume} {96}},\
  \bibinfo {pages} {165106} (\bibinfo {year} {2017}{\natexlab{a}})}\BibitemShut
  {NoStop}%
\bibitem [{\citenamefont {Hal\'asz}\ \emph {et~al.}(2017)\citenamefont
  {Hal\'asz}, \citenamefont {Hsieh},\ and\ \citenamefont {Balents}}]{balents}%
  \BibitemOpen
  \bibfield  {author} {\bibinfo {author} {\bibfnamefont {G.~B.}\ \bibnamefont
  {Hal\'asz}}, \bibinfo {author} {\bibfnamefont {T.~H.}\ \bibnamefont {Hsieh}},
  \ and\ \bibinfo {author} {\bibfnamefont {L.}~\bibnamefont {Balents}},\ }\href
  {\doibase 10.1103/PhysRevLett.119.257202} {\bibfield  {journal} {\bibinfo
  {journal} {Phys. Rev. Lett.}\ }\textbf {\bibinfo {volume} {119}},\ \bibinfo
  {pages} {257202} (\bibinfo {year} {2017})}\BibitemShut {NoStop}%
\bibitem [{\citenamefont {Pai}\ \emph {et~al.}(2019)\citenamefont {Pai},
  \citenamefont {Pretko},\ and\ \citenamefont {Nandkishore}}]{PaiLocalization}%
  \BibitemOpen
  \bibfield  {author} {\bibinfo {author} {\bibfnamefont {S.}~\bibnamefont
  {Pai}}, \bibinfo {author} {\bibfnamefont {M.}~\bibnamefont {Pretko}}, \ and\
  \bibinfo {author} {\bibfnamefont {R.~M.}\ \bibnamefont {Nandkishore}},\
  }\href {\doibase 10.1103/PhysRevX.9.021003} {\bibfield  {journal} {\bibinfo
  {journal} {Phys. Rev. X}\ }\textbf {\bibinfo {volume} {9}},\ \bibinfo {pages}
  {021003} (\bibinfo {year} {2019})}\BibitemShut {NoStop}%
\bibitem [{\citenamefont {{Sala}}\ \emph {et~al.}(2019)\citenamefont {{Sala}},
  \citenamefont {{Rakovszky}}, \citenamefont {{Verresen}}, \citenamefont
  {{Knap}},\ and\ \citenamefont {{Pollmann}}}]{SalaErgodicity}%
  \BibitemOpen
  \bibfield  {author} {\bibinfo {author} {\bibfnamefont {P.}~\bibnamefont
  {{Sala}}}, \bibinfo {author} {\bibfnamefont {T.}~\bibnamefont {{Rakovszky}}},
  \bibinfo {author} {\bibfnamefont {R.}~\bibnamefont {{Verresen}}}, \bibinfo
  {author} {\bibfnamefont {M.}~\bibnamefont {{Knap}}}, \ and\ \bibinfo {author}
  {\bibfnamefont {F.}~\bibnamefont {{Pollmann}}},\ }\href@noop {} {\  (\bibinfo
  {year} {2019})},\ \Eprint {http://arxiv.org/abs/1904.04266}
  {arXiv:1904.04266} \BibitemShut {NoStop}%
\bibitem [{\citenamefont {{Moudgalya}}\ \emph {et~al.}(2019)\citenamefont
  {{Moudgalya}}, \citenamefont {{Prem}}, \citenamefont {{Nandkishore}},
  \citenamefont {{Regnault}},\ and\ \citenamefont
  {{Bernevig}}}]{moudgalya2019}%
  \BibitemOpen
  \bibfield  {author} {\bibinfo {author} {\bibfnamefont {S.}~\bibnamefont
  {{Moudgalya}}}, \bibinfo {author} {\bibfnamefont {A.}~\bibnamefont {{Prem}}},
  \bibinfo {author} {\bibfnamefont {R.}~\bibnamefont {{Nandkishore}}}, \bibinfo
  {author} {\bibfnamefont {N.}~\bibnamefont {{Regnault}}}, \ and\ \bibinfo
  {author} {\bibfnamefont {B.~A.}\ \bibnamefont {{Bernevig}}},\ }\href@noop {}
  {\  (\bibinfo {year} {2019})},\ \Eprint {http://arxiv.org/abs/1910.14048}
  {arXiv:1910.14048} \BibitemShut {NoStop}%
\bibitem [{\citenamefont {{Nandkishore}}\ and\ \citenamefont
  {{Hermele}}(2019)}]{fractonreview}%
  \BibitemOpen
  \bibfield  {author} {\bibinfo {author} {\bibfnamefont {R.~M.}\ \bibnamefont
  {{Nandkishore}}}\ and\ \bibinfo {author} {\bibfnamefont {M.}~\bibnamefont
  {{Hermele}}},\ }\href {\doibase 10.1146/annurev-conmatphys-031218-013604}
  {\bibfield  {journal} {\bibinfo  {journal} {Annual Review of Condensed Matter
  Physics}\ }\textbf {\bibinfo {volume} {10}},\ \bibinfo {pages} {295}
  (\bibinfo {year} {2019})},\ \Eprint {http://arxiv.org/abs/1803.11196}
  {arXiv:1803.11196} \BibitemShut {NoStop}%
\bibitem [{\citenamefont {{Pretko}}\ \emph {et~al.}(2020)\citenamefont
  {{Pretko}}, \citenamefont {{Chen}},\ and\ \citenamefont
  {{You}}}]{fractonreview2}%
  \BibitemOpen
  \bibfield  {author} {\bibinfo {author} {\bibfnamefont {M.}~\bibnamefont
  {{Pretko}}}, \bibinfo {author} {\bibfnamefont {X.}~\bibnamefont {{Chen}}}, \
  and\ \bibinfo {author} {\bibfnamefont {Y.}~\bibnamefont {{You}}},\
  }\href@noop {} {\  (\bibinfo {year} {2020})},\ \Eprint
  {http://arxiv.org/abs/2001.01722} {arXiv:2001.01722} \BibitemShut {NoStop}%
\bibitem [{\citenamefont {Wen}(2017)}]{wenreview}%
  \BibitemOpen
  \bibfield  {author} {\bibinfo {author} {\bibfnamefont {X.-G.}\ \bibnamefont
  {Wen}},\ }\href {\doibase 10.1103/RevModPhys.89.041004} {\bibfield  {journal}
  {\bibinfo  {journal} {Rev. Mod. Phys.}\ }\textbf {\bibinfo {volume} {89}},\
  \bibinfo {pages} {041004} (\bibinfo {year} {2017})}\BibitemShut {NoStop}%
\bibitem [{\citenamefont {Moore}\ and\ \citenamefont
  {Seiberg}(1989)}]{Moore1989}%
  \BibitemOpen
  \bibfield  {author} {\bibinfo {author} {\bibfnamefont {G.}~\bibnamefont
  {Moore}}\ and\ \bibinfo {author} {\bibfnamefont {N.}~\bibnamefont
  {Seiberg}},\ }\href {\doibase 10.1007/BF01238857} {\bibfield  {journal}
  {\bibinfo  {journal} {Communications in Mathematical Physics}\ }\textbf
  {\bibinfo {volume} {123}},\ \bibinfo {pages} {177} (\bibinfo {year}
  {1989})}\BibitemShut {NoStop}%
\bibitem [{\citenamefont {{Kitaev}}(2006)}]{kitaev}%
  \BibitemOpen
  \bibfield  {author} {\bibinfo {author} {\bibfnamefont {A.}~\bibnamefont
  {{Kitaev}}},\ }\href {\doibase 10.1016/j.aop.2005.10.005} {\bibfield
  {journal} {\bibinfo  {journal} {Annals of Physics}\ }\textbf {\bibinfo
  {volume} {321}},\ \bibinfo {pages} {2} (\bibinfo {year} {2006})},\ \Eprint
  {http://arxiv.org/abs/cond-mat/0506438} {arXiv:cond-mat/0506438} \BibitemShut
  {NoStop}%
\bibitem [{\citenamefont {Levin}\ and\ \citenamefont {Wen}(2005)}]{Levin2005}%
  \BibitemOpen
  \bibfield  {author} {\bibinfo {author} {\bibfnamefont {M.~A.}\ \bibnamefont
  {Levin}}\ and\ \bibinfo {author} {\bibfnamefont {X.-G.}\ \bibnamefont
  {Wen}},\ }\href {\doibase 10.1103/PhysRevB.71.045110} {\bibfield  {journal}
  {\bibinfo  {journal} {Phys. Rev. B}\ }\textbf {\bibinfo {volume} {71}},\
  \bibinfo {pages} {045110} (\bibinfo {year} {2005})}\BibitemShut {NoStop}%
\bibitem [{\citenamefont {Lan}\ \emph {et~al.}(2018)\citenamefont {Lan},
  \citenamefont {Kong},\ and\ \citenamefont {Wen}}]{wen2018prx}%
  \BibitemOpen
  \bibfield  {author} {\bibinfo {author} {\bibfnamefont {T.}~\bibnamefont
  {Lan}}, \bibinfo {author} {\bibfnamefont {L.}~\bibnamefont {Kong}}, \ and\
  \bibinfo {author} {\bibfnamefont {X.-G.}\ \bibnamefont {Wen}},\ }\href
  {\doibase 10.1103/PhysRevX.8.021074} {\bibfield  {journal} {\bibinfo
  {journal} {Phys. Rev. X}\ }\textbf {\bibinfo {volume} {8}},\ \bibinfo {pages}
  {021074} (\bibinfo {year} {2018})}\BibitemShut {NoStop}%
\bibitem [{\citenamefont {Zhu}\ \emph {et~al.}(2019)\citenamefont {Zhu},
  \citenamefont {Lan},\ and\ \citenamefont {Wen}}]{wen2019prb}%
  \BibitemOpen
  \bibfield  {author} {\bibinfo {author} {\bibfnamefont {C.}~\bibnamefont
  {Zhu}}, \bibinfo {author} {\bibfnamefont {T.}~\bibnamefont {Lan}}, \ and\
  \bibinfo {author} {\bibfnamefont {X.-G.}\ \bibnamefont {Wen}},\ }\href
  {\doibase 10.1103/PhysRevB.100.045105} {\bibfield  {journal} {\bibinfo
  {journal} {Phys. Rev. B}\ }\textbf {\bibinfo {volume} {100}},\ \bibinfo
  {pages} {045105} (\bibinfo {year} {2019})}\BibitemShut {NoStop}%
\bibitem [{\citenamefont {{Walker}}\ and\ \citenamefont
  {{Wang}}(2012)}]{walker2012}%
  \BibitemOpen
  \bibfield  {author} {\bibinfo {author} {\bibfnamefont {K.}~\bibnamefont
  {{Walker}}}\ and\ \bibinfo {author} {\bibfnamefont {Z.}~\bibnamefont
  {{Wang}}},\ }\href {\doibase 10.1007/s11467-011-0194-z} {\bibfield  {journal}
  {\bibinfo  {journal} {Frontiers of Physics}\ }\textbf {\bibinfo {volume}
  {7}},\ \bibinfo {pages} {150} (\bibinfo {year} {2012})},\ \Eprint
  {http://arxiv.org/abs/1104.2632} {arXiv:1104.2632} \BibitemShut {NoStop}%
\bibitem [{\citenamefont {Wan}\ \emph {et~al.}(2015)\citenamefont {Wan},
  \citenamefont {Wang},\ and\ \citenamefont {He}}]{wan2015twisted}%
  \BibitemOpen
  \bibfield  {author} {\bibinfo {author} {\bibfnamefont {Y.}~\bibnamefont
  {Wan}}, \bibinfo {author} {\bibfnamefont {J.~C.}\ \bibnamefont {Wang}}, \
  and\ \bibinfo {author} {\bibfnamefont {H.}~\bibnamefont {He}},\ }\href
  {\doibase 10.1103/PhysRevB.92.045101} {\bibfield  {journal} {\bibinfo
  {journal} {Phys. Rev. B}\ }\textbf {\bibinfo {volume} {92}},\ \bibinfo
  {pages} {045101} (\bibinfo {year} {2015})}\BibitemShut {NoStop}%
\bibitem [{\citenamefont {{Williamson}}\ and\ \citenamefont
  {{Wang}}(2017)}]{williamson2017ham}%
  \BibitemOpen
  \bibfield  {author} {\bibinfo {author} {\bibfnamefont {D.~J.}\ \bibnamefont
  {{Williamson}}}\ and\ \bibinfo {author} {\bibfnamefont {Z.}~\bibnamefont
  {{Wang}}},\ }\href {\doibase 10.1016/j.aop.2016.12.018} {\bibfield  {journal}
  {\bibinfo  {journal} {Annals of Physics}\ }\textbf {\bibinfo {volume}
  {377}},\ \bibinfo {pages} {311} (\bibinfo {year} {2017})},\ \Eprint
  {http://arxiv.org/abs/1606.07144} {arXiv:1606.07144} \BibitemShut {NoStop}%
\bibitem [{\citenamefont {He}\ \emph {et~al.}(2018)\citenamefont {He},
  \citenamefont {Zheng}, \citenamefont {Bernevig},\ and\ \citenamefont
  {Regnault}}]{regnault2}%
  \BibitemOpen
  \bibfield  {author} {\bibinfo {author} {\bibfnamefont {H.}~\bibnamefont
  {He}}, \bibinfo {author} {\bibfnamefont {Y.}~\bibnamefont {Zheng}}, \bibinfo
  {author} {\bibfnamefont {B.~A.}\ \bibnamefont {Bernevig}}, \ and\ \bibinfo
  {author} {\bibfnamefont {N.}~\bibnamefont {Regnault}},\ }\href {\doibase
  10.1103/PhysRevB.97.125102} {\bibfield  {journal} {\bibinfo  {journal} {Phys.
  Rev. B}\ }\textbf {\bibinfo {volume} {97}},\ \bibinfo {pages} {125102}
  (\bibinfo {year} {2018})}\BibitemShut {NoStop}%
\bibitem [{\citenamefont {Ma}\ \emph {et~al.}(2018{\natexlab{b}})\citenamefont
  {Ma}, \citenamefont {Schmitz}, \citenamefont {Parameswaran}, \citenamefont
  {Hermele},\ and\ \citenamefont {Nandkishore}}]{han2}%
  \BibitemOpen
  \bibfield  {author} {\bibinfo {author} {\bibfnamefont {H.}~\bibnamefont
  {Ma}}, \bibinfo {author} {\bibfnamefont {A.~T.}\ \bibnamefont {Schmitz}},
  \bibinfo {author} {\bibfnamefont {S.~A.}\ \bibnamefont {Parameswaran}},
  \bibinfo {author} {\bibfnamefont {M.}~\bibnamefont {Hermele}}, \ and\
  \bibinfo {author} {\bibfnamefont {R.~M.}\ \bibnamefont {Nandkishore}},\
  }\href {\doibase 10.1103/PhysRevB.97.125101} {\bibfield  {journal} {\bibinfo
  {journal} {Phys. Rev. B}\ }\textbf {\bibinfo {volume} {97}},\ \bibinfo
  {pages} {125101} (\bibinfo {year} {2018}{\natexlab{b}})}\BibitemShut
  {NoStop}%
\bibitem [{\citenamefont {Schmitz}\ \emph {et~al.}(2018)\citenamefont
  {Schmitz}, \citenamefont {Ma}, \citenamefont {Nandkishore},\ and\
  \citenamefont {Parameswaran}}]{albert}%
  \BibitemOpen
  \bibfield  {author} {\bibinfo {author} {\bibfnamefont {A.~T.}\ \bibnamefont
  {Schmitz}}, \bibinfo {author} {\bibfnamefont {H.}~\bibnamefont {Ma}},
  \bibinfo {author} {\bibfnamefont {R.~M.}\ \bibnamefont {Nandkishore}}, \ and\
  \bibinfo {author} {\bibfnamefont {S.~A.}\ \bibnamefont {Parameswaran}},\
  }\href {\doibase 10.1103/PhysRevB.97.134426} {\bibfield  {journal} {\bibinfo
  {journal} {Phys. Rev. B}\ }\textbf {\bibinfo {volume} {97}},\ \bibinfo
  {pages} {134426} (\bibinfo {year} {2018})}\BibitemShut {NoStop}%
\bibitem [{\citenamefont {Williamson}\ \emph
  {et~al.}(2019{\natexlab{b}})\citenamefont {Williamson}, \citenamefont {Dua},\
  and\ \citenamefont {Cheng}}]{spurious}%
  \BibitemOpen
  \bibfield  {author} {\bibinfo {author} {\bibfnamefont {D.~J.}\ \bibnamefont
  {Williamson}}, \bibinfo {author} {\bibfnamefont {A.}~\bibnamefont {Dua}}, \
  and\ \bibinfo {author} {\bibfnamefont {M.}~\bibnamefont {Cheng}},\ }\href
  {\doibase 10.1103/PhysRevLett.122.140506} {\bibfield  {journal} {\bibinfo
  {journal} {Phys. Rev. Lett.}\ }\textbf {\bibinfo {volume} {122}},\ \bibinfo
  {pages} {140506} (\bibinfo {year} {2019}{\natexlab{b}})}\BibitemShut
  {NoStop}%
\bibitem [{\citenamefont {Schmitz}\ \emph {et~al.}(2019)\citenamefont
  {Schmitz}, \citenamefont {Huang},\ and\ \citenamefont {Prem}}]{albertespec}%
  \BibitemOpen
  \bibfield  {author} {\bibinfo {author} {\bibfnamefont {A.~T.}\ \bibnamefont
  {Schmitz}}, \bibinfo {author} {\bibfnamefont {S.-J.}\ \bibnamefont {Huang}},
  \ and\ \bibinfo {author} {\bibfnamefont {A.}~\bibnamefont {Prem}},\ }\href
  {\doibase 10.1103/PhysRevB.99.205109} {\bibfield  {journal} {\bibinfo
  {journal} {Phys. Rev. B}\ }\textbf {\bibinfo {volume} {99}},\ \bibinfo
  {pages} {205109} (\bibinfo {year} {2019})}\BibitemShut {NoStop}%
\bibitem [{\citenamefont {{Dua}}\ \emph
  {et~al.}(2019{\natexlab{a}})\citenamefont {{Dua}}, \citenamefont {{Sarkar}},
  \citenamefont {{Williamson}},\ and\ \citenamefont {{Cheng}}}]{dua2019bifur}%
  \BibitemOpen
  \bibfield  {author} {\bibinfo {author} {\bibfnamefont {A.}~\bibnamefont
  {{Dua}}}, \bibinfo {author} {\bibfnamefont {P.}~\bibnamefont {{Sarkar}}},
  \bibinfo {author} {\bibfnamefont {D.~J.}\ \bibnamefont {{Williamson}}}, \
  and\ \bibinfo {author} {\bibfnamefont {M.}~\bibnamefont {{Cheng}}},\
  }\href@noop {} {\  (\bibinfo {year} {2019}{\natexlab{a}})},\ \Eprint
  {http://arxiv.org/abs/1909.12304} {arXiv:1909.12304} \BibitemShut {NoStop}%
\bibitem [{\citenamefont {Shi}\ and\ \citenamefont
  {Lu}(2018)}]{ShiEntanglement}%
  \BibitemOpen
  \bibfield  {author} {\bibinfo {author} {\bibfnamefont {B.}~\bibnamefont
  {Shi}}\ and\ \bibinfo {author} {\bibfnamefont {Y.-M.}\ \bibnamefont {Lu}},\
  }\href {\doibase 10.1103/PhysRevB.97.144106} {\bibfield  {journal} {\bibinfo
  {journal} {Phys. Rev. B}\ }\textbf {\bibinfo {volume} {97}},\ \bibinfo
  {pages} {144106} (\bibinfo {year} {2018})}\BibitemShut {NoStop}%
\bibitem [{\citenamefont {Slagle}\ and\ \citenamefont
  {Kim}(2017{\natexlab{b}})}]{SlagleXcubeQFT}%
  \BibitemOpen
  \bibfield  {author} {\bibinfo {author} {\bibfnamefont {K.}~\bibnamefont
  {Slagle}}\ and\ \bibinfo {author} {\bibfnamefont {Y.~B.}\ \bibnamefont
  {Kim}},\ }\href {\doibase 10.1103/PhysRevB.96.195139} {\bibfield  {journal}
  {\bibinfo  {journal} {Phys. Rev. B}\ }\textbf {\bibinfo {volume} {96}},\
  \bibinfo {pages} {195139} (\bibinfo {year} {2017}{\natexlab{b}})}\BibitemShut
  {NoStop}%
\bibitem [{\citenamefont {Slagle}\ and\ \citenamefont {Kim}(2018)}]{slagle3}%
  \BibitemOpen
  \bibfield  {author} {\bibinfo {author} {\bibfnamefont {K.}~\bibnamefont
  {Slagle}}\ and\ \bibinfo {author} {\bibfnamefont {Y.~B.}\ \bibnamefont
  {Kim}},\ }\href {\doibase 10.1103/PhysRevB.97.165106} {\bibfield  {journal}
  {\bibinfo  {journal} {Phys. Rev. B}\ }\textbf {\bibinfo {volume} {97}},\
  \bibinfo {pages} {165106} (\bibinfo {year} {2018})}\BibitemShut {NoStop}%
\bibitem [{\citenamefont {Shirley}\ \emph {et~al.}(2018)\citenamefont
  {Shirley}, \citenamefont {Slagle}, \citenamefont {Wang},\ and\ \citenamefont
  {Chen}}]{shirleygeneral}%
  \BibitemOpen
  \bibfield  {author} {\bibinfo {author} {\bibfnamefont {W.}~\bibnamefont
  {Shirley}}, \bibinfo {author} {\bibfnamefont {K.}~\bibnamefont {Slagle}},
  \bibinfo {author} {\bibfnamefont {Z.}~\bibnamefont {Wang}}, \ and\ \bibinfo
  {author} {\bibfnamefont {X.}~\bibnamefont {Chen}},\ }\href {\doibase
  10.1103/PhysRevX.8.031051} {\bibfield  {journal} {\bibinfo  {journal} {Phys.
  Rev. X}\ }\textbf {\bibinfo {volume} {8}},\ \bibinfo {pages} {031051}
  (\bibinfo {year} {2018})}\BibitemShut {NoStop}%
\bibitem [{\citenamefont {Prem}\ \emph {et~al.}(2019)\citenamefont {Prem},
  \citenamefont {Huang}, \citenamefont {Song},\ and\ \citenamefont
  {Hermele}}]{cagenet}%
  \BibitemOpen
  \bibfield  {author} {\bibinfo {author} {\bibfnamefont {A.}~\bibnamefont
  {Prem}}, \bibinfo {author} {\bibfnamefont {S.-J.}\ \bibnamefont {Huang}},
  \bibinfo {author} {\bibfnamefont {H.}~\bibnamefont {Song}}, \ and\ \bibinfo
  {author} {\bibfnamefont {M.}~\bibnamefont {Hermele}},\ }\href {\doibase
  10.1103/PhysRevX.9.021010} {\bibfield  {journal} {\bibinfo  {journal} {Phys.
  Rev. X}\ }\textbf {\bibinfo {volume} {9}},\ \bibinfo {pages} {021010}
  (\bibinfo {year} {2019})}\BibitemShut {NoStop}%
\bibitem [{\citenamefont {{Slagle}}\ \emph
  {et~al.}(2019{\natexlab{a}})\citenamefont {{Slagle}}, \citenamefont
  {{Prem}},\ and\ \citenamefont {{Pretko}}}]{symmetric}%
  \BibitemOpen
  \bibfield  {author} {\bibinfo {author} {\bibfnamefont {K.}~\bibnamefont
  {{Slagle}}}, \bibinfo {author} {\bibfnamefont {A.}~\bibnamefont {{Prem}}}, \
  and\ \bibinfo {author} {\bibfnamefont {M.}~\bibnamefont {{Pretko}}},\ }\href
  {\doibase 10.1016/j.aop.2019.167910} {\bibfield  {journal} {\bibinfo
  {journal} {Annals of Physics}\ }\textbf {\bibinfo {volume} {410}},\ \bibinfo
  {eid} {167910} (\bibinfo {year} {2019}{\natexlab{a}})},\ \Eprint
  {http://arxiv.org/abs/1807.00827} {arXiv:1807.00827} \BibitemShut {NoStop}%
\bibitem [{\citenamefont {{Slagle}}\ \emph
  {et~al.}(2019{\natexlab{b}})\citenamefont {{Slagle}}, \citenamefont
  {{Aasen}},\ and\ \citenamefont {{Williamson}}}]{SlagleSMN}%
  \BibitemOpen
  \bibfield  {author} {\bibinfo {author} {\bibfnamefont {K.}~\bibnamefont
  {{Slagle}}}, \bibinfo {author} {\bibfnamefont {D.}~\bibnamefont {{Aasen}}}, \
  and\ \bibinfo {author} {\bibfnamefont {D.}~\bibnamefont {{Williamson}}},\
  }\href {\doibase 10.21468/SciPostPhys.6.4.043} {\bibfield  {journal}
  {\bibinfo  {journal} {SciPost Physics}\ }\textbf {\bibinfo {volume} {6}},\
  \bibinfo {eid} {043} (\bibinfo {year} {2019}{\natexlab{b}})}\BibitemShut
  {NoStop}%
\bibitem [{\citenamefont {Gromov}(2019{\natexlab{b}})}]{gromov2}%
  \BibitemOpen
  \bibfield  {author} {\bibinfo {author} {\bibfnamefont {A.}~\bibnamefont
  {Gromov}},\ }\href {\doibase 10.1103/PhysRevX.9.031035} {\bibfield  {journal}
  {\bibinfo  {journal} {Phys. Rev. X}\ }\textbf {\bibinfo {volume} {9}},\
  \bibinfo {pages} {031035} (\bibinfo {year} {2019}{\natexlab{b}})}\BibitemShut
  {NoStop}%
\bibitem [{\citenamefont {{Tian}}\ \emph {et~al.}(2018)\citenamefont {{Tian}},
  \citenamefont {{Samperton}},\ and\ \citenamefont {{Wang}}}]{tian2018}%
  \BibitemOpen
  \bibfield  {author} {\bibinfo {author} {\bibfnamefont {K.~T.}\ \bibnamefont
  {{Tian}}}, \bibinfo {author} {\bibfnamefont {E.}~\bibnamefont {{Samperton}}},
  \ and\ \bibinfo {author} {\bibfnamefont {Z.}~\bibnamefont {{Wang}}},\
  }\href@noop {} {\  (\bibinfo {year} {2018})},\ \Eprint
  {http://arxiv.org/abs/1812.02101} {arXiv:1812.02101} \BibitemShut {NoStop}%
\bibitem [{\citenamefont {Shirley}\ \emph
  {et~al.}(2019{\natexlab{a}})\citenamefont {Shirley}, \citenamefont {Slagle},\
  and\ \citenamefont {Chen}}]{chenfoliatedent}%
  \BibitemOpen
  \bibfield  {author} {\bibinfo {author} {\bibfnamefont {W.}~\bibnamefont
  {Shirley}}, \bibinfo {author} {\bibfnamefont {K.}~\bibnamefont {Slagle}}, \
  and\ \bibinfo {author} {\bibfnamefont {X.}~\bibnamefont {Chen}},\ }\href
  {\doibase 10.21468/SciPostPhys.6.1.015} {\bibfield  {journal} {\bibinfo
  {journal} {SciPost Phys.}\ }\textbf {\bibinfo {volume} {6}},\ \bibinfo
  {pages} {15} (\bibinfo {year} {2019}{\natexlab{a}})}\BibitemShut {NoStop}%
\bibitem [{\citenamefont {Shirley}\ \emph
  {et~al.}(2019{\natexlab{b}})\citenamefont {Shirley}, \citenamefont {Slagle},\
  and\ \citenamefont {Chen}}]{foliatedcb}%
  \BibitemOpen
  \bibfield  {author} {\bibinfo {author} {\bibfnamefont {W.}~\bibnamefont
  {Shirley}}, \bibinfo {author} {\bibfnamefont {K.}~\bibnamefont {Slagle}}, \
  and\ \bibinfo {author} {\bibfnamefont {X.}~\bibnamefont {Chen}},\ }\href
  {\doibase 10.1103/PhysRevB.99.115123} {\bibfield  {journal} {\bibinfo
  {journal} {Phys. Rev. B}\ }\textbf {\bibinfo {volume} {99}},\ \bibinfo
  {pages} {115123} (\bibinfo {year} {2019}{\natexlab{b}})}\BibitemShut
  {NoStop}%
\bibitem [{\citenamefont {{Shirley}}\ \emph {et~al.}(2019)\citenamefont
  {{Shirley}}, \citenamefont {{Slagle}},\ and\ \citenamefont
  {{Chen}}}]{TwistedFoliated}%
  \BibitemOpen
  \bibfield  {author} {\bibinfo {author} {\bibfnamefont {W.}~\bibnamefont
  {{Shirley}}}, \bibinfo {author} {\bibfnamefont {K.}~\bibnamefont {{Slagle}}},
  \ and\ \bibinfo {author} {\bibfnamefont {X.}~\bibnamefont {{Chen}}},\
  }\href@noop {} {\  (\bibinfo {year} {2019})},\ \Eprint
  {http://arxiv.org/abs/1907.09048} {arXiv:1907.09048} \BibitemShut {NoStop}%
\bibitem [{\citenamefont {Pai}\ and\ \citenamefont
  {Hermele}(2019)}]{hermelefusion}%
  \BibitemOpen
  \bibfield  {author} {\bibinfo {author} {\bibfnamefont {S.}~\bibnamefont
  {Pai}}\ and\ \bibinfo {author} {\bibfnamefont {M.}~\bibnamefont {Hermele}},\
  }\href {\doibase 10.1103/PhysRevB.100.195136} {\bibfield  {journal} {\bibinfo
   {journal} {Phys. Rev. B}\ }\textbf {\bibinfo {volume} {100}},\ \bibinfo
  {pages} {195136} (\bibinfo {year} {2019})}\BibitemShut {NoStop}%
\bibitem [{\citenamefont {Bulmash}\ and\ \citenamefont
  {Barkeshli}(2019)}]{bulmashgauging}%
  \BibitemOpen
  \bibfield  {author} {\bibinfo {author} {\bibfnamefont {D.}~\bibnamefont
  {Bulmash}}\ and\ \bibinfo {author} {\bibfnamefont {M.}~\bibnamefont
  {Barkeshli}},\ }\href {\doibase 10.1103/PhysRevB.100.155146} {\bibfield
  {journal} {\bibinfo  {journal} {Phys. Rev. B}\ }\textbf {\bibinfo {volume}
  {100}},\ \bibinfo {pages} {155146} (\bibinfo {year} {2019})}\BibitemShut
  {NoStop}%
\bibitem [{\citenamefont {Prem}\ and\ \citenamefont
  {Williamson}(2019)}]{premgauging}%
  \BibitemOpen
  \bibfield  {author} {\bibinfo {author} {\bibfnamefont {A.}~\bibnamefont
  {Prem}}\ and\ \bibinfo {author} {\bibfnamefont {D.~J.}\ \bibnamefont
  {Williamson}},\ }\href {\doibase 10.21468/SciPostPhys.7.5.068} {\bibfield
  {journal} {\bibinfo  {journal} {SciPost Phys.}\ }\textbf {\bibinfo {volume}
  {7}},\ \bibinfo {pages} {68} (\bibinfo {year} {2019})}\BibitemShut {NoStop}%
\bibitem [{\citenamefont {Aasen}\ \emph {et~al.}()\citenamefont {Aasen},
  \citenamefont {He}, \citenamefont {Prem}, \citenamefont {Slagle},\ and\
  \citenamefont {Williamson}}]{GeneralizedSMN}%
  \BibitemOpen
  \bibfield  {author} {\bibinfo {author} {\bibfnamefont {D.}~\bibnamefont
  {Aasen}}, \bibinfo {author} {\bibfnamefont {H.}~\bibnamefont {He}}, \bibinfo
  {author} {\bibfnamefont {A.}~\bibnamefont {Prem}}, \bibinfo {author}
  {\bibfnamefont {K.}~\bibnamefont {Slagle}}, \ and\ \bibinfo {author}
  {\bibfnamefont {D.}~\bibnamefont {Williamson}},\ }\href@noop {} {\bibinfo
  {journal} {in preparation}\ }\BibitemShut {NoStop}%
\bibitem [{\citenamefont {{Haah}}(2018)}]{Haah2018b}%
  \BibitemOpen
\bibfield  {journal} {  }\bibfield  {author} {\bibinfo {author} {\bibfnamefont
  {J.}~\bibnamefont {{Haah}}},\ }\href@noop {} {\  (\bibinfo {year} {2018})},\
  \Eprint {http://arxiv.org/abs/1812.11193} {arXiv:1812.11193} \BibitemShut
  {NoStop}%
\bibitem [{\citenamefont {{Dua}}\ \emph
  {et~al.}(2019{\natexlab{b}})\citenamefont {{Dua}}, \citenamefont {{Kim}},
  \citenamefont {{Cheng}},\ and\ \citenamefont {{Williamson}}}]{DuaSorting}%
  \BibitemOpen
  \bibfield  {author} {\bibinfo {author} {\bibfnamefont {A.}~\bibnamefont
  {{Dua}}}, \bibinfo {author} {\bibfnamefont {I.~H.}\ \bibnamefont {{Kim}}},
  \bibinfo {author} {\bibfnamefont {M.}~\bibnamefont {{Cheng}}}, \ and\
  \bibinfo {author} {\bibfnamefont {D.~J.}\ \bibnamefont {{Williamson}}},\
  }\href {\doibase 10.1103/PhysRevB.100.155137} {\bibfield  {journal} {\bibinfo
   {journal} {\prb}\ }\textbf {\bibinfo {volume} {100}},\ \bibinfo {eid}
  {155137} (\bibinfo {year} {2019}{\natexlab{b}})}\BibitemShut {NoStop}%
\bibitem [{\citenamefont {Devakul}\ \emph {et~al.}(2018)\citenamefont
  {Devakul}, \citenamefont {Williamson},\ and\ \citenamefont
  {You}}]{strongsspt}%
  \BibitemOpen
  \bibfield  {author} {\bibinfo {author} {\bibfnamefont {T.}~\bibnamefont
  {Devakul}}, \bibinfo {author} {\bibfnamefont {D.~J.}\ \bibnamefont
  {Williamson}}, \ and\ \bibinfo {author} {\bibfnamefont {Y.}~\bibnamefont
  {You}},\ }\href {\doibase 10.1103/PhysRevB.98.235121} {\bibfield  {journal}
  {\bibinfo  {journal} {Phys. Rev. B}\ }\textbf {\bibinfo {volume} {98}},\
  \bibinfo {pages} {235121} (\bibinfo {year} {2018})}\BibitemShut {NoStop}%
\bibitem [{\citenamefont {{Morrison}}\ and\ \citenamefont
  {{Walker}}(2010)}]{morrison2010}%
  \BibitemOpen
  \bibfield  {author} {\bibinfo {author} {\bibfnamefont {S.}~\bibnamefont
  {{Morrison}}}\ and\ \bibinfo {author} {\bibfnamefont {K.}~\bibnamefont
  {{Walker}}},\ }\href@noop {} {\  (\bibinfo {year} {2010})},\ \Eprint
  {http://arxiv.org/abs/1009.5025} {arXiv:1009.5025} \BibitemShut {NoStop}%
\bibitem [{\citenamefont {{Kapustin}}\ \emph {et~al.}(2010)\citenamefont
  {{Kapustin}}, \citenamefont {{Setter}},\ and\ \citenamefont
  {{Vyas}}}]{kapustin2010}%
  \BibitemOpen
  \bibfield  {author} {\bibinfo {author} {\bibfnamefont {A.}~\bibnamefont
  {{Kapustin}}}, \bibinfo {author} {\bibfnamefont {K.}~\bibnamefont
  {{Setter}}}, \ and\ \bibinfo {author} {\bibfnamefont {K.}~\bibnamefont
  {{Vyas}}},\ }\href@noop {} {\  (\bibinfo {year} {2010})},\ \Eprint
  {http://arxiv.org/abs/1002.0385} {arXiv:1002.0385} \BibitemShut {NoStop}%
\bibitem [{\citenamefont {{Davydov}}\ \emph {et~al.}(2011)\citenamefont
  {{Davydov}}, \citenamefont {{Kong}},\ and\ \citenamefont
  {{Runkel}}}]{dkm2011}%
  \BibitemOpen
  \bibfield  {author} {\bibinfo {author} {\bibfnamefont {A.}~\bibnamefont
  {{Davydov}}}, \bibinfo {author} {\bibfnamefont {L.}~\bibnamefont {{Kong}}}, \
  and\ \bibinfo {author} {\bibfnamefont {I.}~\bibnamefont {{Runkel}}},\
  }\href@noop {} {\  (\bibinfo {year} {2011})},\ \Eprint
  {http://arxiv.org/abs/1107.0495} {arXiv:1107.0495} \BibitemShut {NoStop}%
\bibitem [{\citenamefont {{Carqueville}}(2016)}]{carqueville2016}%
  \BibitemOpen
  \bibfield  {author} {\bibinfo {author} {\bibfnamefont {N.}~\bibnamefont
  {{Carqueville}}},\ }\href@noop {} {\  (\bibinfo {year} {2016})},\ \Eprint
  {http://arxiv.org/abs/1607.05747} {arXiv:1607.05747} \BibitemShut {NoStop}%
\bibitem [{\citenamefont {{Carqueville}}\ \emph {et~al.}(2017)\citenamefont
  {{Carqueville}}, \citenamefont {{Runkel}},\ and\ \citenamefont
  {{Schaumann}}}]{carqueville2017}%
  \BibitemOpen
  \bibfield  {author} {\bibinfo {author} {\bibfnamefont {N.}~\bibnamefont
  {{Carqueville}}}, \bibinfo {author} {\bibfnamefont {I.}~\bibnamefont
  {{Runkel}}}, \ and\ \bibinfo {author} {\bibfnamefont {G.}~\bibnamefont
  {{Schaumann}}},\ }\href@noop {} {\  (\bibinfo {year} {2017})},\ \Eprint
  {http://arxiv.org/abs/1710.10214} {arXiv:1710.10214} \BibitemShut {NoStop}%
\bibitem [{\citenamefont {{Carqueville}}\ \emph {et~al.}(2016)\citenamefont
  {{Carqueville}}, \citenamefont {{Meusburger}},\ and\ \citenamefont
  {{Schaumann}}}]{cms2016}%
  \BibitemOpen
  \bibfield  {author} {\bibinfo {author} {\bibfnamefont {N.}~\bibnamefont
  {{Carqueville}}}, \bibinfo {author} {\bibfnamefont {C.}~\bibnamefont
  {{Meusburger}}}, \ and\ \bibinfo {author} {\bibfnamefont {G.}~\bibnamefont
  {{Schaumann}}},\ }\href@noop {} {\  (\bibinfo {year} {2016})},\ \Eprint
  {http://arxiv.org/abs/1603.01171} {arXiv:1603.01171} \BibitemShut {NoStop}%
\bibitem [{\citenamefont {Song}\ \emph {et~al.}(2017)\citenamefont {Song},
  \citenamefont {Huang}, \citenamefont {Fu},\ and\ \citenamefont
  {Hermele}}]{song2017defect}%
  \BibitemOpen
  \bibfield  {author} {\bibinfo {author} {\bibfnamefont {H.}~\bibnamefont
  {Song}}, \bibinfo {author} {\bibfnamefont {S.-J.}\ \bibnamefont {Huang}},
  \bibinfo {author} {\bibfnamefont {L.}~\bibnamefont {Fu}}, \ and\ \bibinfo
  {author} {\bibfnamefont {M.}~\bibnamefont {Hermele}},\ }\href {\doibase
  10.1103/PhysRevX.7.011020} {\bibfield  {journal} {\bibinfo  {journal} {Phys.
  Rev. X}\ }\textbf {\bibinfo {volume} {7}},\ \bibinfo {pages} {011020}
  (\bibinfo {year} {2017})}\BibitemShut {NoStop}%
\bibitem [{\citenamefont {Huang}\ \emph {et~al.}(2017)\citenamefont {Huang},
  \citenamefont {Song}, \citenamefont {Huang},\ and\ \citenamefont
  {Hermele}}]{huang2017defect}%
  \BibitemOpen
  \bibfield  {author} {\bibinfo {author} {\bibfnamefont {S.-J.}\ \bibnamefont
  {Huang}}, \bibinfo {author} {\bibfnamefont {H.}~\bibnamefont {Song}},
  \bibinfo {author} {\bibfnamefont {Y.-P.}\ \bibnamefont {Huang}}, \ and\
  \bibinfo {author} {\bibfnamefont {M.}~\bibnamefont {Hermele}},\ }\href
  {\doibase 10.1103/PhysRevB.96.205106} {\bibfield  {journal} {\bibinfo
  {journal} {Phys. Rev. B}\ }\textbf {\bibinfo {volume} {96}},\ \bibinfo
  {pages} {205106} (\bibinfo {year} {2017})}\BibitemShut {NoStop}%
\bibitem [{\citenamefont {Else}\ and\ \citenamefont
  {Thorngren}(2019)}]{else2019crystalline}%
  \BibitemOpen
  \bibfield  {author} {\bibinfo {author} {\bibfnamefont {D.~V.}\ \bibnamefont
  {Else}}\ and\ \bibinfo {author} {\bibfnamefont {R.}~\bibnamefont
  {Thorngren}},\ }\href {\doibase 10.1103/PhysRevB.99.115116} {\bibfield
  {journal} {\bibinfo  {journal} {Phys. Rev. B}\ }\textbf {\bibinfo {volume}
  {99}},\ \bibinfo {pages} {115116} (\bibinfo {year} {2019})}\BibitemShut
  {NoStop}%
\bibitem [{\citenamefont {{Bravyi}}\ and\ \citenamefont
  {{Kitaev}}(1998)}]{bravyi1998}%
  \BibitemOpen
  \bibfield  {author} {\bibinfo {author} {\bibfnamefont {S.~B.}\ \bibnamefont
  {{Bravyi}}}\ and\ \bibinfo {author} {\bibfnamefont {A.~Y.}\ \bibnamefont
  {{Kitaev}}},\ }\href@noop {} {\  (\bibinfo {year} {1998})},\ \Eprint
  {http://arxiv.org/abs/quant-ph/9811052} {arXiv:quant-ph/9811052} \BibitemShut
  {NoStop}%
\bibitem [{\citenamefont {Bombin}(2010)}]{BombinDualityDefect}%
  \BibitemOpen
  \bibfield  {author} {\bibinfo {author} {\bibfnamefont {H.}~\bibnamefont
  {Bombin}},\ }\href {\doibase 10.1103/PhysRevLett.105.030403} {\bibfield
  {journal} {\bibinfo  {journal} {Phys. Rev. Lett.}\ }\textbf {\bibinfo
  {volume} {105}},\ \bibinfo {pages} {030403} (\bibinfo {year}
  {2010})}\BibitemShut {NoStop}%
\bibitem [{\citenamefont {Bombin}\ and\ \citenamefont
  {Martin-Delgado}(2008)}]{bombin2008}%
  \BibitemOpen
  \bibfield  {author} {\bibinfo {author} {\bibfnamefont {H.}~\bibnamefont
  {Bombin}}\ and\ \bibinfo {author} {\bibfnamefont {M.~A.}\ \bibnamefont
  {Martin-Delgado}},\ }\href {\doibase 10.1103/PhysRevB.78.115421} {\bibfield
  {journal} {\bibinfo  {journal} {Phys. Rev. B}\ }\textbf {\bibinfo {volume}
  {78}},\ \bibinfo {pages} {115421} (\bibinfo {year} {2008})}\BibitemShut
  {NoStop}%
\bibitem [{\citenamefont {{Kitaev}}\ and\ \citenamefont
  {{Kong}}(2012)}]{kitaevkong}%
  \BibitemOpen
  \bibfield  {author} {\bibinfo {author} {\bibfnamefont {A.}~\bibnamefont
  {{Kitaev}}}\ and\ \bibinfo {author} {\bibfnamefont {L.}~\bibnamefont
  {{Kong}}},\ }\href {\doibase 10.1007/s00220-012-1500-5} {\bibfield  {journal}
  {\bibinfo  {journal} {Communications in Mathematical Physics}\ }\textbf
  {\bibinfo {volume} {313}},\ \bibinfo {pages} {351} (\bibinfo {year}
  {2012})},\ \Eprint {http://arxiv.org/abs/1104.5047} {arXiv:1104.5047}
  \BibitemShut {NoStop}%
\bibitem [{\citenamefont {{Beigi}}\ \emph {et~al.}(2011)\citenamefont
  {{Beigi}}, \citenamefont {{Shor}},\ and\ \citenamefont
  {{Whalen}}}]{beigi2011bdry}%
  \BibitemOpen
  \bibfield  {author} {\bibinfo {author} {\bibfnamefont {S.}~\bibnamefont
  {{Beigi}}}, \bibinfo {author} {\bibfnamefont {P.~W.}\ \bibnamefont {{Shor}}},
  \ and\ \bibinfo {author} {\bibfnamefont {D.}~\bibnamefont {{Whalen}}},\
  }\href {\doibase 10.1007/s00220-011-1294-x} {\bibfield  {journal} {\bibinfo
  {journal} {Communications in Mathematical Physics}\ }\textbf {\bibinfo
  {volume} {306}},\ \bibinfo {pages} {663} (\bibinfo {year} {2011})},\ \Eprint
  {http://arxiv.org/abs/1006.5479} {arXiv:1006.5479} \BibitemShut {NoStop}%
\bibitem [{\citenamefont {Barkeshli}\ \emph {et~al.}(2019)\citenamefont
  {Barkeshli}, \citenamefont {Bonderson}, \citenamefont {Cheng},\ and\
  \citenamefont {Wang}}]{barkeshli2014symmetry}%
  \BibitemOpen
  \bibfield  {author} {\bibinfo {author} {\bibfnamefont {M.}~\bibnamefont
  {Barkeshli}}, \bibinfo {author} {\bibfnamefont {P.}~\bibnamefont
  {Bonderson}}, \bibinfo {author} {\bibfnamefont {M.}~\bibnamefont {Cheng}}, \
  and\ \bibinfo {author} {\bibfnamefont {Z.}~\bibnamefont {Wang}},\ }\href
  {\doibase 10.1103/PhysRevB.100.115147} {\bibfield  {journal} {\bibinfo
  {journal} {Phys. Rev. B}\ }\textbf {\bibinfo {volume} {100}},\ \bibinfo
  {pages} {115147} (\bibinfo {year} {2019})}\BibitemShut {NoStop}%
\bibitem [{\citenamefont {{Cong}}\ \emph {et~al.}(2016)\citenamefont {{Cong}},
  \citenamefont {{Cheng}},\ and\ \citenamefont {{Wang}}}]{cong2016gapped}%
  \BibitemOpen
  \bibfield  {author} {\bibinfo {author} {\bibfnamefont {I.}~\bibnamefont
  {{Cong}}}, \bibinfo {author} {\bibfnamefont {M.}~\bibnamefont {{Cheng}}}, \
  and\ \bibinfo {author} {\bibfnamefont {Z.}~\bibnamefont {{Wang}}},\
  }\href@noop {} {\  (\bibinfo {year} {2016})},\ \Eprint
  {http://arxiv.org/abs/1609.02037} {arXiv:1609.02037} \BibitemShut {NoStop}%
\bibitem [{\citenamefont {{Yoshida}}(2017)}]{yoshida2017gapped}%
  \BibitemOpen
  \bibfield  {author} {\bibinfo {author} {\bibfnamefont {B.}~\bibnamefont
  {{Yoshida}}},\ }\href {\doibase 10.1016/j.aop.2016.12.014} {\bibfield
  {journal} {\bibinfo  {journal} {Annals of Physics}\ }\textbf {\bibinfo
  {volume} {377}},\ \bibinfo {pages} {387} (\bibinfo {year} {2017})},\ \Eprint
  {http://arxiv.org/abs/1509.03626} {arXiv:1509.03626} \BibitemShut {NoStop}%
\bibitem [{\citenamefont {{Wang}}\ \emph {et~al.}(2018)\citenamefont {{Wang}},
  \citenamefont {{Li}}, \citenamefont {{Hu}},\ and\ \citenamefont
  {{Wan}}}]{wang2018gapped}%
  \BibitemOpen
  \bibfield  {author} {\bibinfo {author} {\bibfnamefont {H.}~\bibnamefont
  {{Wang}}}, \bibinfo {author} {\bibfnamefont {Y.}~\bibnamefont {{Li}}},
  \bibinfo {author} {\bibfnamefont {Y.}~\bibnamefont {{Hu}}}, \ and\ \bibinfo
  {author} {\bibfnamefont {Y.}~\bibnamefont {{Wan}}},\ }\href {\doibase
  10.1007/JHEP10(2018)114} {\bibfield  {journal} {\bibinfo  {journal} {Journal
  of High Energy Physics}\ }\textbf {\bibinfo {volume} {2018}},\ \bibinfo {eid}
  {114} (\bibinfo {year} {2018})},\ \Eprint {http://arxiv.org/abs/1807.11083}
  {arXiv:1807.11083} \BibitemShut {NoStop}%
\bibitem [{\citenamefont {{Chen}}\ \emph {et~al.}(2010)\citenamefont {{Chen}},
  \citenamefont {{Gu}},\ and\ \citenamefont {{Wen}}}]{ChenWenLU}%
  \BibitemOpen
  \bibfield  {author} {\bibinfo {author} {\bibfnamefont {X.}~\bibnamefont
  {{Chen}}}, \bibinfo {author} {\bibfnamefont {Z.-C.}\ \bibnamefont {{Gu}}}, \
  and\ \bibinfo {author} {\bibfnamefont {X.-G.}\ \bibnamefont {{Wen}}},\ }\href
  {\doibase 10.1103/PhysRevB.82.155138} {\bibfield  {journal} {\bibinfo
  {journal} {\prb}\ }\textbf {\bibinfo {volume} {82}},\ \bibinfo {eid} {155138}
  (\bibinfo {year} {2010})}\BibitemShut {NoStop}%
\bibitem [{\citenamefont {Castelnovo}\ and\ \citenamefont
  {Chamon}(2008)}]{chamonfiniteT}%
  \BibitemOpen
  \bibfield  {author} {\bibinfo {author} {\bibfnamefont {C.}~\bibnamefont
  {Castelnovo}}\ and\ \bibinfo {author} {\bibfnamefont {C.}~\bibnamefont
  {Chamon}},\ }\href {\doibase 10.1103/PhysRevB.78.155120} {\bibfield
  {journal} {\bibinfo  {journal} {Phys. Rev. B}\ }\textbf {\bibinfo {volume}
  {78}},\ \bibinfo {pages} {155120} (\bibinfo {year} {2008})}\BibitemShut
  {NoStop}%
\bibitem [{\citenamefont {Dijkgraaf}\ and\ \citenamefont
  {Witten}(1990)}]{dijkgraaf}%
  \BibitemOpen
  \bibfield  {author} {\bibinfo {author} {\bibfnamefont {R.}~\bibnamefont
  {Dijkgraaf}}\ and\ \bibinfo {author} {\bibfnamefont {E.}~\bibnamefont
  {Witten}},\ }\href {\doibase 10.1007/BF02096988} {\bibfield  {journal}
  {\bibinfo  {journal} {Communications in Mathematical Physics}\ }\textbf
  {\bibinfo {volume} {129}},\ \bibinfo {pages} {393} (\bibinfo {year}
  {1990})}\BibitemShut {NoStop}%
\bibitem [{\citenamefont {Lan}\ \emph {et~al.}(2015)\citenamefont {Lan},
  \citenamefont {Wang},\ and\ \citenamefont {Wen}}]{Lan2015}%
  \BibitemOpen
  \bibfield  {author} {\bibinfo {author} {\bibfnamefont {T.}~\bibnamefont
  {Lan}}, \bibinfo {author} {\bibfnamefont {J.~C.}\ \bibnamefont {Wang}}, \
  and\ \bibinfo {author} {\bibfnamefont {X.-G.}\ \bibnamefont {Wen}},\ }\href
  {\doibase 10.1103/PhysRevLett.114.076402} {\bibfield  {journal} {\bibinfo
  {journal} {Phys. Rev. Lett.}\ }\textbf {\bibinfo {volume} {114}},\ \bibinfo
  {pages} {076402} (\bibinfo {year} {2015})}\BibitemShut {NoStop}%
\bibitem [{\citenamefont {Chen}\ \emph {et~al.}(2011)\citenamefont {Chen},
  \citenamefont {Gu},\ and\ \citenamefont {Wen}}]{Chen2011}%
  \BibitemOpen
  \bibfield  {author} {\bibinfo {author} {\bibfnamefont {X.}~\bibnamefont
  {Chen}}, \bibinfo {author} {\bibfnamefont {Z.-C.}\ \bibnamefont {Gu}}, \ and\
  \bibinfo {author} {\bibfnamefont {X.-G.}\ \bibnamefont {Wen}},\ }\href
  {\doibase 10.1103/PhysRevB.83.035107} {\bibfield  {journal} {\bibinfo
  {journal} {Phys. Rev. B}\ }\textbf {\bibinfo {volume} {83}},\ \bibinfo
  {pages} {035107} (\bibinfo {year} {2011})}\BibitemShut {NoStop}%
\bibitem [{\citenamefont {Schuch}\ \emph {et~al.}(2011)\citenamefont {Schuch},
  \citenamefont {P\'erez-Garc\'{\i}a},\ and\ \citenamefont
  {Cirac}}]{Schuch2011}%
  \BibitemOpen
  \bibfield  {author} {\bibinfo {author} {\bibfnamefont {N.}~\bibnamefont
  {Schuch}}, \bibinfo {author} {\bibfnamefont {D.}~\bibnamefont
  {P\'erez-Garc\'{\i}a}}, \ and\ \bibinfo {author} {\bibfnamefont
  {I.}~\bibnamefont {Cirac}},\ }\href {\doibase 10.1103/PhysRevB.84.165139}
  {\bibfield  {journal} {\bibinfo  {journal} {Phys. Rev. B}\ }\textbf {\bibinfo
  {volume} {84}},\ \bibinfo {pages} {165139} (\bibinfo {year}
  {2011})}\BibitemShut {NoStop}%
\bibitem [{\citenamefont {Williamson}\ and\ \citenamefont
  {Cheng}()}]{GaugedLayers}%
  \BibitemOpen
  \bibfield  {author} {\bibinfo {author} {\bibfnamefont {D.}~\bibnamefont
  {Williamson}}\ and\ \bibinfo {author} {\bibfnamefont {M.}~\bibnamefont
  {Cheng}},\ }\href@noop {} {\bibinfo  {journal} {in preparation}\
  }\BibitemShut {NoStop}%
\bibitem [{\citenamefont {{Wen}}(2020)}]{Wen2020}%
  \BibitemOpen
\bibfield  {journal} {  }\bibfield  {author} {\bibinfo {author} {\bibfnamefont
  {X.-G.}\ \bibnamefont {{Wen}}},\ }\href@noop {} {\  (\bibinfo {year}
  {2020})},\ \Eprint {http://arxiv.org/abs/2002.02433} {arXiv:2002.02433}
  \BibitemShut {NoStop}%
\bibitem [{\citenamefont {{Baez}}(1996)}]{BaezBF}%
  \BibitemOpen
  \bibfield  {author} {\bibinfo {author} {\bibfnamefont {J.~C.}\ \bibnamefont
  {{Baez}}},\ }\href {\doibase 10.1007/BF00398315} {\bibfield  {journal}
  {\bibinfo  {journal} {Letters in Mathematical Physics}\ }\textbf {\bibinfo
  {volume} {38}},\ \bibinfo {pages} {129} (\bibinfo {year} {1996})},\ \Eprint
  {http://arxiv.org/abs/q-alg/9507006} {arXiv:q-alg/9507006} \BibitemShut
  {NoStop}%
\bibitem [{foo()}]{foot:BFtheory}%
  \BibitemOpen
  \href@noop {} {}\bibinfo {note} {See also appendix A and B of
  \refcite{SlagleXcubeQFT}.}\BibitemShut {Stop}%
\bibitem [{\citenamefont {Wang}\ and\ \citenamefont
  {Chen}(2017)}]{wang2017twisted}%
  \BibitemOpen
  \bibfield  {author} {\bibinfo {author} {\bibfnamefont {Z.}~\bibnamefont
  {Wang}}\ and\ \bibinfo {author} {\bibfnamefont {X.}~\bibnamefont {Chen}},\
  }\href {\doibase 10.1103/PhysRevB.95.115142} {\bibfield  {journal} {\bibinfo
  {journal} {Phys. Rev. B}\ }\textbf {\bibinfo {volume} {95}},\ \bibinfo
  {pages} {115142} (\bibinfo {year} {2017})}\BibitemShut {NoStop}%
\end{thebibliography}%


\newpage 

\appendix

\section{Field Theory}
\label{sec:field theory}

In this appendix, we briefly describe the field theory description of the defects shown in \figref{fig:SMNdefect} and \ref{fig:SMNexcitations}.
The field theory is essentially the same as the foliated field theory in \refcite{SlagleSMN},
  except in \refcite{SlagleSMN} the defect layers were infinitesimally close together.
In this appendix, we show how the field theory is modified when the defects have a finite spacing between them.

Each 3-strata is described by a 3+1D toric code.
The field theory for 3+1D $\mathbb{Z}_N$ toric code is 3+1D BF theory \cite{BaezBF,foot:BFtheory}:
\begin{equation}
  L_\text{3} = \frac{2\pi}{N} b \wedge d a.
\end{equation}
Here $b$ is a 2-form gauge field while $a$ is a 1-form gauge field.
$db$ is the density of charge excitations, while $da$ is the density of flux strings.

Each 2-stratum is described by a 2d defect that is similar to 2d toric code, but has different local excitations.
The 1-strata connect the four neighboring 2-strata such that they behave like a pair of independent layers;
  so we will define the Lagrangian on these layers now.
A 2d toric code can be described by 2+1D BF theory with $L = \frac{2\pi}{N} B \wedge d A$ where $A$ and $B$ are 1-form gauge fields.
In order to obtain the modified toric code describing the defect shown in Figs.~\ref{fig:SMNdefect} and \ref{fig:SMNexcitations}, the Lagrangian needs an additional term coupling it to the 3+1D theory:
\begin{equation}
  L_\text{2}^{(\ell)} = \frac{2\pi}{N} B^{(\ell)} \wedge d A^{(\ell)} - \frac{2\pi}{N} b \wedge A^{(\ell)}.
\end{equation}
where $\ell$ indexes different layers.
For example, an X-cube model on an $L \times L \times L$ cubic lattice would be described by $3L$ layers.
See \refcite{SlagleSMN} for more details on how the last term affects the physics.

Note that $L_\text{2}^{(\ell)}$ is a 2+1D Lagrangian.
The action for the entire theory is
\begin{equation}
  S = \int L_\text{3} + \sum_{\ell} \int_{\ell} L_\text{2}^{(\ell)}
\end{equation}
where $\sum_{\ell}$ sums over every layer and $\int_{\ell}$ integrates over the 2+1 dimensional spacetime of the layer $\ell$.

\section{Fractal-like Lineon Model}
\label{app:fractallineonmodel}

In this appendix, we briefly explain a simple construction to obtain a lineon model where lineons are created at corners of fractal operators.

\begin{figure}
    \centering
        \includegraphics[width=0.95\linewidth]{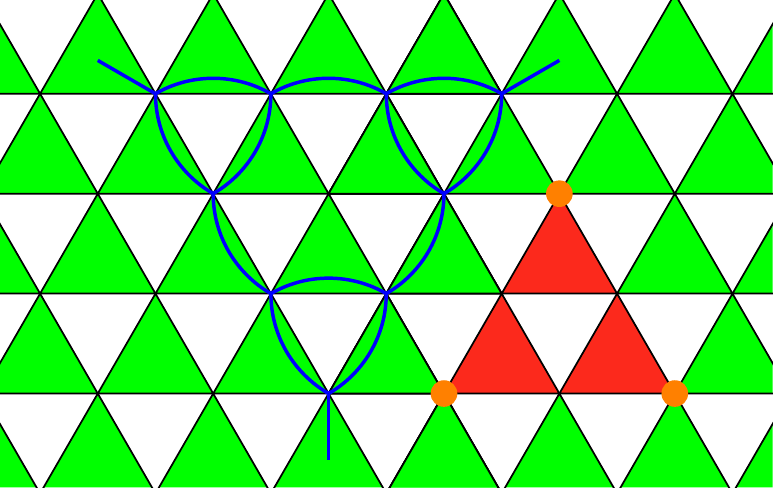}
    \caption{
    A 2d slice of an XY-plane showing electric and magnetic lineon excitations created at corners of fractal operators (blue and red). Each triangle represents a triangular prism of 3d toric code that extends into the Z-axis.
    }\label{fig:triangles}
\end{figure}

The construction begins with a grid of 3+1D toric codes on the colored triangular prisms shown in \figref{fig:triangles}.
Nothing is placed in the white regions.
Similar to the X-Cube construction in \secref{sec:Xcube defect condensation}, flux excitations are condensed along the faces of the 3+1D toric codes.
Along the 1D corner where three triangular prisms touch, pairs of fluxes or triples of charges are condensed; a generating set of condensed excitations is
\begin{align} 
\{ m_1 m_2, m_2 m_3, e_1 e_2 e_3 \}
\end{align}
where 1, 2, and 3 label the three different 3-strata.
(The ordering does not matter since $m_1 m_3$ is also condensed as a result.)

The model hosts electric and magnetic lineon excitations which are created at corners of fractal operators.
For example, three electric lineons are created at the ends of the blue electric string operator in \figref{fig:triangles} (which is a product of $Z$ operators using the conventions of \secref{sec:3D toric code}), and three magnetic lineons are created at the corners (orange) of the red flux membrane operator (built from a product of $X$ operators).

\section{\texorpdfstring{$\mathbb{Z}_3$}{Z3} type-I fracton model from coupled 2+1D layers}
\label{app:trivbulk}

In this appendix we present a construction of a type-I model from coupled 2+1D $\mathbb{Z}_3$ gauge theory. This example illustrates that \emph{cubic} lattice topological defect networks with trivial 3-strata may produce non-trivial fracton models.

Our starting point is $\mathbb{Z}_3$ gauge theory, written in the particle basis that decomposes into chiral and antichiral $\mathbb{Z}_3$ layers. The particles are labelled $ij$ with $i,j\in{0,1,2}$, fusion is given by separate additions mod 3 for the $i$ and $j$ entries, all $F$-symbols are trivial and the braidings are specified by the $S$-matrix
\begin{align}
    S_{ij,kl} = \omega^{ik-jl} ,
\end{align}
for $\omega = e^{2 \pi i / 3}$. 

\begin{figure}[t]
    \centering
    \subfloat[]{\includegraphics[height=0.4\columnwidth]{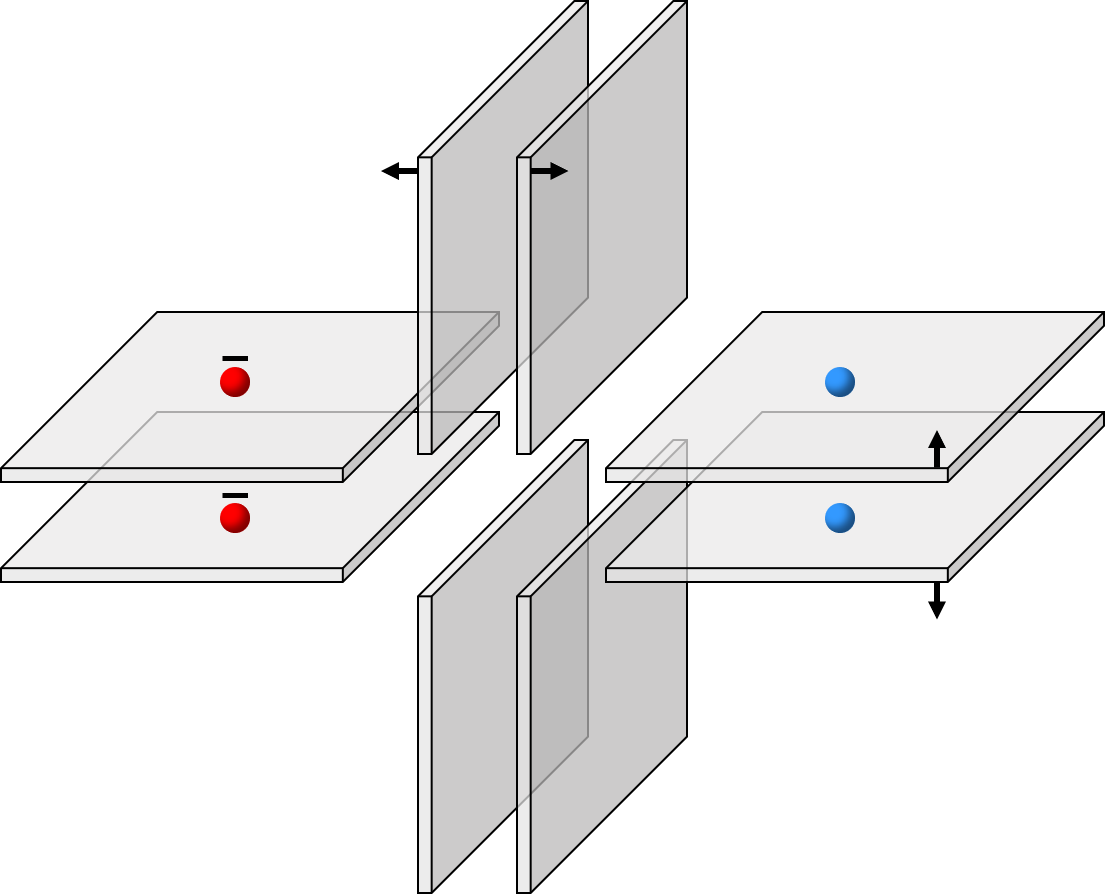}}
    \subfloat[\label{fig:Z3model2}]{\includegraphics[height=0.4\columnwidth]{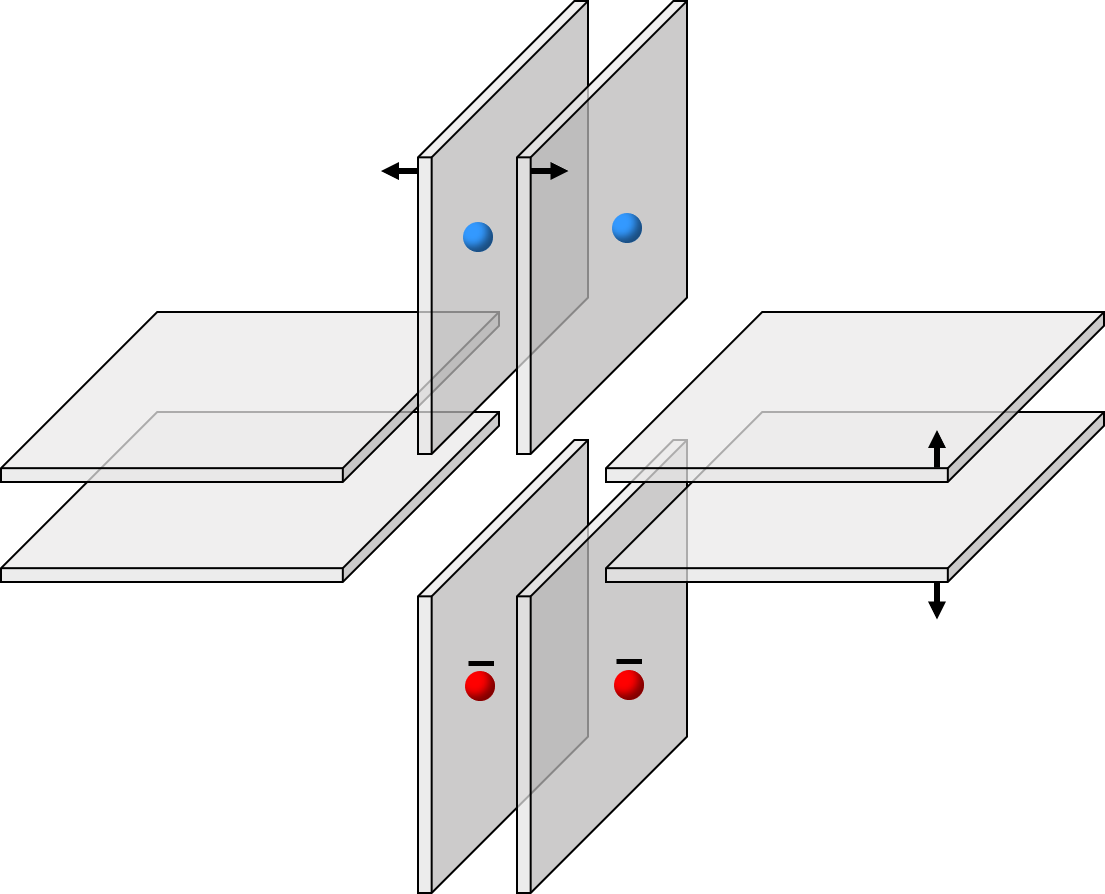}}
    \\
    \subfloat[\label{fig:Z3model13}]{\includegraphics[height=0.4\columnwidth]{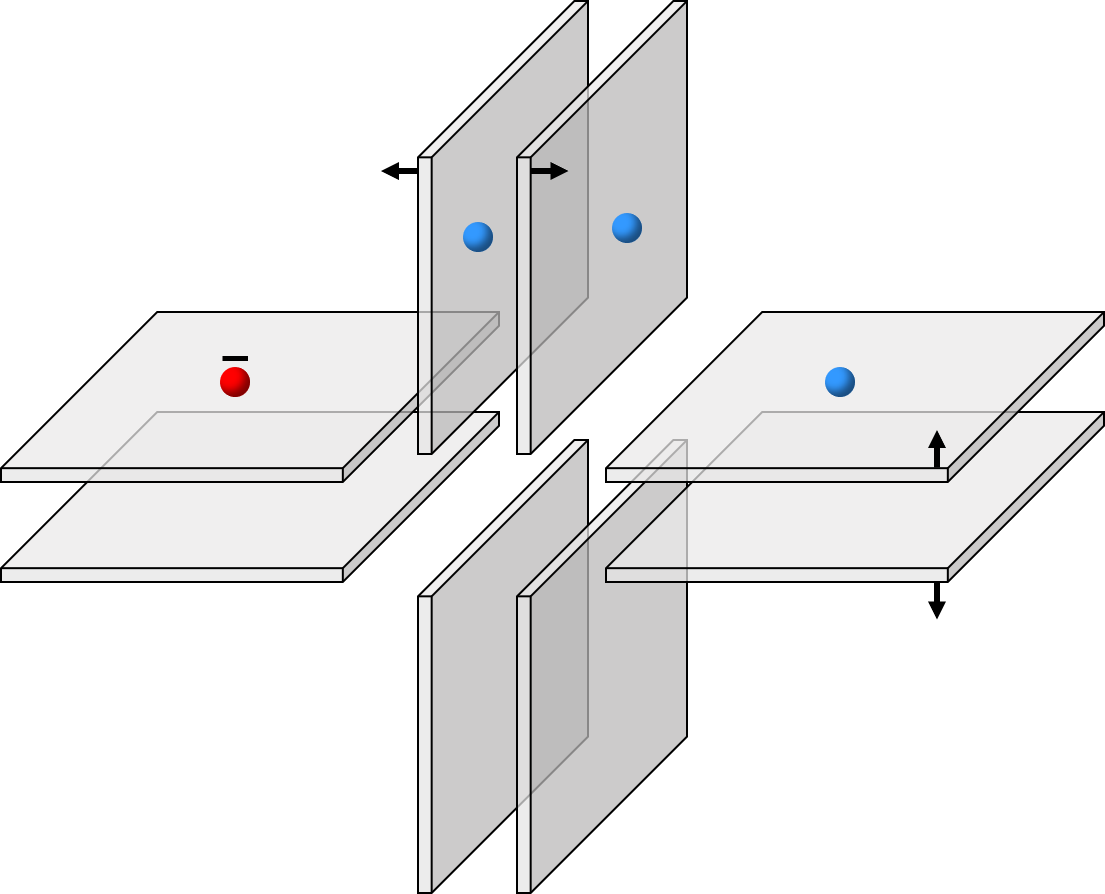}}
    \subfloat[\label{fig:Z3model14}]{\includegraphics[height=0.4\columnwidth]{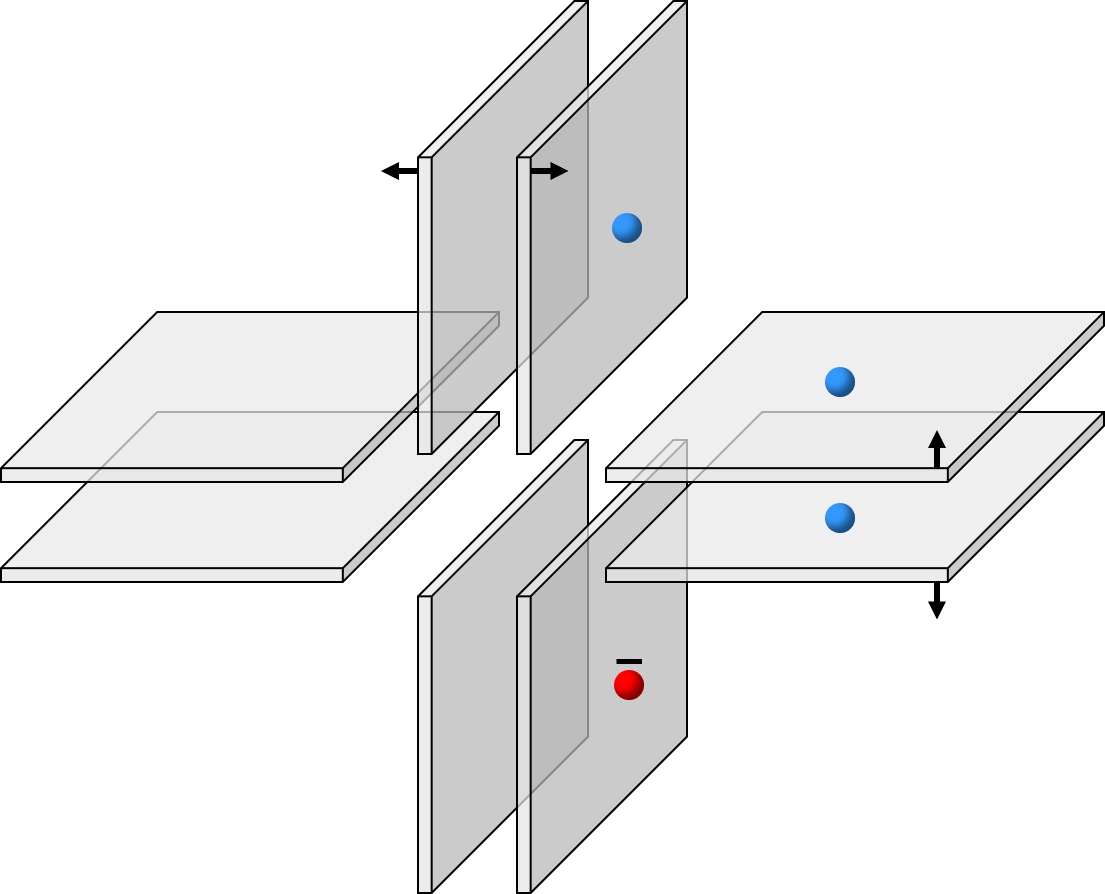}}
    \caption{Particles that condense at the 1-strata of the $\mathbb{Z}_3$ defect network. Blue spheres denote a $1$ particle and red spheres with a bar denote its antiparticle $2$. The arrows indicate the chirality of $\mathbb{Z}_3$ layers. }
    \label{fig:Z3model}
\end{figure}

To construct a type-I fracton topological defect network we chose the 3-strata to be trivial, and the 2-strata to hold $\mathbb{Z}_3$ gauge theories. 
The 1-strata are specified by a Lagrangian algebra for the four layers of $\mathbb{Z}_3$ gauge theory meeting at an edge that is generated by 
\begin{align}
    \{ 11\, 00\, 22\, 00, \, 00\, 11\, 00\, 22, \, 10\, 11\, 02\, 00, \, 11\, 01\, 00\, 20  \}. 
\end{align}
The condensate on the 1-strata is shown in Fig.~\ref{fig:Z3model} for the $\hat{x}$-axis orientation, and specified similarly for $\hat{y}$- and $\hat{z}$-axis orientations. 
The 0-strata are specified by allowing no further condensation, and having trivial statistics with all braiding processes around them.  

\begin{figure}[t]
    \centering
    \subfloat[]{\includegraphics[width=0.3\columnwidth]{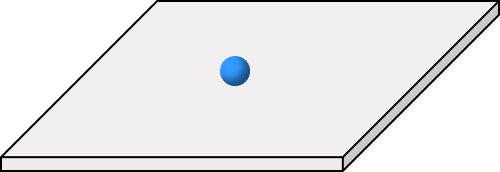}}
    \hspace{.1cm}
    \subfloat[\label{fig:Z3ex3}]{\includegraphics[width=0.3\columnwidth]{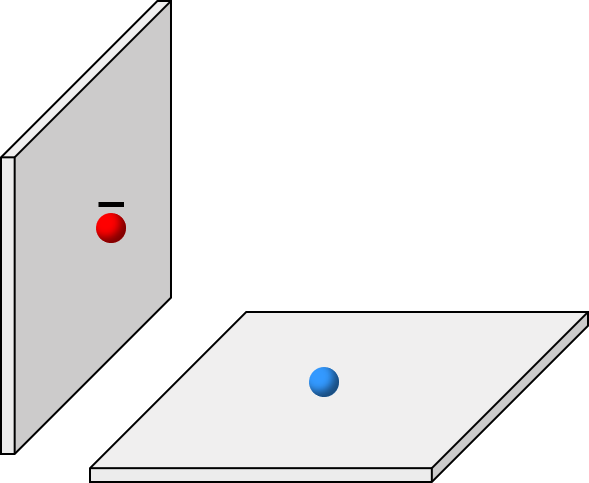}}
    \hspace{.1cm}
    \subfloat[\label{fig:Z3ex2}]{\includegraphics[width=0.35\columnwidth]{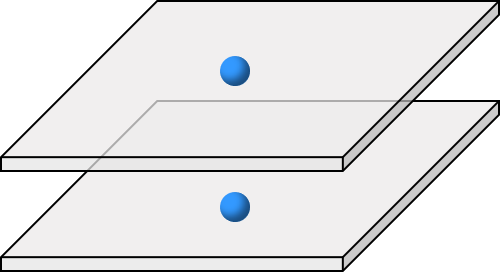}}
    \caption{Excitations in the $\mathbb{Z}_3$ defect network corresponding to a fracton~(a), lineon~(b) and planon~(c).  }
    \label{fig:Z3ex}
\end{figure}

The phase of matter produced by this defect network construction is a type-I fracton topological order. One can check that single topological charges on the 2-strata become fractons, while pairs on adjacent 2-strata may be either lineons or planons depending upon how they are configured, see Fig.~\ref{fig:Z3ex}. 

\section{X-Cube from defects lattice model}
\label{defectXCube}

In this Appendix we write out the explicit lattice model coming from the defect network discussed in Sec.~\ref{sec:Xcube}.
For simplicity we work on the cubic lattice.
We will take the 2-strata defects to be the $xy-$, $yz-$, and $zx-$planes, and the 1-strata defects to be the $x-$, $y-$ and $z-$axes.
We also take the fine grained lattice to be a cubic lattice, see \figref{fig:3DTClattice}.
The Hilbert space has one qubit for every plaquette which does not reside in the 2-strata defects: 
\begin{align}
    \mathcal{H} = \bigotimes_{p \in S_3}^{p \not \in S_2} \mathbb{C}^2_p,
\end{align}
where $\mathbb{C}_p^2$ is the two-dimensional Hilbert space associated with a qubit on plaquette $p$.
In the following we will use the notation $S_j$ for $j-$strata.

The Hamiltonian consists of fluctuation and constraint terms. 
These terms depend on whether or not their support overlaps with a defect.
The Hamiltonian is written schematically as \begin{align}
    H =H_3 + H_2 + H_1 + H_0,
\end{align}
where $H_j$ is associated with terms in the Hamiltonian near the $j-$strata.
For convenience and clarity, we will write out these terms diagrammatically.
Light orange faces are identified with Pauli-X operators on a plaquette, and light purple faces are identified with Pauli-Z operators. 
We only draw a representative set of terms;
  the other terms are found by symmetry and in the following will be denoted with $+\cdots$.

The terms in the interior of the 3-strata are given by
\begin{align}
    H_3 = &-J_{c,3}\sum_{\substack{c \in S_3 \\ \partial c \cap S_2=\emptyset}} \cubethree \nonumber \\
    &- J_{e,3} \sum_{\substack{e \in S_3 \\ e \cap( S_2\cup S_1)=\emptyset}} \edgethree+\cdots.
\end{align}
The cube term appearing in the first sum is given by a product of six Pauli-X operators on the plaquettes surrounding a cube $c$, and the sum only contains cubes which don't overlap with the 2-strata.
The edge term is given by a product of four Pauli-Z operators on the plaquettes which terminate on an edge $e$, and the sum is over edges which do not reside in the 2- or 1-strata.

The 2-strata Hamiltonian takes the form
\begin{align}
    H_2 = -J_{c,2}\sum_{\substack{f \in S_2 \\ \partial f \cap S_1=\emptyset}} \cubetwo+\cdots.
\end{align}
For each cube $c$ that has exactly one face on a 2-strata defect,
  $H_2$ includes a term given by the product of five Pauli-X operators on the five faces around the cube $c$ that are not on the 2-strata defect.
The green face is a face on the 2-strata defect
  and does not have any spin degree of freedom on it.
There are two terms for each face on a 2-strata.

The 1-strata terms are given by,
\begin{align}
    H_1 = &-J_{c,1}\sum_{\substack{e\in S_1 \\ \partial e \cap S_0=\emptyset}} \cubeone+\cdots \nonumber\\
    &- J_{v,1} \sum_{\substack{v \in S_1 \\ v \not\in S_0}} \vertexone+\cdots.
\end{align}
The first line in $H_1$ sums over each edge on a 1-strata that does not neighbor a 0-strata,
  and is given by a product of eight Pauli-X operators over the eight orange faces of two adjacent cubes. There are four such terms for each edge $e$, all related by $90\degree$ rotation symmetry about the edge $e$.
The second line is a product of four Pauli-Z operators on the faces terminating on a given vertex in the interior of a 1-strata. 

Finally, the 0-strata terms are given by
\begin{align}
    H_0 = -J_{c,0}\sum_{v \in S_0}\vertexzero+\cdots.
\end{align}
The term is given by a product of twelve Pauli-X operators on the faces (that aren't on a 2-strata) of four cubes sharing a 0-strata vertex.
There are six such terms for each vertex given by $90\degree$ rotation symmetry.


\section{Lattice model for B code from defects}
\label{app:BCodeLatticeModel}

In this appendix, we construct an explicit lattice model for the defect network in Sec.~\ref{sec:HaahB}. As in Appendix~\ref{defectXCube}, we take a cubic lattice of defects, where 2-strata defects are parallel to the $xy-$, $yz-$, and $zx-$planes and the 1-strata defects are parallel to the $x-$, $y-$ and $z-$axes.
We also take the fine grained lattice to be a cubic lattice, as in \figref{fig:3DTClattice}.
The Hilbert space has two qubits for every plaquette which does not reside on the 2-strata defects: 
\begin{align}
    \mathcal{H} = \bigotimes_{p \in S_3}^{p \not \in S_2} \mathbb{C}^2_{p,A} \otimes \mathbb{C}^2_{p,B}
\end{align}
Here the two qubits on plaquette $p$ are labeled with an $A$ and a $B$ ``layer" and $\mathbb{C}_{p,A/B}^2$ is the two-dimensional Hilbert space associated with the $A/B$ qubit on $p$.
In the following we will use the notation $S_j$ for the set of $j-$strata.

As in Sec.~\ref{defectXCube}, the Hamiltonian has the form
\begin{equation}
    H=H_3+H_2+H_1+H_0
\end{equation}
where $H_j$ is associated with terms near the $j$-strata. We will write these terms diagrammatically. A plaquette that is light orange, blue, or pink, respectively, means that the Pauli operator $X_A$, $X_B$, or $X_AX_B$ acts on the qubits on that plaquette. Similarly, a plaquette that is light purple, grey, or brown, respectively, means that the Pauli operator $Z_A$, $Z_B$, or $Z_AZ_B$ acts on the qubits on that plaquette.

The terms in the interior of the 3-strata are given by two decoupled copies of the 3+1D toric code:
\begin{widetext}
\begin{align}
    H_3 = &-J_{c,3}\sum_{\substack{c \in S_3 \\ \partial c \cap S_2=\emptyset}} \cubethree + \cubethreeblue - J_{e,3} \sum_{\substack{e \in S_3 \\ e \cap( S_2\cup S_1)=\emptyset}} \edgethree+ \edgethreegrey + \cdots.
\end{align}
\end{widetext}

A cube term appearing in the first sum is given by a product of six Pauli $X$ operators in the same layer on the plaquettes surrounding a cube $c$. There is one term per layer, and the sum only contains cubes which do not overlap with the 2-strata.

Each edge term is given by a product of four Pauli $Z$ operators on the plaquettes which terminate on an edge $e$, and the sum is over edges which do not reside in the 2- or 1-strata.

The 2-strata Hamiltonian takes the form  
\begin{align}
    H_2 = -J_{c,2}\sum_{\substack{f \in S_2 \\ \partial f \cap S_1=\emptyset}} \cubetwosmall+\cubetwobluesmall+\cdots.
\end{align}
For each cube $c$ that has exactly one face on a 2-strata defect,
  $H_2$ includes one term for each layer, given by the product of five Pauli $X$ operators in that layer on the five faces around the cube $c$ that are not on the 2-stratum defect.
The solid green face is a face on the 2-strata defect
  and does not have any spin degree of freedom on it.
There are two terms for each face on a 2-strata, one for each cube on either side.

There are six terms per edge on a 1-strata and two terms per vertex in the interior of a 1-stratum. They are given by:
\begin{widetext}
\begin{align}
H_1 = &-J_{c,1} \sum_{\substack{e\in S_1 \\ \partial e\cap S_0 = \emptyset}} \BCodecubeoneone + \BCodecubeonetwo + \BCodecubeonethree +  \nonumber \\
& + \BCodecubeonefour + \BCodecubeonefive + \BCodecubeonesix + -J_{v,1} \sum_{\substack{v\in S_1 \\ v \notin S_0}} \BCodevertexoneone + \BCodevertexonetwo + \cdots
\label{eqn:BCode1StrataHamiltonian}
\end{align}
\end{widetext}
The 1-strata edge terms consist of a product of four, eight, or twelve Pauli $X$ operators on the indicated cube faces. The 1-strata vertex terms consist of products of four Pauli $Z$ operators acting on the faces terminating on the given vertex. Only the terms for the $z$-oriented (in the convention of Fig.~\ref{fig:BCode_CubeLabels}) are given here; the $\cdots$ mean the terms for the other strata. Note that the Hamiltonian terms correspond to condensed objects in the sense that $m_1^Am_6^B$ being condensed on the $z$-oriented 1-strata corresponds to the first term in Eq.~\eqref{eqn:BCode1StrataHamiltonian}, that is, $X_A$ operators appear on faces of cube 1 and $X_B$ operators appear on cube 6. Similarly, the fact that $e_1^Ae_1^Be_2^Ae_6^B$ is  condensed on this 1-strata corresponds to the first vertex term in Eq.~\eqref{eqn:BCode1StrataHamiltonian}, in that $Z_AZ_B$, $Z_A$, and $Z_B$ operators, respectively, appear on a plaquette in positions 1, 2, and 6 relative to the 1-strata. The terms for the other orientations of the 1-strata can be obtained using the same schematic correspondence.

Finally, the 0-strata terms are given by 
\begin{widetext}
\begin{align}
H_0 = &-J_{c,0} \sum_{v \in S_0} \BCodevertexzeroone + \BCodevertexzerotwo + \BCodevertexzerothree + \BCodevertexzerofour + \nonumber \\ 
&+ \BCodevertexzerofive + \BCodevertexzerosix + \BCodevertexzeroseven + \BCodevertexzeroeight
\label{eqn:BCode0StrataHamiltonian}
\end{align}
\end{widetext}
The first six terms are products of twelve Pauli $X$ operators, and the last two are products of three Pauli $X$ operators. The 1-strata are shown in dark purple, and plaquettes on the 2-strata are shown in green in the first six terms. The black dots in the last three terms indicate which plaquettes are acted on, and the plaquettes which are part of the 2-strata are not colored because they would cover the plaquettes which are acted on.


\section{Lattice Hamiltonian for non-Abelian defect model}
\label{app:d4lattice}

In this Appendix, we provide an explicit realization of the non-Abelian defect network discussed in the main text (see Sec.~\ref{sec:nonabelian}). We start with a brief review of exactly solvable models describing discrete gauge theories in three dimensions, which are straightforward generalizations of Kitaev's quantum double models in two dimensions~\cite{kitaev} and correspond to the Hamiltonian formulation of untwisted Dijkgraaf-Witten TQFTs~\cite{dijkgraaf,wan2015twisted,williamson2017ham}.

\subsection{Brief review of 3+1D Lattice Gauge Theory}
\label{app:d4rev}

\begin{figure}[t]
    \centering
    \includegraphics[width=0.2\textwidth]{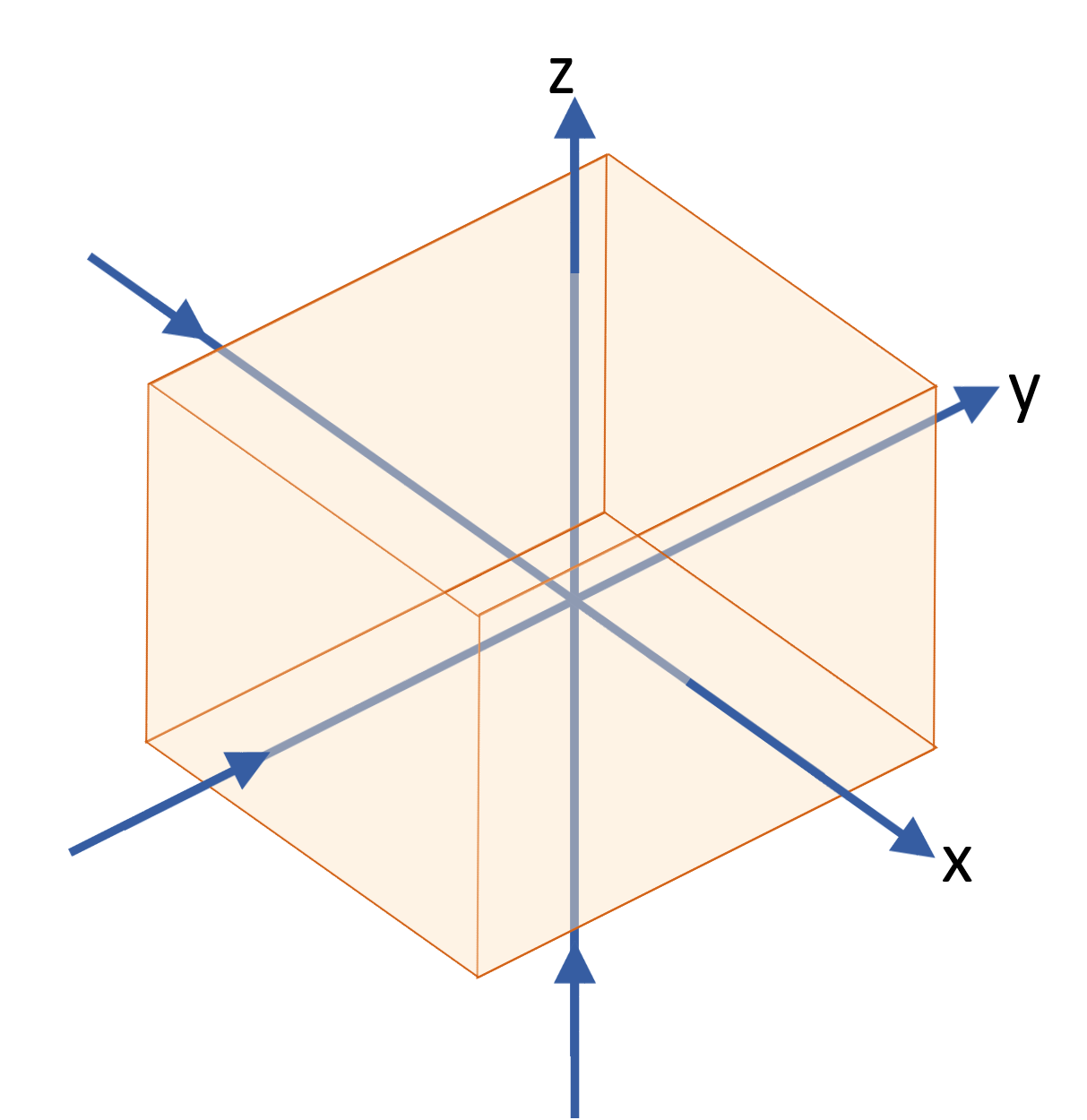}
    \caption{We choose a right-handed frame such that all links of the dual lattice (in blue) are oriented with respect to these directions. This defines a natural orientation for the plaquettes $p$ (in orange) of the original lattice.}
    \label{fig:orientate}
\end{figure}

Consider a cubic lattice with a local Hilbert space $\mathcal{H} = \mathbb{C}[G]$ on each plaquette $p$ of the lattice, with $G$ a finite group. A natural orientation for the plaquettes is defined by assigning links of the dual cubic lattice an orientation as shown in Fig.~\ref{fig:orientate}, such that plaquettes forming a cube $c$ carry a positive (negative) orientation if the corresponding dual link is entering (exiting) the cube. Each plaquette is further endowed with a natural orthonormal basis $\{ \ket{g}_p : g \in G \}$, referred to as the group or ``flux" orthonormal basis.

For each plaquette, we define the following operators which act on the local Hilbert space as follows:
\begin{align}
L_+^g = \sum_k \ket{gk}\bra{k} \, , \quad & T_+^g = \ket{g}\bra{g} \, ,  \nonumber \\
L_-^g = \sum_k \ket{k g^{-1}}\bra{k} \, , \quad & T_-^g  =\ket{g^{-1}} \bra{g^{-1}} \, ,
\end{align}
where $L_+, L_-$ are shift operators (left and right multiplication respectively) and $T_+, T_-$ are group projectors. For each cube $c$ of the lattice, we can further define the operator
\beq
A^h(c) = \prod_{p \in \partial c^+} L_+^h \prod_{p \in \partial c^-} L_-^h \,
\eeq
where $\partial c^{+ (-)}$ denotes faces of the cube $c$ with positive (negative) orientations. The action of $A^h(c)$ on a particular configuration is given by
\begin{align}
\label{eq:lgtcube}
A^h(c) \, \Bigg | \bmm \includegraphics[height=1in]{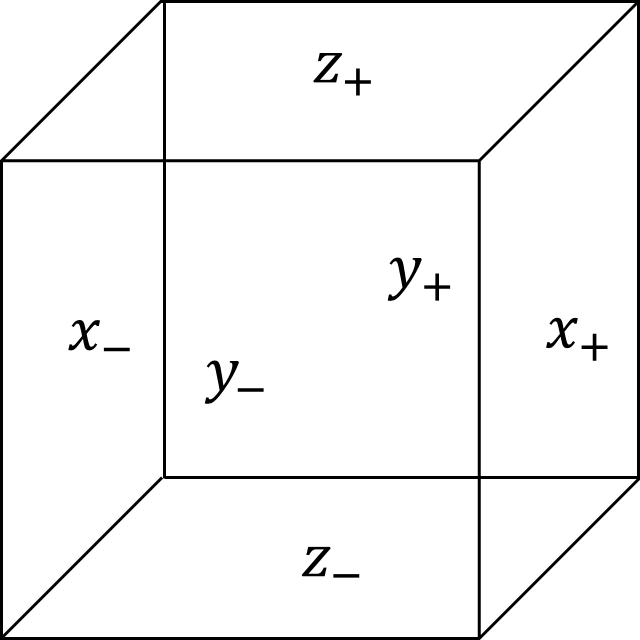} \emm \, \Bigg \rangle  = & \, \Bigg | \bmm \includegraphics[height=1in]{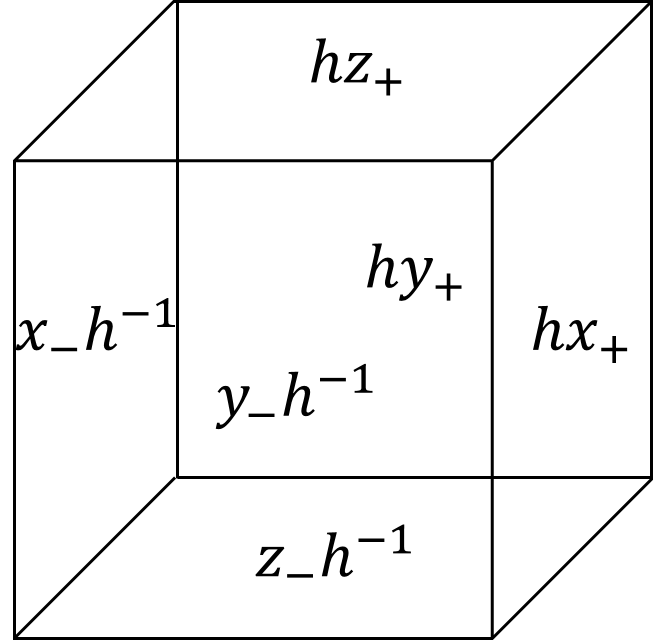} \emm \, \Bigg \rangle \, .
\end{align}
On each link $\ell$, we define an operator $B^h(\ell)$, whose action on e.g., a $y$-oriented link is given by
\begin{align}
\label{eq:lgtedge}
B^h (\ell) \, \Bigg | \bmm \includegraphics[height=1in]{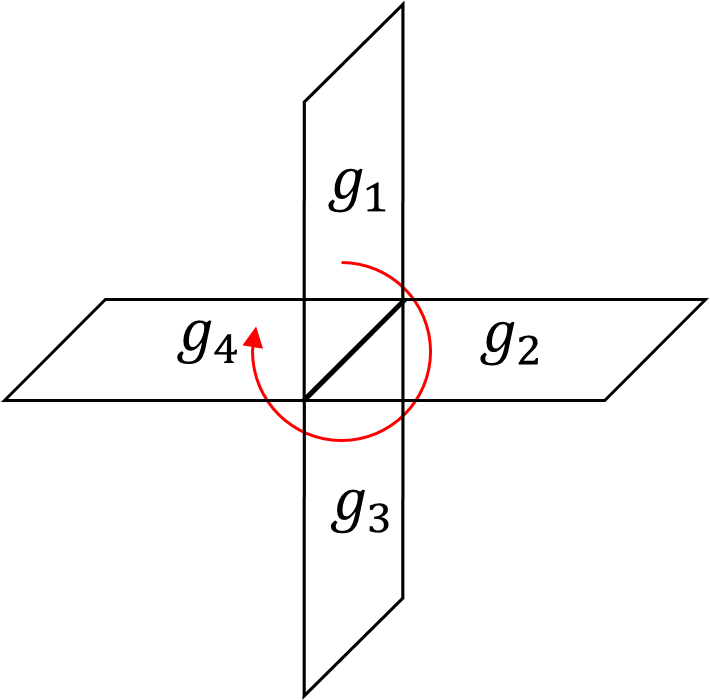} \emm \, \Bigg \rangle = & \delta\left(g_1 g_2^{-1} g_3^{-1} g_4,h \right) \nonumber \\
& \times \Bigg | \bmm \includegraphics[height=1in]{Bp.png} \emm \, \Bigg \rangle \, ,
\end{align}
and similarly for $x$- and $z$-oriented links. Unless $h = e$, the order of multiplication (indicated by the red arrow) matters for non-Abelian groups. 

We can then define the following operators:
\beq
A^\mu (c) = \frac{|d_\mu|}{|G|} \sum_{g \in G} \chi_\mu^g A^g (c), \quad B^{[C]}(\ell) = \sum_{h \in \{C\} } B^h (\ell) \, ,
\eeq
where $\chi_\mu^g$ is the character of $g$ in irrep $\mu$ and where $\{C\}$ denotes a conjugacy class of $G$. The Hamiltonian for the 3+1D generalization of Kitaev's quantum double model is then given by
\beq
\label{eq:H3D}
H_{3D} = -\sum_c A(c) - \sum_l B(\ell) \, ,
\eeq
where $A(c) = A^{1}(c)$ and $B(\ell) \equiv B^e(\ell)$, with $e$ the identity element of $G$ and $1$ the trivial irrep. The operator $A(c)$ is a projector onto gauge-invariant states, while the operator $B(\ell)$ projects onto zero flux configurations.

It is straightforward to check that all terms in the Hamiltonian commute with each other \textit{i.e.,}
\begin{align}
    [A(c),A(c')] &= 0, \quad \forall \, c,c',\nonumber \\
    [B(\ell), B(\ell')] &= 0, \quad \forall \, \ell,\ell', \nonumber \\
    [A(c), B(\ell)] &= 0, \quad \forall \, c,\ell \, .
\end{align}
Hence, the ground states $\ket{\Psi_{GS}}$ of this gapped Hamiltonian satisfy:
\beq
A(c) \ket{\Psi_{GS}} = \ket{\Psi_{GS}} \, \forall c, \quad B(\ell) \ket{\Psi_{GS}} = \ket{\Psi_{GS}} \, \forall \ell.
\eeq
On the three-torus, the Hamiltonian in Eq.~\eqref{eq:H3D} exhibits a finite non-trivial ground state degeneracy~\cite{wan2015twisted}, indicative of its intrinsic topological order. Excited states correspond to violations of the local gauge constraint $A(c)$, of the no-flux condition enforced by $B(\ell)$, or of both, and are denoted charges, fluxes, and dyons respectively. In $d=3$ spatial dimensions, gauge charges are point-like and are described by Rep($G$) while the gauge-fluxes are flux loops and are locally labelled by conjugacy classes $|C|$ of $G$. Details regarding the braiding and fusion of excitations in these theories can be found in Refs.[~\onlinecite{wan2015twisted,williamson2017ham,wang2017twisted}].

Finally, we note that the gauge theory models Eq.~\eqref{eq:H3D} describing topological ordered phases in 3+1D have been extended to cases where the system has spatial boundaries~\cite{wang2018gapped}. In particular, when the system has a two-dimensional boundary, a subset of gapped boundary Hamiltonians compatible with the bulk topological order have been classified: for an untwisted $G$ gauge theory in the bulk, its two-dimensional gapped boundaries are labelled by the pair $(K,\omega)$, where $K \leq G$ is a subgroup of $G$ and $\omega \in H^3(K,U(1))$ is a 3-cocycle. 

\subsection{\texorpdfstring{$D_4$}{D4} defect Hamiltonian}
\label{sec:d4ham}

\begin{figure}[t]
\includegraphics[width=0.3\textwidth]{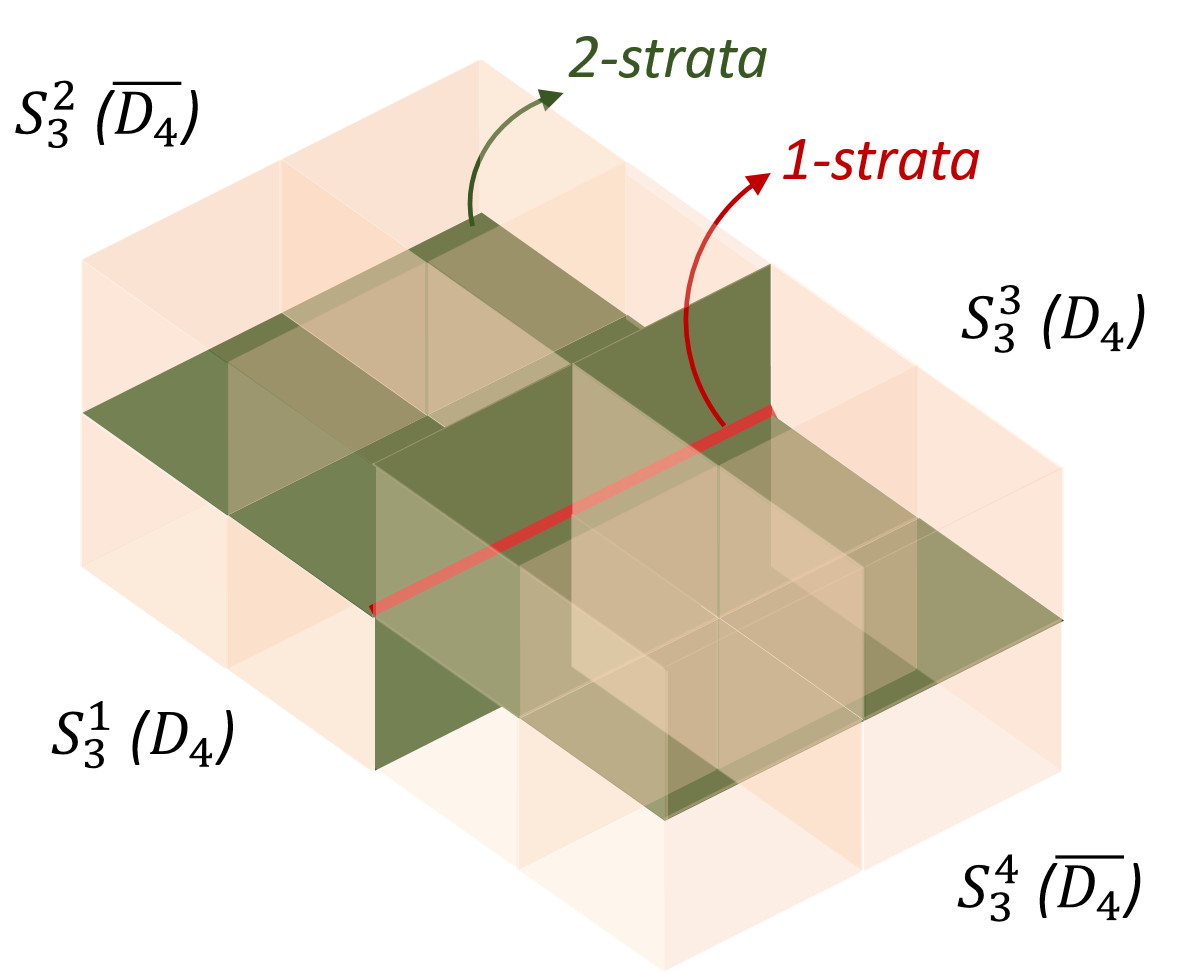}
\caption{Pairs of $D_4$ gauge theory meet at 2-strata defects (green plaquettes) with four of them meeting at a 1-strata defect (red links). Plaquettes belonging to the two $\bar{D_4}$ 3-strata have orientations opposite to those belonging to the $D_4$ 3-strata.}
\label{fig:fourd4s}
\end{figure}

With the notation set, we are now ready to construct an explicit Hamiltonian which realizes the defect network discussed in Sec.~\ref{sec:nonabelian}. We work on the cubic lattice for simplicity, with the $xy-$, $yz-$, and $xz-$planes specifying the 2-strata defects while the $x-$, $y-$, and $z-$axes specify the 1-strata defects (see Fig.~\ref{fig:3DTClattice}). As in prior sections, we will use the notation $S_j$ to refer to $j$-strata.  For each plaquette not residing on any of the 2-strata defects, the local Hilbert space is defined as $\mathbb{C}[G]$, where we now set to $G = D_4$. The total Hilbert space of the system is hence 
\beq
\mathcal{H} = \bigotimes_{p\in S_3}^{p \notin S_2} \mathbb{C}[G] \, .
\eeq
The Hamiltonian consists of terms which locally constrain the membrane configurations as well as terms which give dynamics to the set of allowed configurations. In analogy with the X-Cube defect Hamiltonian (see Appendix~\ref{defectXCube}), we write the Hamiltonian as a sum of terms 
\beq
H = \sum_{j=0}^3 H_j \, ,
\eeq
where $H_j$ acts on degrees of freedom in the vicinity of the $j$-strata.

\begin{figure}[t]
    \centering
    \includegraphics[width=0.4\textwidth]{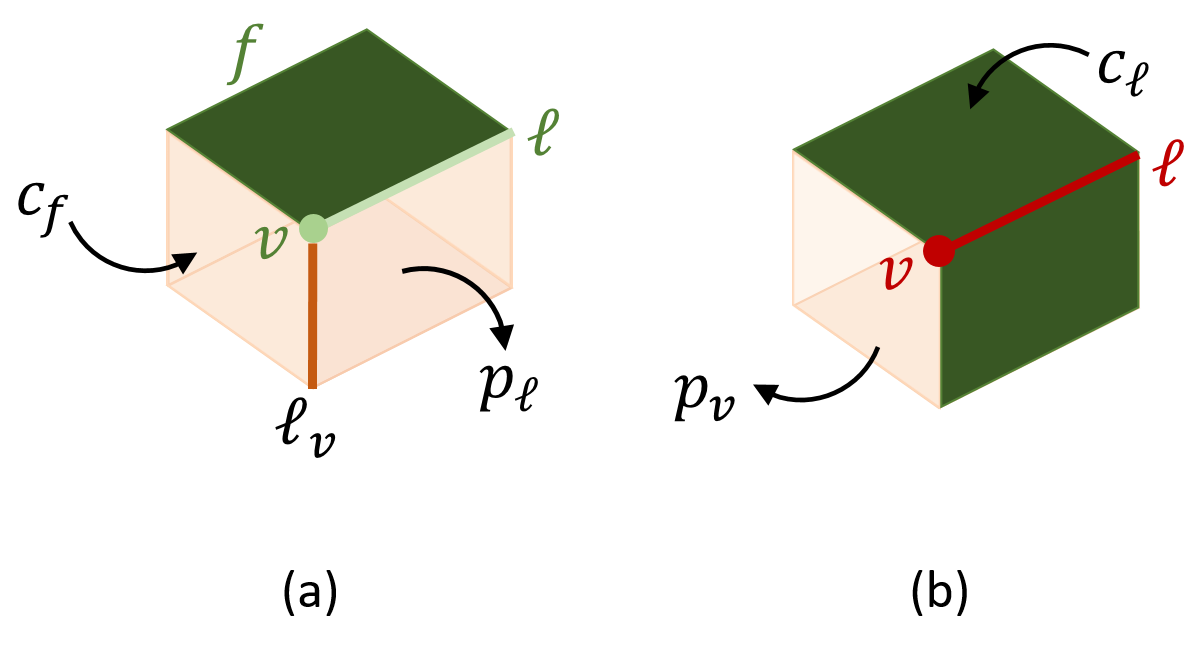}
    \caption{(a) Cubes, plaquettes, and links sharing a face ($f$), link ($\ell$), or vertex ($v$) with a 2-strata are denoted by $c_f,p_\ell$, and $\ell_v$ respectively. (b) Cubes and plaquettes sharing a link ($\ell$) or vertex ($v$) with a 1-strata are labelled $c_\ell$ and $p_v$ respectively. The 3-strata index $\alpha$ has been suppressed for clarity.}
    \label{fig:strataops}
\end{figure}

We index the four 3+1D $D_4$ 3-strata meeting at a 1-strata defect by $\alpha = 1,2,3,4$ (mod 4) as shown in Fig.~\ref{fig:fourd4s}. Further, to distinguish between operators with support near distinct strata, we introduce the following notation (see Fig.~\ref{fig:strataops}): $c^\alpha,p^\alpha,\ell^\alpha,v^\alpha$ refer to cubes, plaquettes, links, and vertices which lie entirely within the 3-strata $S_3^\alpha$. $c_f^\alpha$, $p_\ell^\alpha$, and $\ell_v^\alpha$ refer to cubes, plaquettes, and links in $S_3^\alpha$ which share only a single face ($f$), link ($\ell$), or vertex ($v$) with a 2-strata $S_2^{\alpha,\alpha+1}$ defect between two adjacent 3-strata. Finally, $c_\ell^\alpha$ and $p_v^\alpha$ refer to cubes and plaquettes in $S_3^\alpha$ which share only a single link ($\ell$) or vertex ($v$) with the 1-strata defect $S_1 \equiv S_1^{1,2,3,4}$ between four 3-strata.

The 3-strata Hamiltonian is given by
\begin{align}
H_3 & = \sum_{\alpha = 1}^4 H_{3D}^\alpha, \nonumber \\
H_{3D}^\alpha & = -\sum_{c^\alpha \in S_3^\alpha} \sum_{c^\alpha} A(c^\alpha) - \sum_{\ell^\alpha \in S_3^\alpha} \sum_{\ell^\alpha} B(\ell^\alpha) \, ,
\end{align}
with the terms $A(c)$ and $B(\ell)$ defined earlier. This term simply states that the bulk of each 3-strata is specified by $D_4$ gauge theory.

\begin{figure}[t]
    \centering
    \includegraphics[width=0.5\textwidth]{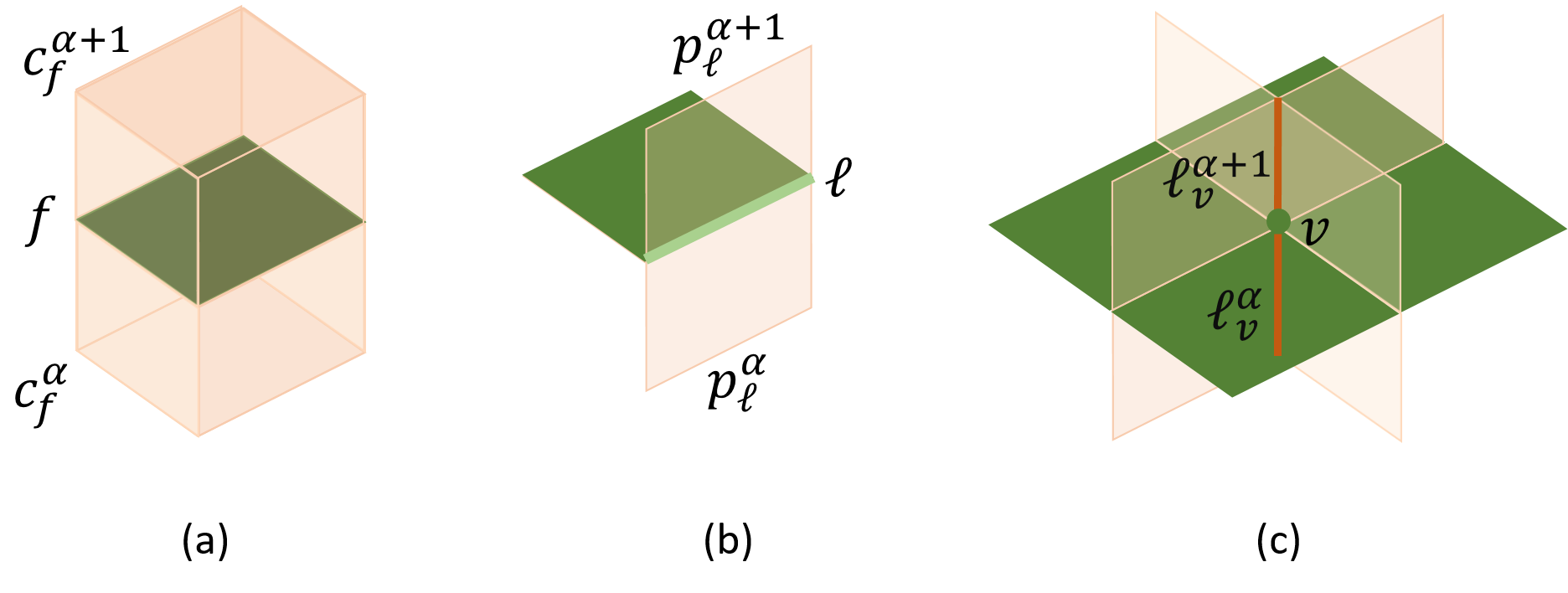}
    \caption{Terms in $H_2$ Eq.~\eqref{eq:2strat} act as follows: (a) $A_2^K$ acts on two cubes sharing a 2-strata face $f$, (b) $T_2^K$ and $L_2^K$ act on two plaquettes sharing a 2-strata link $l$, and (c) $B_2^K$ acts on two links sharing a 2-strata vertex $v$.}
    \label{fig:schematic1}
\end{figure}

Now, each 2-strata $S_2^{\alpha,\alpha+1}$ defines an interface between two adjacent 3-strata $S_3^\alpha$ and $S_3^{\alpha+1}$. This gapped interface can be mapped onto the two-dimensional gapped boundary to vacuum of a 3+1D $D_4 \times D_4$ gauge theory---as mentioned earlier, such boundaries are labelled by the pair $(K \leq G, \omega \in H^3(K,U(1)))$. We pick
\beq
K = \langle g_{\alpha} g_{\alpha+1},s_\alpha^2| g\in D_4 \rangle \cong D_4 \times \mathbb{Z}_2  \, ,
\eeq
along with the trivial 3-cocycle for each 2-strata $S_2^{\alpha,\alpha+1}$. This completely specifies the 2-strata Hamiltonian~\cite{cong2016gapped,wang2018gapped}, which is given by 
\begin{align}
\label{eq:2strat}
    H_{2}  = -\sum_{\alpha = 1}^4 & \left[\frac{1}{|K|} \sum_{f} A_2^K(c_f^\alpha,c_f^{\alpha+1})  + \sum_{v} B_2^K (\ell_v^\alpha , \ell_v^{\alpha+1}) \right. \nonumber \\
    + & \left. \sum_{\ell} T_2^K(p_\ell^\alpha,p_\ell^\alpha+1) + \frac{1}{|K|} \sum_{\ell} L_2^K(p_\ell^\alpha,p_\ell^\alpha+1) \right] \, ,
\end{align}
where $f,\ell,v \in  S_2^{\alpha,\alpha+1}$ and
\begin{align}
    A_2^K(c_f^\alpha,c_f^{\alpha+1}) &= \sum_{g \in {D_4}} \left(A^{g_\alpha}(c_f^{\alpha}) + A^{g_\alpha s_\alpha^2}(c_f^\alpha) \right) A^{g_{\alpha+1}}(c_f^{\alpha+1}) , \nonumber \\
    B_2^K(\ell_v^\alpha,\ell_v^{\alpha+1}) & = \sum_{g \in D_4} \left(B^{g_\alpha}(\ell_v^\alpha) + B^{g_\alpha s^2_\alpha}(\ell_v^\alpha) \right) B^{g_{\alpha+1}}(\ell_v^{\alpha+1}), \nonumber \\
    L_2^K(p_\ell^\alpha,p_\ell^{\alpha+1}) & = \sum_{g \in D_4} \left(L_+^{g_\alpha}(p_\ell^\alpha) + L_+^{g_\alpha s^2_\alpha}(p_\ell^\alpha) \right) L_+^{g_{\alpha+1}}(p_\ell^{\alpha+1}) , \nonumber \\
    T_2^K(p_\ell^\alpha,p_\ell^{\alpha+1}) & = \sum_{g \in D_4} \left(T_+^{g_\alpha}(p_\ell^\alpha) + T_+^{g_\alpha s^2_\alpha}(p_\ell^\alpha) \right) T_+^{g_{\alpha+1}}(p_\ell^{\alpha+1}) .
\end{align}

\noindent Here, $g_\alpha$ refers to local degrees of freedom in $S_3^\alpha$. The configurations on which these operators act are schematically shown in Fig.~\ref{fig:schematic1}. The terms $A_2^K$ and $L_2^K$ project onto a trivial sector of the representation of $K$ while the terms $B_2^K$ and $T_2^K$ restrict the flux on the 2-strata to elements of $K$. Following the procedure delineated in Refs.[\onlinecite{bombin2008,cong2016gapped,wang2018gapped}], one can show that the bulk excitations condensing to vacuum on the 2-strata are indeed given by Eq.~\eqref{eq:cond2strat}.

\begin{figure}[b]
    \centering
    \includegraphics[width=0.5\textwidth]{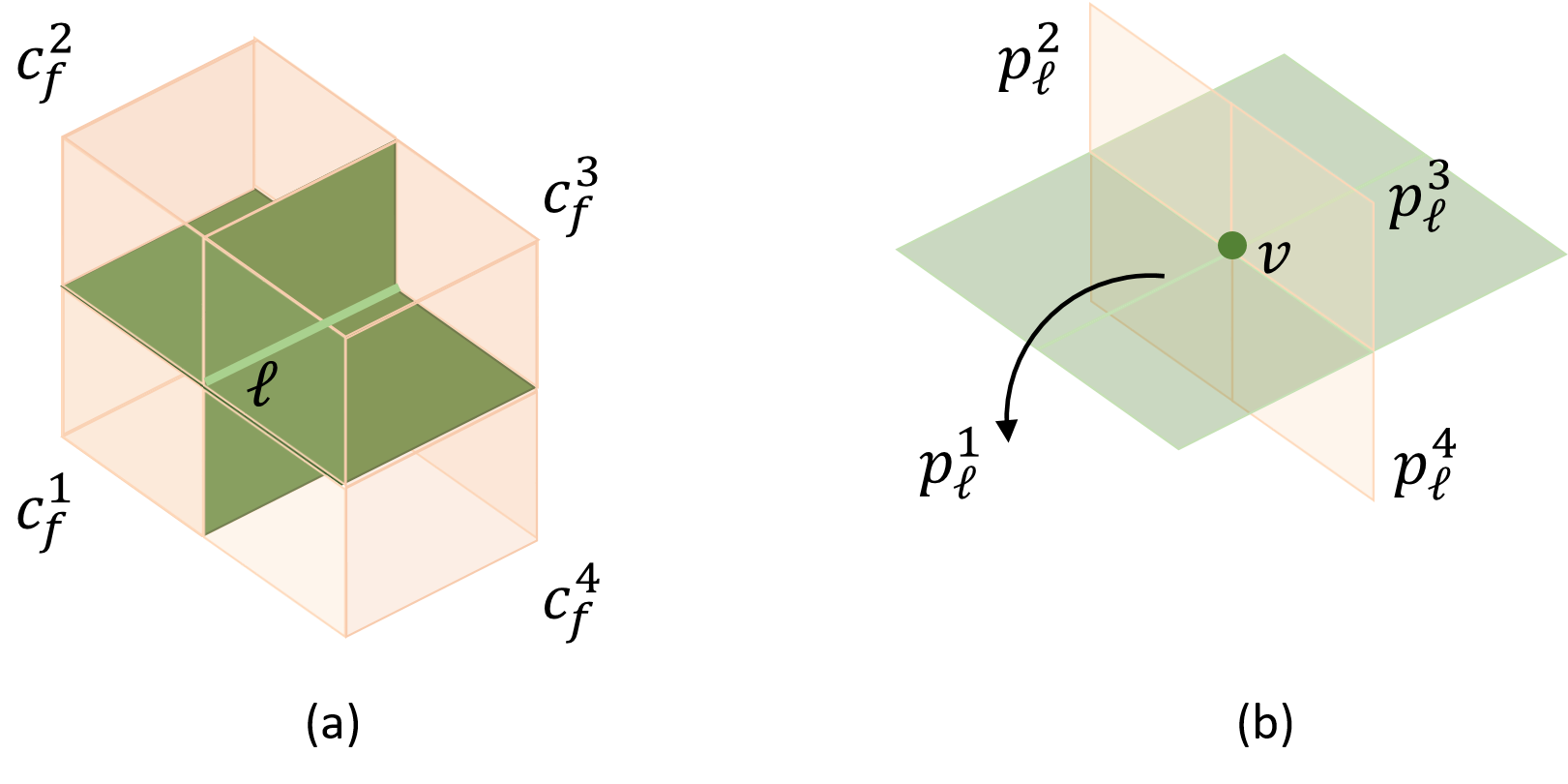}
    \caption{Terms in $H_1$ Eq.~\eqref{eq:1strat} act as follows: (a) $A_1^K$ acts on four cubes sharing a 1-strata link $l$, and (b) $T_1^K$ and $L_1^K$ act on four plaquettes sharing a 1-strata vertex $v$.}
    \label{fig:schematic2}
\end{figure}

With the 2-strata defect specified, we pick the following subgroup to specify the 1-strata defect:
\beq
M = \langle g_1 g_2 g_3 g_4, s_1^2 s_2^2, s_2^2 s_3^2 | \forall g \in D_4 \rangle \cong D_4 \times \mathbb{Z}_2^2 \, ,
\eeq
which is generated by the diagonal subgroup of $(D_4)^{\times 4}$ and pairs $s_\alpha^2 s_\beta^2$ for $\alpha \neq \beta$. This choice of 1-strata defect enforces the constraint that quadruples of $\sigma$ charges are condensed on the 1-strata, thereby causing isolated $\sigma$ charges (which are non-Abelian) to behave as fractons. Explicitly, the 1-strata Hamiltonian is given by:
\begin{align}
\label{eq:1strat}
    H_1 = & - \frac{1}{|M|}\sum_{\ell} A^M_1(c_\ell^1,c_\ell^2,c_\ell^3,c_\ell^4) - \sum_{v} T_1^M(p_v^1,p_v^2,p_v^3,p_v^4) \nonumber \\
     & - \frac{1}{|M|}\sum_{v} L_1^M(p_v^1,p_v^2,p_v^3,p_v^4) \, ,
\end{align}
where $\ell,v \in S_1$. Here,
\begin{widetext}
\begin{align}
    A^M_1(c_\ell^1,c_\ell^2,c_\ell^3,c_\ell^4) & =\sum_{g\in D_4} \left( A^{g_1}(c_l^1) A^{g_2}(c_l^2) A^{g_3}(c_l^3) + A^{g_1 s_1^2}(c_l^1) A^{g_2 s_2^2}(c_l^2) A^{g_3}(c_l^3) + A^{g_1}(c_l^1) A^{g_2 s_2^2}(c_l^2) A^{g_3 s_3^2}(c_l^3) \right) A^{g_4}(c_l^4) \, , \nonumber \\
    L_1^M(p_v^1,p_v^2,p_v^3,p_v^4) & = \sum_{g\in D_4} \left( L_+^{g_1}(p_v^1) L_+^{g_2}(p_v^2) L_+^{g_3}(p_v^3) + L_+^{g_1 s_1^2}(p_v^1) L_+^{g_2 s_2^2}(p_v^2) L_+^{g_3}(p_v^3) + L_+^{g_1}(p_v^1) L_+^{g_2 s_2^2}(p_v^2) L_+^{g_3 s_3^2}(p_v^3) \right) L_+^{g_4}(p_v^4) \, ,\nonumber \\
    T_1^M(p_v^1,p_v^2,p_v^3,p_v^4) & = \sum_{g\in D_4} \left( T_+^{g_1}(p_v^1) T_+^{g_2}(p_v^2) T_+^{g_3}(p_v^3) + T_+^{g_1 s_1^2}(p_v^1) T_+^{g_2 s_2^2}(p_v^2) T_+^{g_3}(p_v^3) + T_+^{g_1}(p_v^1) T_+^{g_2 s_2^2}(p_v^2) T_+^{g_3 s_3^2}(p_v^3) \right) T_+^{g_4}(p_v^4)\, ,
\end{align}
\end{widetext}
where the configurations on which these operators act are schematically sketched in Fig.~\ref{fig:schematic2}.

\begin{figure}[t]
    \centering
    \includegraphics[width=0.5\textwidth]{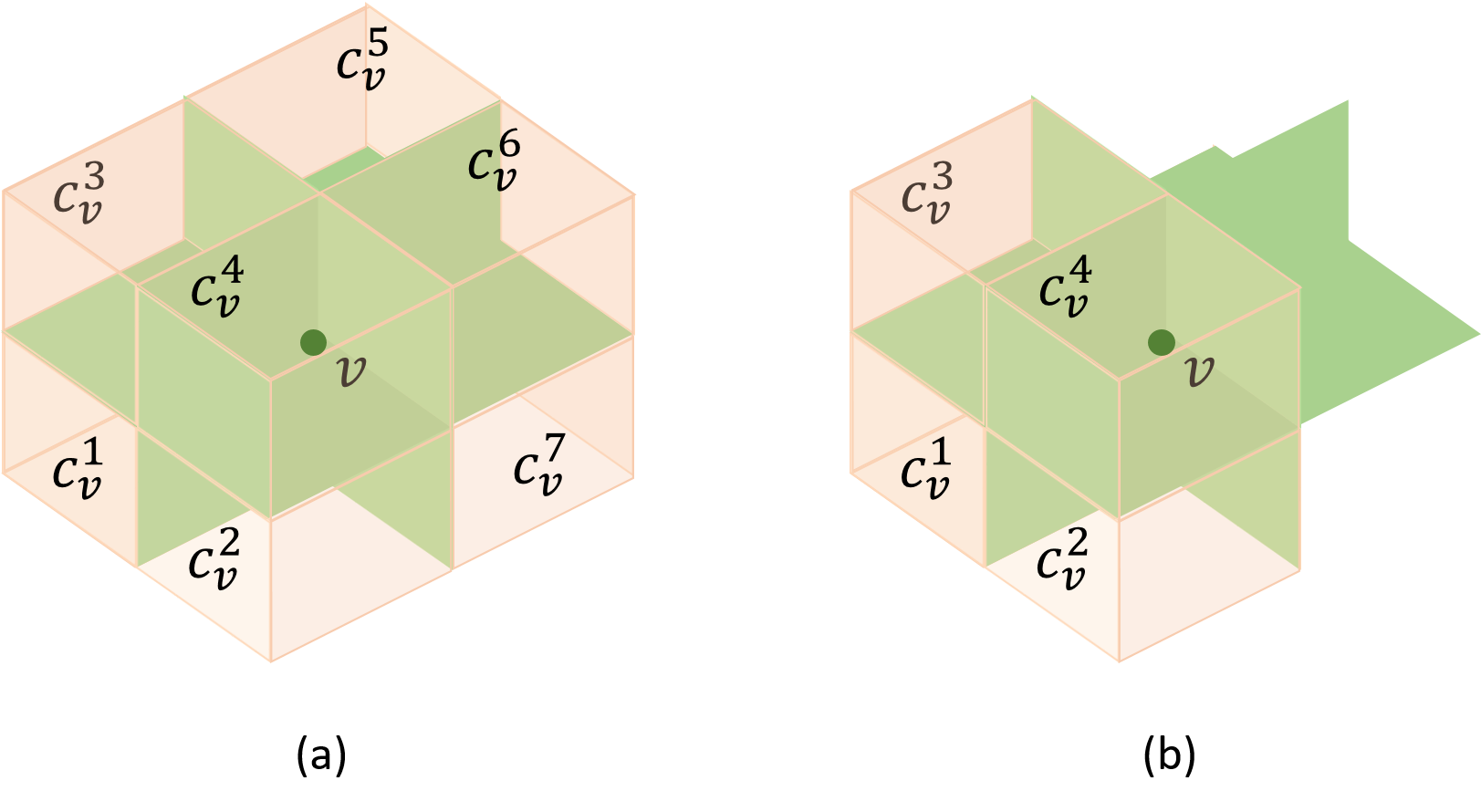}
    \caption{Eight $D_4$ 3-strata meeting at a 0-strata defect. The 3-strata are indexed as shown (with the hidden cube labelled $c_v^8$). The first term in $H_0$ Eq.~\eqref{eq:0strat} acts on all eight cubes meeting at $v \in S_0$ while the remaining terms act on four adjacent cubes as shown in (b). There are six such terms generated by $H_0$ $\forall v \in S_0$, given by 90$\degree$ rotation symmetry. }
    \label{fig:schematic3}
\end{figure}

 Finally, we have eight different $D_4$ 3-strata meeting at a 0-strata $S_0$. The Hamiltonian $H_0$ excludes configurations with membranes ending on 0-strata and is specified by
\begin{align}
\label{eq:0strat}
H_0 = & - \sum_{\substack{v \in S_0 \\ g=e,s^2}} \prod_{\alpha=1}^8 A^g_\alpha(c_v^\alpha) -\sum_{\substack{v \in S_0 \\ g=e,s^2}} \left( \prod_{\alpha=1}^4 + \prod_{\alpha=3}^6 \right) A^{g_\alpha}(c_v^\alpha)\nonumber \\
& -\sum_{\substack{v \in S_0 \\ g=e,s^2}} \left( A^{g_2}(c_v^2)A^{g_4}(c_v^4) A^{g_6}(c_v^6) A^{g_7}(c_v^7) \right) \, ,
\end{align}
where $c_v^\alpha$ corresponds to a cube from $S_3^\alpha$ which shares only a single vertex $v$ with $S_0$ and where the eight cubes meeting at $S_0$ are indexed as shown in Fig.~\ref{fig:schematic3}. The first term in $H_0$ corresponds to the product over all eight cubes meeting at $v \in S_0$ and allows closed membrane configurations with any group labels around $S_0$; the other three terms correspond to the product over four cubes meeting at $v \in S_0$ as depicted in Fig.~\ref{fig:schematic3}. Note that there are a total of six such four-cube terms for every $v \in S_0$ generated by $H_0$, permitting configurations with hemispheres labelled by $s^2$ over four 3-strata adjacent to a single 1-strata.

It is straightforward to check that the defect $H$ is a sum of mutually commuting terms and represents an exactly soluble realization of the non-Abelian defect network discussed in Sec.~\ref{sec:nonabelian}. 

\subsection{The 1-strata defect in the net-basis}

This section is aimed squarely at cognoscenti familiar with Levin-Wen string-nets~\cite{Levin2005} and even the brave reader who has made it this far into the Appendices may wish to skip it. Specifically, we show explicitly why four $\sigma$ charges get condensed on the 1-strata by going to the representation basis, obtained by performing a Fourier transform:
\beq
\label{eq:fourier}
\ket{\mu; i j} = \sum_{g \in G} \sqrt{\frac{|d_\mu|}{|G|}}[\Gamma^\mu(g)]_{ij} \, \ket{g} \, .
\eeq
Here, basis states are labelled by the matrix elements of unitary irreducible representations (irreps) $\mu$ of $G$ and $[\Gamma^\mu(g)]_{ij}$ is given by the $(i,j)^{th}$ matrix element of $g$ in irrep $\mu$, whose dimension is $|d_\mu|$.

In this basis, the local Hilbert space on each plaquette is spanned by $\ket{\mu; i,j}$. The following operator 
\beq
\mP^k(p) = \sum_{J \in C_G} \frac{X_0^0 X_k^0}{X_0^J X_k^J} W^J(p) \, ,
\eeq
forces the irrep label on a plaquette $p$ to equal $k$, where $X_j^J$ are the fusion coefficients of the theory and $W^J(p)$ creates a flux loop corresponding to conjugacy class $\{J\}$ around plaquette $p$. For group $G$, the fusion coefficients are defined as
\beq
X_j^J = \sqrt{\frac{|J|}{|G|}} \chi_j(g^J) \, .
\eeq
where $\chi_j$ are the character functions with respect to conjugacy classes, $g^J$ is a representative from $J$, $|J|$ is the cardinality of the conjugacy class, and $|G|$ is the order of the group. The string-operator $W^J(p)$ is defined by its action on a plaquette $p$ ($j_p$ is the irrep label on $p$)
\beq
W^J (p) \ket{ \, j_p} = \frac{X_0^J X_{j_p}^J}{X_0^0 X_{j_p}^0} \ket{\, j_p} \, ,
\eeq
which only adds a phase and does not change the string-labels on the plaquette $p$, thereby corresponding to a string operator for pure flux excitations. 

Now, let us consider a 1-strata defect between four $D_4$ gauge theories. In the dual representation, we wish to enforce the constraint that only four $\sigma$ strands are allowed to end on 1-strata defects. This can be enforced by adding the following term to the 1-strata Hamiltonian:
\beq
\tilde{H}_1 = - \sum_{v \in S_1} \sum_{\alpha \neq \beta} \mP^{\sigma_\alpha} (p_v^\alpha)  \mP^{\sigma_\beta} (p_v^\beta) \, ,
\eeq
where $\alpha, \beta = 1,2,3,4$ index the 3-strata $S_3^\alpha$ and this term acts on all pairs of plaquettes sharing a single vertex $v$ with the 1-strata defect. Fourier transforming back to the group element basis, we first note that $W^J(p) \mapsto \sum_{g \in \{ J\}} L_+^{g_j}$ (up to overall constants) so that the operator
\beq
\tilde{H}_1 \mapsto \sum_{v \in S_1} \sum_{\alpha \neq \beta} L_+^{s^2_\alpha}(p_v^\alpha)  L_+^{s^2_\beta}(p_v^\beta) + \dots \, ,
\eeq
where the dots represent other terms generated under the mapping. We see that $\tilde{H}_1$ contains precisely the operator responsible for condensing $s^2_\alpha s^2_\beta$ flux-loop pairs on the 1-strata and is included in the 1-strata Hamiltonian Eq.~\eqref{eq:1strat} discussed earlier. Thus, going to the representation basis allows us to explicitly show that condensing pairs of $s^2$ flux loops on the 1-strata confines single $\sigma_\alpha$ charges as well $\sigma_\alpha \sigma_\beta$ pairs but allows $\sigma_1 \sigma_2 \sigma_3 \sigma_4$ to condense on $S_1$, which in turn prevents $\sigma$ particles from leaving their 3-strata. Hence, the $D_4$ defect network hosts non-Abelian $\sigma$ charges which are immobile in isolation \textit{i.e.,} non-Abelian fractons.

\end{document}